\documentclass[aps,twoside,amsmath,amssymb,longbibliography]{revtex4-1}

\usepackage[margin=1in]{geometry} 
\usepackage{graphicx} 
\usepackage{dcolumn} 
\usepackage{bm} 
\usepackage{hyperref} 
\usepackage{setspace} 

\begin{document}

\title{Input-output analysis of Mach 0.9 jet noise}

\author{Jinah Jeun}\thanks{Corresponding author: jeunx002@umn.edu}
\author{Joseph W. Nichols}
\affiliation{Department of Aerospace Engineering and Mechanics, University 
of Minnesota, Minneapolis, MN 55455, USA}
\date{\today}

\setstretch{2}

\begin{abstract}
We examine amplifying behavior of small perturbations about Reynolds-averaged 
Navier-Stokes solutions (RANS) of a $M_j = 0.9$ subsonic turbulent jet 
using input-output (I/O) analysis.  Singular value decomposition (SVD) of the 
resolvent of the linearized Navier-Stokes (LNS) equations forms an 
orthonormal set of I/O mode pairs, sorted in descending order by 
the magnitude of the corresponding singular values.  In this study we design 
a filter to restrict input forcings to be active only in regions where 
turbulent kinetic energy (TKE) of jet turbulence is high.  For an output 
domain, we directly implement the Ffowcs Williams-Hawkings (FWH) method 
within I/O analysis framework to measure far-field sound of 
observers distributed along an arc.  In this way we find that the 
resulting TKE-weighted input modes captures coherent structures in the 
near-field of turbulent jets.  Optimal input modes correspond to wavepackets 
represented by asymmetric pseudo-Gaussian envelope functions at a given 
forcing frequency.  These wavepackets remain similar in shape over a range 
of frequencies for $St > 0.5$, and scale as $St^{-0.5}$.  While the optimal 
mode is a wavepacket, sub-optimal modes represent decoherence of the optimal 
input mode.  By projecting high-fidelity large eddy simulation (LES) data 
onto the basis of input modes, we find that input modes do indeed capture 
the acoustically relevant dynamics in the jet.  The far-field acoustics 
predicted by the LES are recovered using only a few number of I/O modes.
\end{abstract}

\maketitle

\section{Introduction}
\label{sec:intro}
Since Lighthill's pioneering work~\cite{lighthill1952}, researchers
have devoted more than 60 years to identifying the aerodynamic sources
of sound in turbulence.  Lighthill's acoustic analogy rearranges the
exact Navier-Stokes equations into the form of an inhomogeneous wave
equation.  The nonlinearities of the Navier-Stokes equations are
lumped into the forcing terms on the right hand side of this
equation. Corresponding to the acoustic sources in turbulent flows,
these terms arise from deterministic nonlinear dynamics, but they are
often studied only in terms of their
statistics~\cite{curle1955,phillips1960,ffowcs1963,lilley1974,
  goldstein2003}.  In this framework, the acoustic analogy relates the
statistics of the acoustic sources to the statistics of the far-field
sound.  This suggests that the sources of sound in turbulence may be
viewed as stochastic and fine-scale: a quasi-random collection of
point sources. In the early 1960s, however, Mollo-Christensen observed
highly-organized vortical flow features within seemingly chaotic
turbulent flow in jets~\cite{mollo1963,mollo1967}.  His work inspired
fellow researchers to model aerodynamic sound generation associated
with turbulence in terms of spatio-temporally coherent flow
structures, modulated by growing and decaying
envelopes~\cite{crow1971,michalke1970,michalke1972,mattingly1974,
  crighton1976,michalke1984,tam1984,suzuki2006}.  In recent years it
has been found that such organized structures, now more popularly
known as wavepackets, are closely connected to instability waves in
jets~\cite{jordan2013}.  

Experiments
\cite{crow1971,fuchs1972,fuchs1972b,seiner1974,armstrong1977,
  moore1977,suzuki2006,reba2010,cavalieri2012} and
simulations~\cite{freund2001, suzuki2006,reba2010,cavalieri2012} have
confirmed the presence of wavepackets in the near-field of high-speed
turbulent jets.  For a given a base flow, wavepackets can be
efficiently computed using reduced-order methods based on the
parabolized stability equations (PSE)~\cite{cheung2007,
  gudmundsson2011,rodriguez2013,sinha2014,jordan2013}, global mode
analysis~\cite{nichols2011}, and optimal forcing
approaches~\cite{schmid2001, garnaud2013,nichols2014,jeun2016}.  In
addition, several studies have developed theoretical models of
wavepackets based on experimental and numerical
data~\cite{michalke1977,crighton1990,reba2010,cavalieri2012,
  baqui2015,serre2015}.  The theoretical models use analytic envelope
functions to modulate an underlying instability wave.  Both the
compactness and the asymmetry of the wavepacket envelope affect its
acoustic efficiency and directivity~\cite{serre2015}.

In contrast to the fine-scale view of acoustic sources, wavepackets 
represent large-scale, coherent sources of sound.  Given these two 
different views of aerodynamic production of sound in jets, it is natural to 
suppose that both mechanisms (large-scale and fine-scale) may be active 
simultaneously in high-speed jets.  In support of this, experimental 
measurements of far-field acoustic spectra of jets over a wide range of 
operating conditions seem to be well-represented by the superposition of two 
similarity spectra~\cite{tam1996}.  Acoustic radiation at small angles to 
the jet follows the Large-Scale Similarity (LSS) spectrum, whereas sideline 
noise radiation aligns with the Fine-Scale Similarity (FSS) spectrum. 
Compared to the LSS spectrum, the FSS spectrum tends to have lower peak 
levels, but is active over a broader range of frequencies.  It is important 
to note, however, that the LSS/FSS spectra are constructed only from the 
statistics of the far-field acoustics, and thus are not directly connected 
to the near-field dynamics.  Even within the statistical framework, however, 
refined statistical models of jet acoustic sources that account for 
decoherence and acoustic interference at high radiation angles reveal that 
both the LSS and FSS spectral shapes can be recovered from a single source 
mechanism~\cite{karabasov2010}.  Compared to the two-source model, this 
represents a significant reduction in complexity.  

It has recently been shown, however, that a dynamical model based on
stochastic similarity wavepackets can match experimental measurements
of a Mach 0.9 over a large range of frequencies and observer
angles~\cite{papamoschou2011}.  Compared to statistical models of jet
noise, stochasticity enters into this model in just six places, yet
further reducing the dimensionality and complexity of the acoustic
source system.  Moreover, the stochastic wavepacket model offers
physical insight into the dynamical mechanisms responsible for sound
production.  On the other hand, while the stochastic similarity
wavepacket model matches experimental measurements, the analysis
starts on a surface in the near-field surrounding the jet, and so is
not directly connected to the jet turbulence.  The analysis assumes
the near field of a jet can be modeled as a superposition of
wavepackets appearing and disappearing intermittently in time and
space.  Furthermore, the model is based on a choice of the functional
shape of these wavepackets, although wavepackets at different
frequencies are related to one another by a similarity variable.

In the current paper, we address these issues by applying I/O 
analysis to connect far-field acoustics directly to the 
turbulence-containing regions of a Mach 0.9 jet.  Our previous study showed 
that for supersonic jets, I/O analysis recovers the downstream peak jet noise 
radiation associated with the LSS spectrum, which PSE calculations also 
predict~\cite{jeun2016}.  Our analysis also showed that sideline noise can 
be recovered from a small number of sub-optimal I/O modes, not predicted by 
PSE-based methods.  The fact that sideline noise can be generated by a small 
number of modes suggests that it is generated by large-scale flow features, 
rather than fine-scale flow features that the FSS spectrum assumes.  This 
suggests that jet noise at all observer angles may be connected to 
large-scale flow features, and explains why the stochastic wavepacket model 
works as well as it does.

In the analysis presented below, we examine I/O analysis in the context of 
wavepacket models of jet noise.  Our analysis depends crucially on a careful 
description of how forcing from turbulence drives a linear systems model of 
the jet dynamics, and of how acoustic outputs are measured.  While our 
analysis is based on resolvents, our I/O formulation allows us to 
investigate only those dynamics in the jet that are connected to significant 
noise radiation.  This distinguishes our analysis from other resolvent-based 
approaches~\cite{schmid2001,garnaud2013} which focus upon modes that are 
energetically important in terms of the jet aerodynamics, but which may or 
may not radiate significant noise.  This distinction is important, as it is 
well known that only a small fraction of the total fluctuation energy in a 
jet is radiated as sound~\cite{dowling1983,jordan2013}.

Focusing in this way upon the acoustically dominant dynamics, the
question of physical interpretation then remains: are the acoustically
important dynamics organized into large-scale modes, or can they only
be described by fine-scale fluctuations?  We are specifically
interested in quantifying the dimensionality (or rank) of the
acoustically important dynamics.  Can these dynamics be represented by
a reduced-order model as the stochastic similarity wavepacket model
suggests?  In this paper, we address these questions by comparing the
results of our I/O analysis to high-fidelity simulations to show that
acoustic source dynamics are indeed low-rank.  As a rigorous test, we
focus upon a Mach 0.9 jets, including its sideline noise, which
traditionally is a case that has posed the largest challenge for
PSE-based methods.  Moreover, we show that the low-rank acoustics
source dynamics are organized into similarity wavepackets, even for
sound emitted at large angles to the jet axis.  This provides physical
justification for the theoretical stochastic similarity wavepacket
model of jet noise.

The rest of this paper is organized as follows.  In
Sec.~\ref{sec:methods} we define a base flow and governing equations.
We also describe a hybrid input-output/Ffowcs Williams-Hawkings
(I/O-FWH) method and its numerical parameters in the same section.  In
Sec.~\ref{sec:results} we compare our I/O results to previous
wavepacket models.  We show that the optimal input modes obtained from
I/O analysis are in fact self-similar over a range of frequencies.  In
addition to the optimal modes, the physical origin of sub-optimal
modes is also discussed.  In Sec.~\ref{sec:discussion}, we project
high-fidelity large eddy simulation data onto a set of input modes.
This provides further insight into the physical meaning of our input
modes, and suggests that a particular type of jitter may be
responsible for a significant portion of the sound production in
subsonic jets.  Our conclusions are discussed in
Sec.~\ref{sec:conclusions}.

\section{Methodology}
\label{sec:methods}
\subsection{Base flow}
We consider a Mach 0.9 isothermal, round, turbulent jet matching the
conditions of the experiment used as a basis for the stochastic
similarity wavepacket model~\cite{papamoschou2011}.  Isothermal Mach
0.9 jets have also been the subject of several other experimental and
numerical
investigations~\cite{shur2005,cavalieri2011,cavalieri2011b,bres2015}.
Like supersonic jets, Mach 0.9 jets support instability waves and
wavepackets that can be predicted by stability analysis based on
either the PSE~\cite{jordan2014} or the linearized Euler
equations~\cite{jeun2016}.  In supersonic jets, these wavepackets are
directly connected to the acoustic far-field through the mechanism of
Mach wave radiation~\cite{tam1995}.  In subsonic jets, however,
wavepackets are not as directly connected to the far-field.  In
particular, a subsonic wavepacket does not radiate sound at its
associated peak wavenumber.  In wavenumber space, acoustic radiation
comes instead from the supersonic ``tail" of the
wavepacket~\cite{jordan2013}.  This depends on the modulation and
asymmetry of the wavepacket envelope in physical
space~\cite{serre2015}.  The result is that, in isolation, wavepackets
seem to underpredict the sound generated by subsonic jets.  This
underprediction is partially rectified by introducing either ``jitter"
or a decoherence scale into the model.  The stochastic similarity
wavepacket model incorporates these effects through a randomized
superposition of coherent wavepackets, which, by fitting, can closely
reproduce experimentally measured far-field acoustic spectra. Applying
I/O analysis, we found acoustic source terms may be linked to
several sub-optimal modes in addition to the
wavepacket~\cite{jeun2016}.  The sub-optimal modes were especially
important for subsonic jets.  For this reason, we focus upon a Mach
0.9 jet for the remainder of the paper.  We will provide an
interpretation of the physics behind optimal and sub-optimal acoustic
sources predicted by I/O analysis and use this to explain the
success of the stochastic similarity wavepacket model.
 
\begin{figure}
  \centering
	\includegraphics[width=14cm]{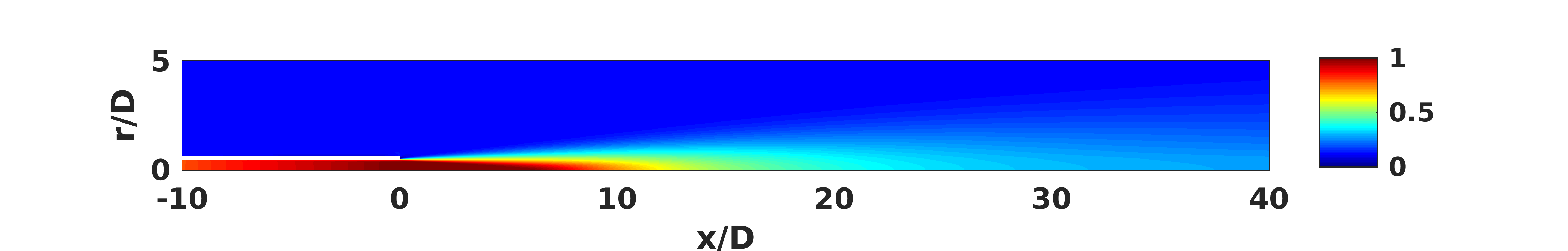} \\
  \caption{Contours of the real part of axial velocity of RANS solutions 
	for the Mach 0.9 subsonic jet. The 	velocity contours are normalized by 
	the velocity at the nozzle exit ($x/D = 0$).}
  \label{fig:base}
\end{figure}

Figure~\ref{fig:base} shows a Reynolds-averaged Navier-Stokes (RANS) solution
of the jet computed using a modified $k - \varepsilon$ turbulence model for 
high-speed jets~\cite{thies1996} as a base flow.  The flow exhausts from a 
straight cylindrical nozzle with inner diameter $D = 1$ and finite 
thickness $t = 0.1 D$.  Along the wall, no-slip boundary conditions are 
employed, allowing boundary layers to grow as the flow travels downstream.  
The upstream boundary conditions are chosen to produce the desired jet exit 
velocity, pressure, and temperature at $x/D = 0$.  In terms of the jet 
diameter $D$, the jet Reynolds number 
$Re = \rho_j u_j D / \mu = 2 \times 10^5$, where $\mu$ is the constant dynamic 
viscosity across the full numerical domain.  Here, $\rho_j$ and $u_j$ are 
respectively the density and the velocity at the nozzle exit.  Throughout this 
paper, the geometry and properties of the jet flow are scaled by the jet 
diameter $D$ and properties measured at the nozzle exit denoted by 
subscripts $j$.  While Fig.~\ref{fig:base} visualizes only part of it, the 
actual numerical domain extends from $x/D = -10$ to $x/D = 40$ and from 
$r/D = 0$ to $r/D = 25$ in the axial and radial directions, respectively.

\subsection{Linearized Navier-Stokes Equations}
In this study the dynamics of high-speed turbulent jet is described by the 
fully compressible Navier-Stokes equations (NSE) for the system state 
$\bm{q} = \left[p; \, \bm{u}^{T}; \, s\right]^T$, where $p$, 
$\bm{u}$ and $s$ are the fluid pressure, velocity, and entropy, 
respectively.  Scaled with respect to the jet diameter and jet exit 
properties, the NSE are written in dimensionless form as:
\begin{subequations}
\begin{align}
  \frac{\partial p}{\partial t} 
  + \bm{u} \bm{\cdot} \bm{\nabla} p 
	+ \rho c^2 \bm{\nabla} \bm{\cdot} \bm{u} 
	&= \frac{1}{Re}
	\left(\frac{1}{M_j^2 Pr} \bm{\nabla}^2 T 
	+ (\gamma-1)\Phi \right), \\
  \frac{\partial \bm{u}}{\partial t} 
  + \frac{1}{\rho} \bm{\nabla} p 
	+ \bm{u} \bm{\cdot} \bm{\nabla} \bm{u} 
	&= \frac{1}{Re} 
	\frac{1}{\rho}	\bm{\nabla} \bm{\cdot} \bm{\tau}, \\
	\frac{\partial s}{\partial t} 
  + \bm{u} \bm{\cdot} \bm{\nabla} s 
  &= \frac{1}{Re} \frac{1}{\rho T} 
	\left( \frac{1}{(\gamma-1)M_j^2 Pr} 
	\bm{\nabla}^2 T + \Phi \right),
\end{align}
\label{eq:nse}
\end{subequations}
and the equation of state for an ideal gas reads $\gamma M_j^2 p = \rho T$, 
where $\gamma = 1.4$ represents the constant ratio of specific heats.  Here, 
the Prandtl number is defined as $Pr = c_p \mu / \lambda$, where $c_p$ is 
the specific heat at constant pressure and $\lambda$ is the thermal 
conductivity, which are assumed to be constant throughout the computational 
domain.  Furthermore, $\Phi$ and $\tau$ denote the dissipation function and 
viscous stress function, respectively.  In the last equation the entropy 
$s$ is set to be zero when $p = 1$ and $T = 1$, by defining 
$s = \ln(T) / \left( (\gamma-1) M_j^2 \right) - \ln(p)/\left( \gamma M_j^2 
\right)$ (see also~\cite{nichols2009,nichols2011}).

Using the standard Reynolds decomposition of the system state such as 
$\bm{q} = \overline{\bm{q}} + \bm{q}'$, where $\overline{\bm{q}}$ represents 
the mean flow part and $\bm{q}'$ is the fluctuating part, and keeping the 
first-order terms only, we write the linearized Navier-Stokes (LNS) equations 
as the governing equations for small fluctuations about which given base 
flows as:
\begin{subequations}
\begin{align}
  &\frac{\partial p'}{\partial t} +
  \bm{\bar{u}} \bm{\cdot} \bm{\nabla} p' +
	\bm{u'} \bm{\cdot} \bm{\nabla} \bar{p} +
	\bar{\rho} \bar{c}^2 \bm{\nabla} \bm{\cdot} \bm{u'} +
	\gamma (\bm{\nabla} \bm{\cdot} \bm{\bar{u}}) p' =
	\nonumber \\
	&\hspace{3mm} \frac{1}{Re}
	\left[ \frac{1}{M_j^2 Pr} \lambda \bm{\nabla}^2 T' + 
	(\gamma-1)\big( \bm{\bar{\tau}} \bm{:} \bm{\nabla} \bm{u'} +
	\bm{\tau'} \bm{:} \bm{\nabla} \bm{\bar{u}} \big) \right], 
	\label{eq:lns_p} \\[8pt]
  &\frac{\partial \bm{u'}}{\partial t} +
  \frac{1}{\bar{\rho}} \bm{\nabla} p' +
	\frac{\rho'}{\bar{\rho}} \bm{\nabla} \bar{p} +
	\bm{\bar{u}} \bm{\cdot} \bm{\nabla} \bm{u'} +
	\bm{u'} \bm{\cdot} \bm{\nabla} \bm{\bar{u}} =
	\frac{1}{Re} \frac{1}{\bar{\rho}} 
	\left( \bm{\nabla} \bm{\cdot} \bm{\tau'} -
	\frac{\rho'}{\bar{\rho}}\bm{\nabla} \bm{\cdot} \bm{\bar{\tau}} \right), 
	\label{eq:lns_u} \\[8pt]
  &\frac{\partial s'}{\partial t} +
  \bm{\bar{u}} \bm{\cdot} \bm{\nabla} s' +
	\bm{u'} \bm{\cdot} \bm{\nabla} \bar{s} =
	\nonumber \\
  &\hspace{3mm}\frac{1}{Re} \frac{1}{\bar{\rho}\bar{T}} 
	\bigg[ \frac{1}{(\gamma-1)M_j^2 Pr} 
	\bigg( \lambda \bm{\nabla}^2 T' -
	\frac{p'}{\bar{p}} \lambda \bm{\nabla}^2 \bar{T} \bigg) +	
	\bm{\bar{\tau}} \bm{:} \bm{\nabla} \bm{u'} +
	\bm{\tau'} \bm{:} \bm{\nabla} \bm{\bar{u}} -
	\frac{p'}{\bar{p}} \bm{\bar{\tau}} \bm{:} \bm{\nabla} \bm{\bar{u}} \bigg]. 
	\label{eq:lns_s}
\end{align}
\end{subequations}

In summary, the resulting linear system is written in matrix form as:
 \begin{equation}
  \frac{\partial \bm{q}}{\partial t} = A \bm{q},
  \label{eq:matrix}
\end{equation}
where the linear operator $A$ corresponding to the governing equations is 
determined uniquely by a base flow.

\subsection{Input-output analysis}
\label{subsec:io}
The high-speed turbulent jets we consider in the present paper are
stable in a global sense but are unstable to convective perturbations.
Such systems behave as selective noise amplifiers of external
perturbations and thus, are best studied by analyzing their
sensitivity to external forcing.  I/O analysis achieves this
by adding an external forcing term $\bm{f}$ to the original linear
system~\eqref{eq:matrix} as follows:
\begin{subequations}
\label{eq:sys}
\begin{align}
  \dot{\bm{q}} &= A \bm{q} + B \bm{f}, \\
  \bm{y} &= C \bm{q},
\end{align}
\end{subequations}
where $\bm{y}$ represents output quantities, $B$ and $C$ are matrices 
determined depending on inputs and outputs of interests, respectively.  For 
example, we specify $B$ to select forcings applied to the velocity equations 
near the jet turbulence while $C$ is chosen to specify noise in the region 
far way from sources.  In this way our analysis investigates how input 
forcings map onto output quantities of interests.  Specifying matrices $B$ 
and $C$ makes our analysis unique compared to other approaches which 
consider the entire system state to evaluate the gain.  

\begin{figure}
  \centering
	\includegraphics[width=8cm]{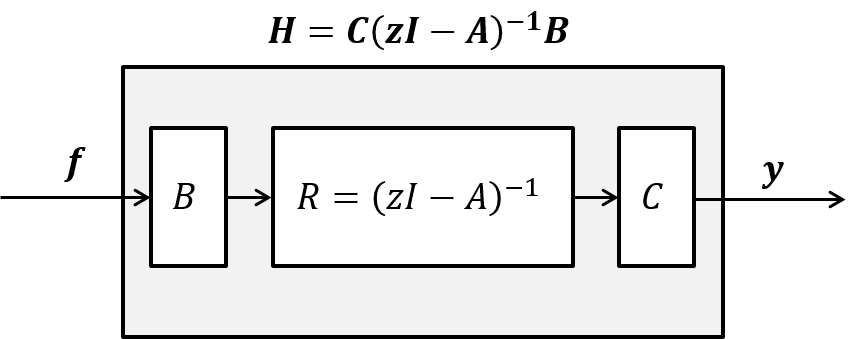} \\
  \caption{A schematic representation of the linear system in 
	Eq.~\eqref{eq:sys} using the transfer function $H$.}
  \label{fig:sys}
\end{figure}

To system (\ref{eq:sys}), we apply the wavepacket ansatz as:
\begin{equation}
  \bm{q}(x, r, \theta, t) = 
	\hat{\bm{q}}(x,r)e^{\mathrm{i}(m\theta-\omega t)},
\end{equation}
where $m$ is an azimuthal wavenumber and $\omega$ is a temporal
frequency.  By assuming similar harmonic forcing functions in this
way, we find a transfer function $H$ from inputs to outputs that
consists of a resolvent operator $R = (z I - A)^{-1}$ as:
\begin{equation}
  y = C (z I - A)^{-1} B f = H f,
\end{equation}
where $z = -\mathrm{i}\omega$.  This is schematically described in 
Fig.~\ref{fig:sys}.  Singular value decomposition of the transfer function 
$H$ forms a set of pairs of input and output vectors, which respectively 
correspond to columns of matrices $V$ and $U$ in:
\begin{equation}
  H = U \Sigma V^*,
\end{equation}
where the superscript $^*$ indicates the complex-conjugate transpose.  
Furthermore, the matrix $\Sigma$ is a diagonal matrix containing a set of 
singular values.  Each singular value represents the amplitude gain from the 
corresponding input to output vectors that is defined as:
\begin{equation}
  \sigma = \frac{\left\| \bm{y} \right\|}
	{\left\| \bm{f} \right\|},
\end{equation}
where $\left\| \cdot \right\|$ denotes the $L^2$ norm.  

Finally, singular value decomposition of the transfer function may be computed 
through the eigen-decomposition of $H^\dagger H$: 
\begin{equation}
  H^\dagger H = B^+ (z^*I-A^\dagger)^{-1} C^\dagger C (zI-A)^{-1} B, \\
\end{equation}
where $H^\dagger$ represents the adjoint of the transfer function, which maps 
outputs back onto inputs.  Hence, the eigenvalues of $H^\dagger H$ are 
squares of the corresponding singular values of $H$.  For more details, 
please refer to~\cite{jeun2016}.

\subsection{TKE-weighted input modes}
\label{subsec:io_tke}
\begin{figure}
  \centering
  \includegraphics[width=14cm]{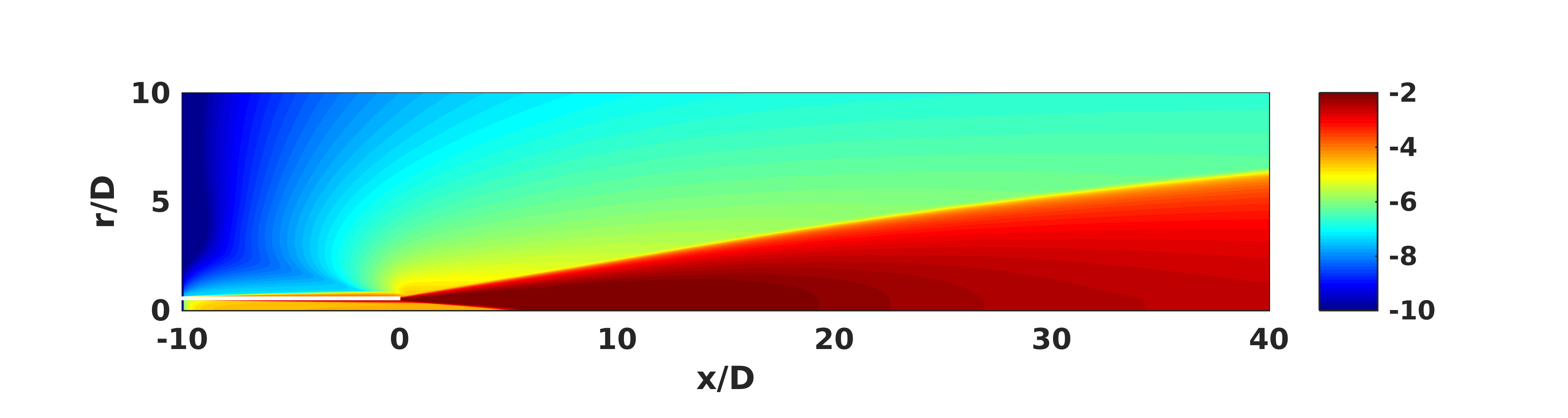}
  \caption{Contours of the TKE for the $M_j = 0.9$ subsonic jet in 
	logarithmic scale normalized by the square of the	jet exit velocity.}
  \label{fig:tke}
\end{figure}

In high-speed jets, the sources of sound are embedded in the jet
turbulence.  Specifically, the nonlinear terms driving our system
(i.e., the Lighthill sources) are proportional to the fluctuating
Reynolds stress tensor.  To model this, we weight input forcings by
the turbulent kinetic energy (TKE) as shown in Fig.~\ref{fig:tke},
which is evaluated using the same RANS model used in computing
the base flow.

\begin{figure}
  \centering
	\includegraphics[width=14cm]{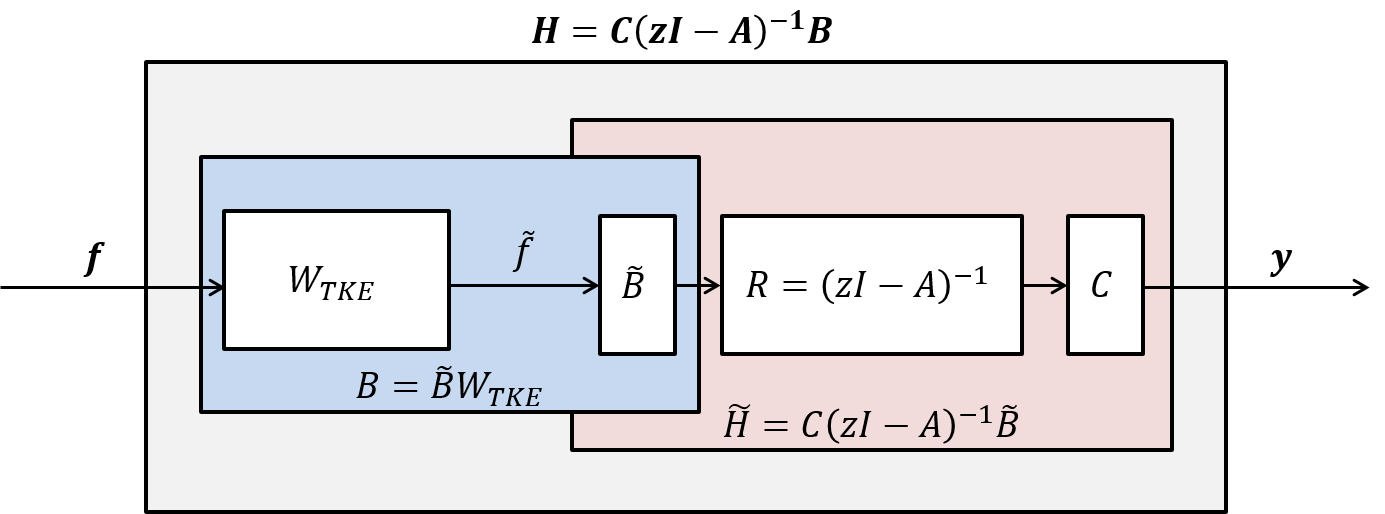} \\
  \caption{A schematic representation of the modified system with the TKE 
	weighting matrix indicated by $W_{TKE}$.}
  \label{fig:sys_tke}
\end{figure}

A schematic description of the modified system with TKE-weighted input 
forcings is given in Fig.~\ref{fig:sys_tke}.  Here, the TKE-weighting filter 
denoted by $W_{TKE}$ is applied to input forcings $\bm{f}$, yielding the 
TKE-weighted input forcings $\bm{\tilde{f}}$ written as:
\begin{equation}
  \bm{\tilde{f}} = W_{TKE} \bm{f}.
  \label{eq:f_tke}
\end{equation}
The resolvent $R$ remains unchanged since it uniquely depends on the base 
flow.  The matrix $C$ also remains the same as before.  In this way the 
original, unweighted system in Eq.~\eqref{eq:sys} becomes:
\begin{subequations}
\begin{align}
  \dot{\bm{q}} &= A \bm{q} + B \bm{f} = 
	A \bm{q} + \tilde{B} \bm{\tilde{f}}, \\
  \bm{y} &= C \bm{q},
	\label{eq:sys_tke}
\end{align}
\end{subequations}
where 
\begin{equation}
  \tilde{B} = B W_{TKE}^{-1}.
\end{equation}
Because we are interested in output produced by the weighted inputs 
$\bm{\tilde{f}}$, we consider a new transfer function $\tilde{H}$ 
corresponding to the red box in Fig.~\ref{fig:sys_tke}.  This new transfer 
function $\tilde{H}$ is related to the transfer function $H$ in the 
original, unweighted system such that:
\begin{equation}
	\tilde{H} = C (z I - A)^{-1} \tilde{B} = H W_{TKE}^{-1}.
	\label{eq:H_tke}
\end{equation}
Consequently, by substituting Eqs.~\eqref{eq:f_tke} and~\eqref{eq:H_tke} 
into the Arnoldi iteration of $H^\dagger H$:
\begin{equation}
  \bm{f}_{new} = H^\dagger H \bm{f} 
  = W_{TKE}\tilde{H}^\dagger\tilde{H}W_{tke}\bm{f}
  = W_{TKE}\tilde{H}^\dagger\tilde{H}\bm{\tilde{f}},
\end{equation}
we obtain that:
\begin{equation}
  \bm{\tilde{f}}_{new} = W_{TKE}^2\tilde{H}^\dagger\tilde{H}\bm{\tilde{f}}.
\end{equation}
Note that without weighting by the TKE, i.e., when $W_{TKE} = I$, the system 
returns to the original, unweighted system described in Sec.~\ref{subsec:io}.  

\subsection{Hybrid I/O-FWH solver}
In this study, we focus upon the jet dynamics that generate the largest 
amount of far-field sound.  These dynamics may be different from those 
having the greatest energy.  The outputs $\bm{y}$ of interest are the 
far-field pressure of observers located along an arc.  Extending the 
numerical domain to the far-field, however, requires a huge amount of 
computational resources.  Instead, a projection method such as the FWH 
method may be employed to compute the far-field pressure from the near-field 
flow data.  Since we consider a linearized system, the source terms to the 
FWH equations are also linear.  We therefore treat them explicitly so that we 
can write the FWH formulation as a linear operator inside the matrix $C$.  
In this way the adjoint of the matrix $C$ takes far-field pressure 
fluctuations and maps these back onto state vectors on the FWH surface, 
which are then injected into the system.  Further details about the 
derivation of the FWH solver and two test cases are provided in the appendix.

\begin{figure}
  \centering
	\includegraphics[width=10cm]{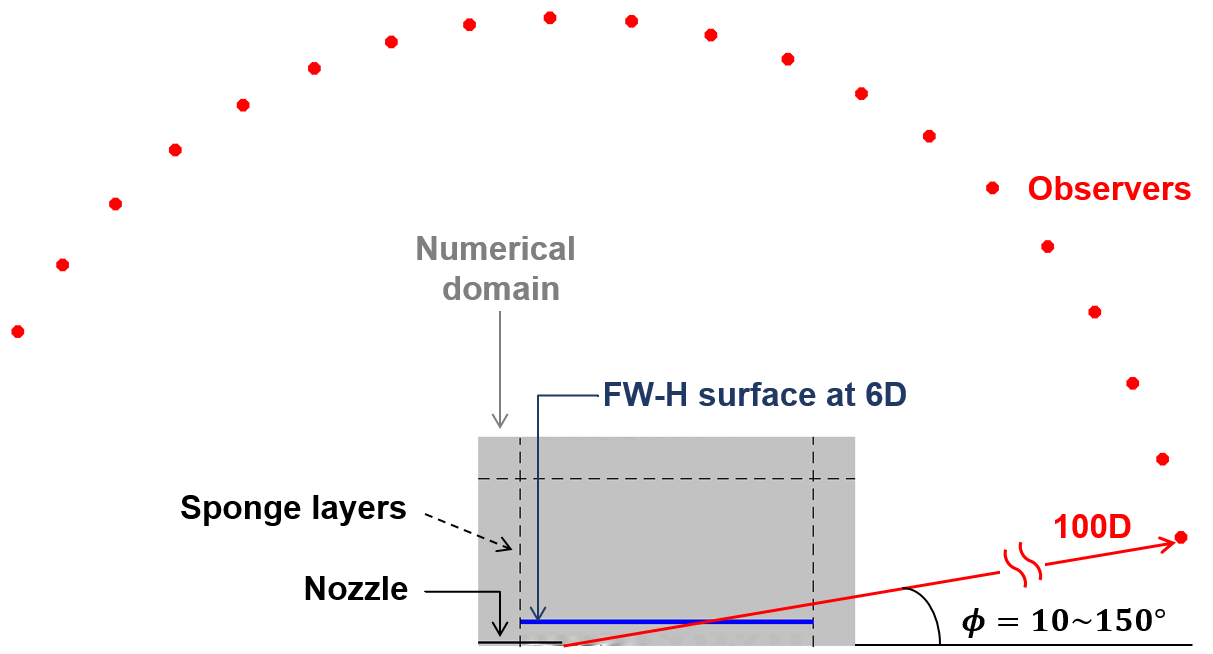}
  \caption{A schematic representation of the hybrid I/O-FWH 
	analysis method.  A straight cylindrical FWH projection surface is located 
	at $x/D = 6$, denoted by a blue straight line.  Red dots indicate 
	observers distributed uniformly along an arc in far-field from 	
	$\phi = 10^\circ$ to $\phi = 150^\circ$.}
  \label{fig:hybrid_fwh_io}
\end{figure}

Since the FWH solver is linearized, a projection surface should be
placed in purely acoustic region that is sufficiently far from the jet
turbulence.  To ensure that we are in the linear regime in LES, we
select a straight cylinder with radius $r/D = 6$ whose axis lies along
the jet centerline as a projection surface.  Using this, we measure
the far-field pressure of probes uniformly distributed along an arc at
a distance of $100$ jet diameters from the nozzle exit.  The observer
angles range from $\phi = 10^\circ$ to $\phi = 150^\circ$ with an
increment of $\Delta \phi = 1^\circ$, where the polar angle $\phi$ is
measured from the downstream jet axis.  These angles are chosen by
considering the extent of the FWH surface and sponge layers to prevent
outgoing waves from reflecting back.  The choice of arc then naturally
excludes regions outside of the sponge layers, avoiding spurious
modes.  A schematic of this hybrid I/O-FWH analysis is given
in Fig.~\ref{fig:hybrid_fwh_io}.

\subsection{Numerical methods}
We use fourth-order centered finite difference scheme to discretize the LNS 
on a stretched mesh.  Since the non-dissipative nature of the centered finite 
difference scheme may introduce unphysical waves for high wavenumbers, a 
scale-selective fourth-order numerical filter is applied to suppress only 
the highest wavenumbers supported by the mesh.  Furthermore, non-reflecting 
boundary conditions are satisfied by employing numerical sponge 
layers~\cite{khalighi2010,mani2012} at the upstream ($x/D = -5$), downstream 
($x/D = 35$), and lateral boundaries ($r/D = 20$).  
  
The convergence of I/O analysis was tested in an earlier 
study~\cite{jeun2016} for various grid resolutions.  As a trade-off between 
the computational cost and the accuracy of the solutions, we choose to use a 
mesh with $801$ and $288$ grid points in the axial and radial directions,  
respectively.  Whereas the grids are distributed uniformly in the axial 
direction, they are clustered in the radial direction along the nozzle 
lipline to resolve the boundary layer and thin shear layer formed near the 
nozzle lip.  

The eigen-decomposition of $\tilde{H}^\dagger\tilde{H}$ is performed 
efficiently using the implicitly restarted Arnoldi method (IRAM) implemented 
by the software package ARPACK~\cite{lehoucq1998}.  An inverse matrix of the 
resolvent operator is computed by the parallel SuperLU 
package~\cite{li2003}.  The adjoints are evaluated using the continuous 
adjoint approach, which derives the equations adjoint to the LNS first and 
discretizes them later to estimate the matrix $A^\dagger$.  In this way we 
obtain one-sided differences consistent with continuous derivatives near to 
nozzle walls.  Since $\tilde{H}^\dagger\tilde{H}$ is Hermitian, the resulting 
input modes are orthonormal to each other, and the Arnoldi method converges 
quickly.

\section{Results}
\label{sec:results}
\subsection{Input-output modes}
\label{sec:results_mode}
\begin{figure}
  \centering
  \includegraphics[width=10cm]{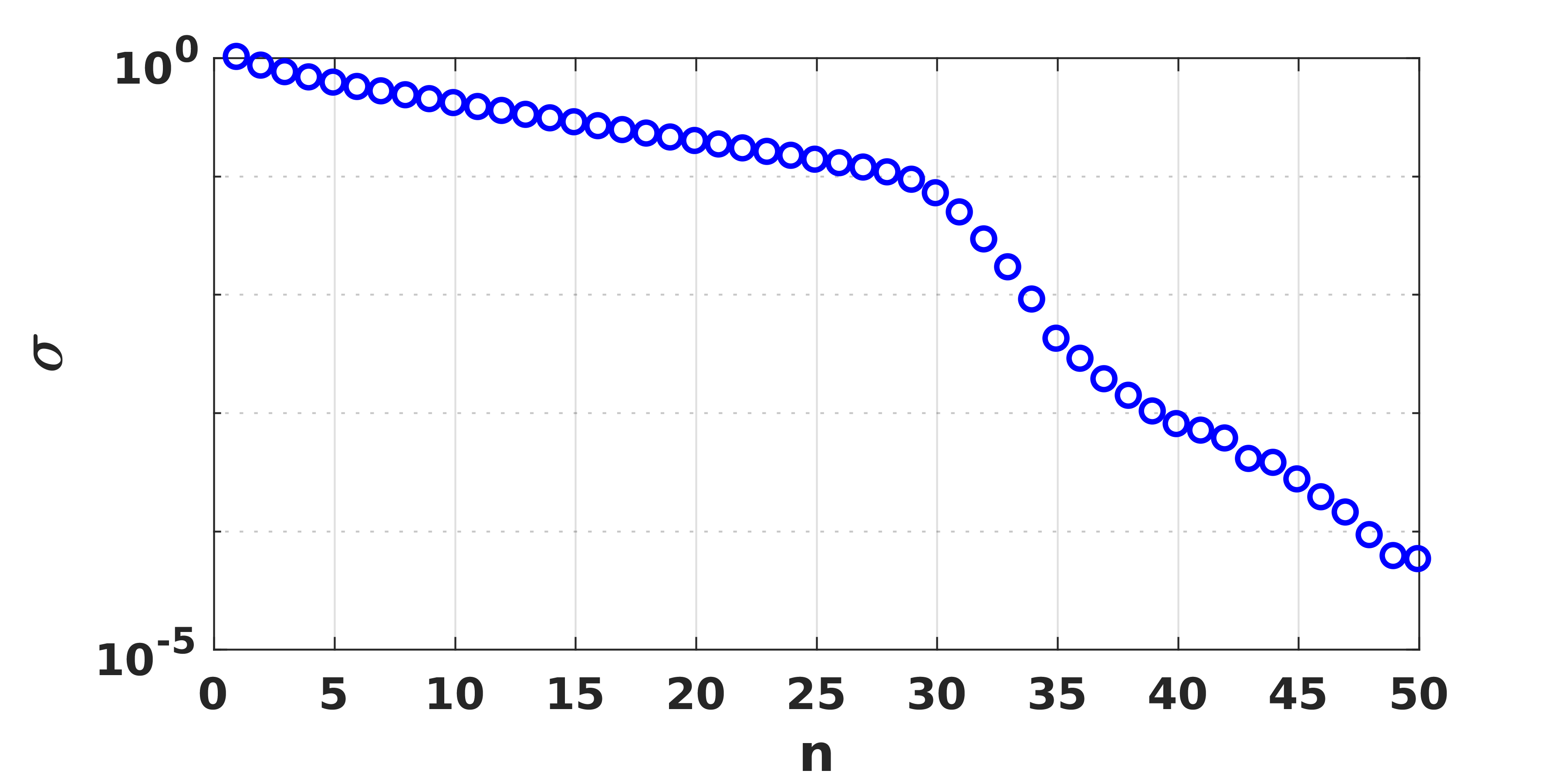}
  \caption{Singular values as a function of the mode number for the Mach 0.9 
  subsonic jet at forcing frequency $St = 0.59$.}
  \label{fig:gain}
\end{figure}

Applying the hybrid I/O-FWH methodology, we obtain a spectrum of
singular values representing gains between inputs and outputs, as
discussed in Sec.~\ref{subsec:io_tke}.  Figure~\ref{fig:gain} shows
the first 50 singular values, ordered by amplification factor, for the
Mach 0.9 jet at a forcing frequency of $St = 0.59$.  For the purposes
of this paper, which focuses upon the physical meaning of the results
of I/O analysis, it is sufficient to consider axisymmetric
disturbances only ($m = 0$).  Our I/O formulation can handle higher
azimuthal wavenumbers, however, which become important especially to
describe noise radiation in the sideline direction \cite{jeun2018}.
The first 29 singular values show a relatively slow decrease in
amplification factor, which means that the first 29 I/O modes produce
approximately the same amount of far-field noise per unit energy of
forcing.  This is consistent with previous results for subsonic jets,
where sub-optimal I/O modes were found to produce nearly the same
amplification as the optimal modes.  This is in contrast, however, to
supersonic jets where the optimal mode becomes
dominant~\cite{jeun2016}.

\begin{figure}
  \centering
	\begin{tabular}{c}
    \includegraphics[width=14cm]{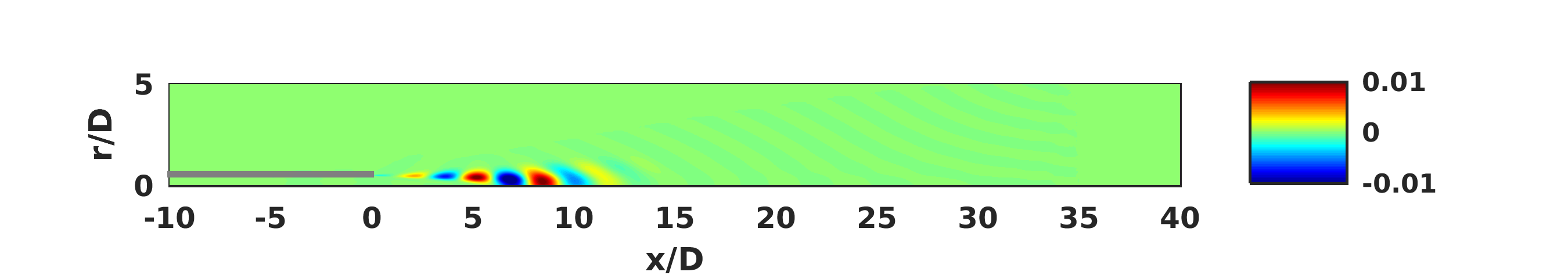} \\
	  (a) $n = 1$ \\
		\includegraphics[width=14cm]{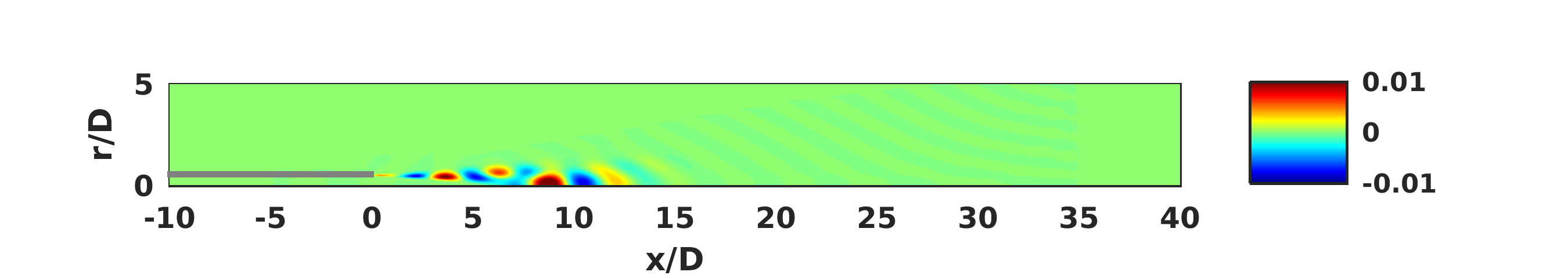} \\
		(b) $n = 2$ \\
		\includegraphics[width=14cm]{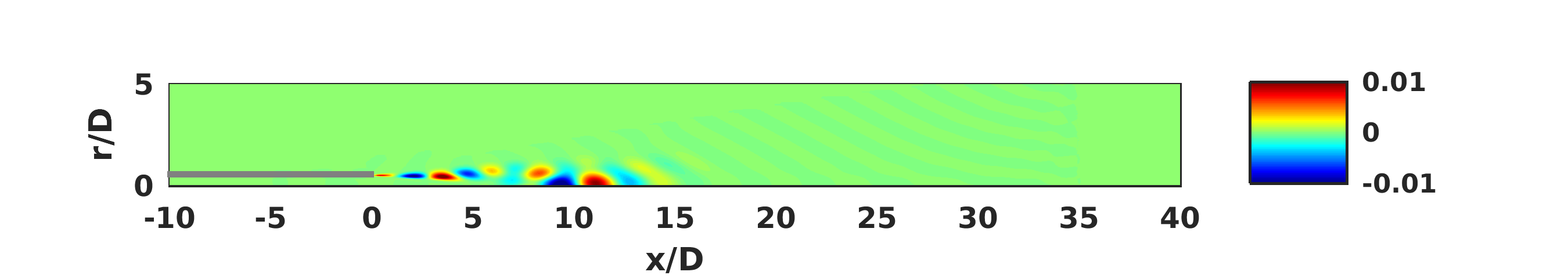} \\
		(c) $n = 3$ \\
		\includegraphics[width=14cm]{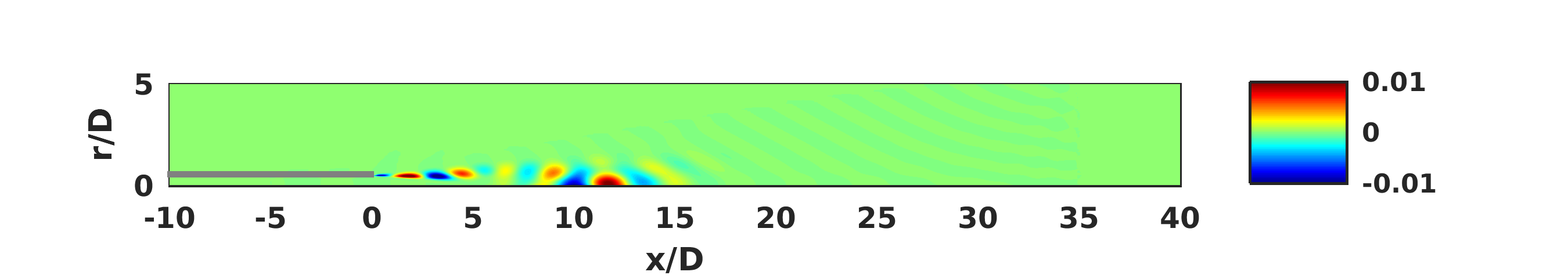} \\
		(d) $n = 4$ \\ 
  \end{tabular}
	\caption{The first four input modes of the Mach 0.9 jet for forcing 
	frequency $St = 0.59$.  Contours visualizes the real part of the normalized 
	axial velocity forcing.}
  \label{fig:input}
\end{figure}

\begin{figure}
  \centering
  \includegraphics[width=10cm]{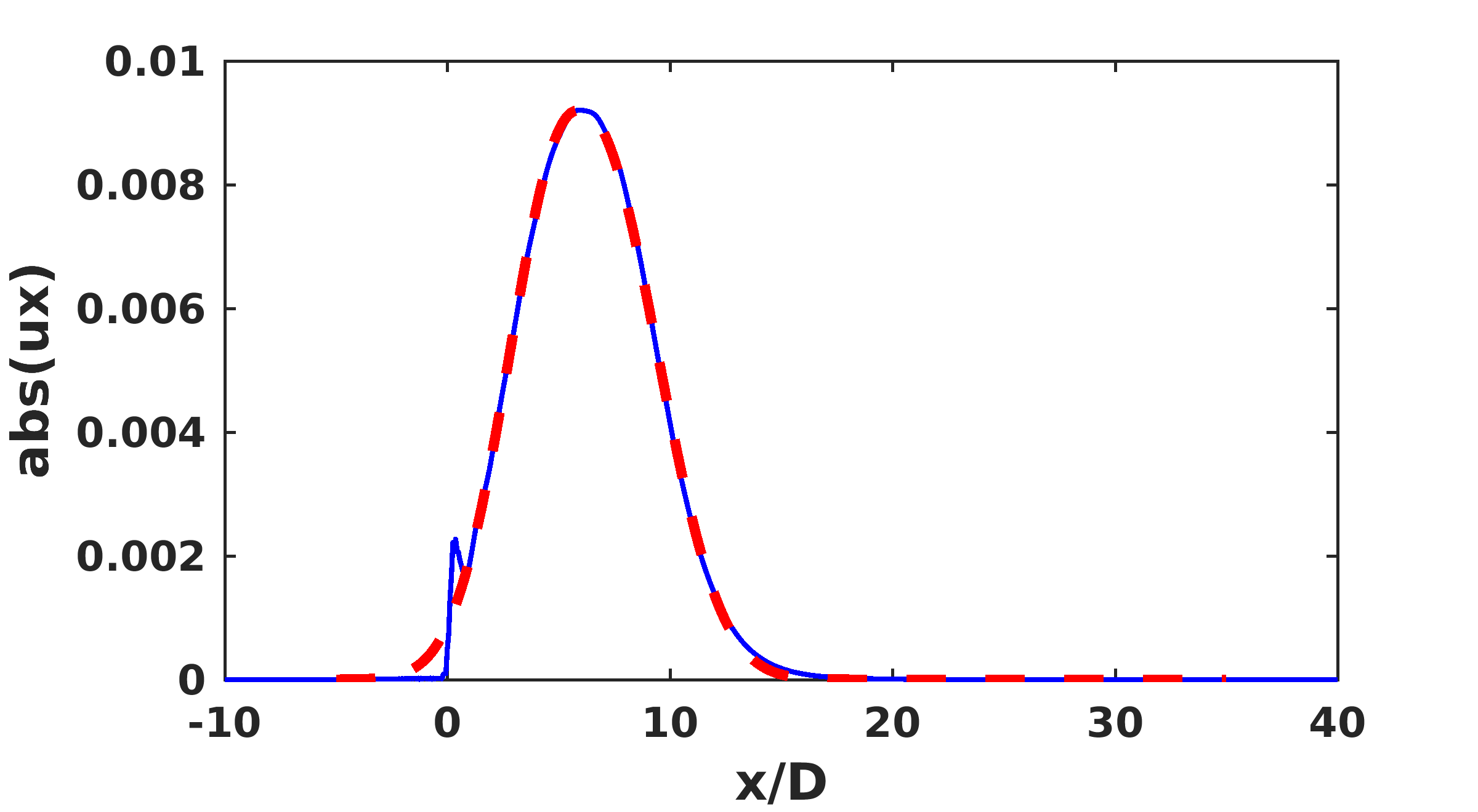}
  \caption{The lipline wavepacket measured at $r/D = 0.5$ (blue solid line) 
	modeled by an asymmetric pseudo Gaussian 	wavepacket (red dashed line) for 
	the optimal mode of the 	$M_j = 0.9$ subsonic jet at forcing frequency 
	$St = 0.59$.}
  \label{fig:lipline_wavepacket}
\end{figure}

Along with the gains, the singular value decomposition also produces 
orthogonal sets of corresponding input and output modes. 
Figure~\ref{fig:input} shows the first four input modes corresponding to the 
four largest singular values shown in Fig.~\ref{fig:gain}.  In 
Fig.~\ref{fig:input}, the gray rectangle represents the nozzle wall, which 
ends at $x/D = 0$.  Contours showing the real part of the axial velocity 
forcing reveal the input modes to have significant structure.  The optimal 
input mode ($n = 1$), in particular, is clearly a wavepacket. At this 
frequency, the wavepacket is centered close to the end of the potential core 
of the jet, and extends several diameters upstream along the jet shear 
layers as well as downstream along the jet centerline.  
To characterize the physics of this wavepacket source, we consider the
(complex) amplitude of the optimal input mode along the nozzle lipine
($r/D = 0.5$).  The blue solid line in
Fig.~\ref{fig:lipline_wavepacket} represents the absolute magnitude of
the x-component of the input forcing for the optimal mode.  Along this
slice, we find a wavepacket that peaks around the end of the potential
core (at $x/D = 6$).  Although the effect is subtle, this wavepacket
grows slightly faster along its upstream edge than it decays
downstream.  Such asymmetric wavepackets have been observed in
experiments and simulations and have been modeled theoretically by a
variety of different functions~\cite{michalke1977,crighton1990,
  reba2010,cavalieri2012,serre2015,baqui2015}.  One of the most
popular functional forms is the following asymmetric pseudo-Gaussian:
\begin{equation}
  u_x(x) = 
	\begin{cases}
  exp\left[-\left(\frac{x-b_1}{c_1}\right)^{p_1}\right], 
	& \text{if x $\leq b_1$}, \\
	exp\left[-\left(\frac{x-b_1}{c_2}\right)^{p_2}\right],
	& \text{if x $\geq b_1$}, \\	
	\end{cases}
  \label{eq:pseudo_Gaussian}
\end{equation}
where $b_1$ locates the peak of the wavepacket envelop, $c_1$ and
$c_2$ respectively determine widths of the amplifying and decaying
parts, and $p_1$ and $p_2$ represent the exponents, respectively.
Applying a nonlinear least squares fitting algorithm, we find that
$b_1 = 5.9375D, c_1 = 4.0599D, p_1 = 2.2545, c_2 = 4.5584D$, and $p_2
= 2.2517$ produce a pseudo-Gaussian curve that almost exactly matches
our wavepacket.  The fact that $p_1 > p_2$ means that the wavepacket
amplifies faster than it decays. This asymmetry is also indicated by
$c_1 < c_2$.  Also, because both $p_1$ and $p_2$ are approximately
equal to two, the shape of our wavepacket is nearly (but not quite)
Gaussian.  The shape of the wavepacket, and in particular its
asymmetry, are important factors determining its efficiency at
generating acoustic radiation~\cite{serre2015}.
 
While the leading input mode represents the optimal way to force a jet
to make noise, Figs.~\ref{fig:input}(b)-(d) show sub-optimal input
modes ($n = 2, 3, 4$), which produce nearly the same amount noise as
the optimal mode.  They are active along upstream and downstream edges
of the wavepacket associated with the optimal mode.  As the mode
number increases, the sub-optimal input modes progressively reach
further upstream and downstream.  To understand the pattern that they
follow, it is helpful to to examine the sound fields they produce.

\begin{figure}
  \centering
  \begin{tabular}{cc}      
    \includegraphics[width=6.5cm]{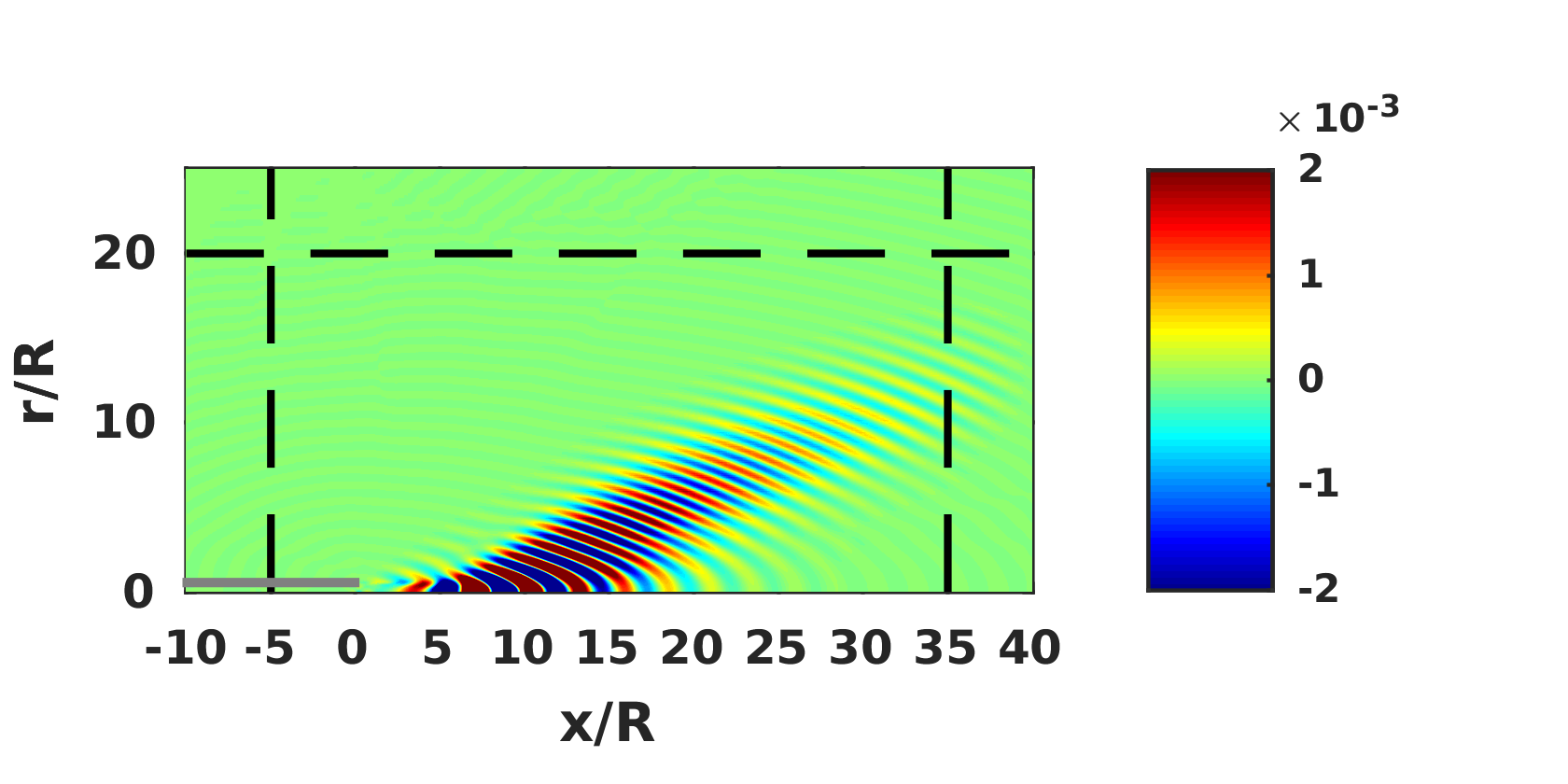} &
    \includegraphics[width=6.5cm]{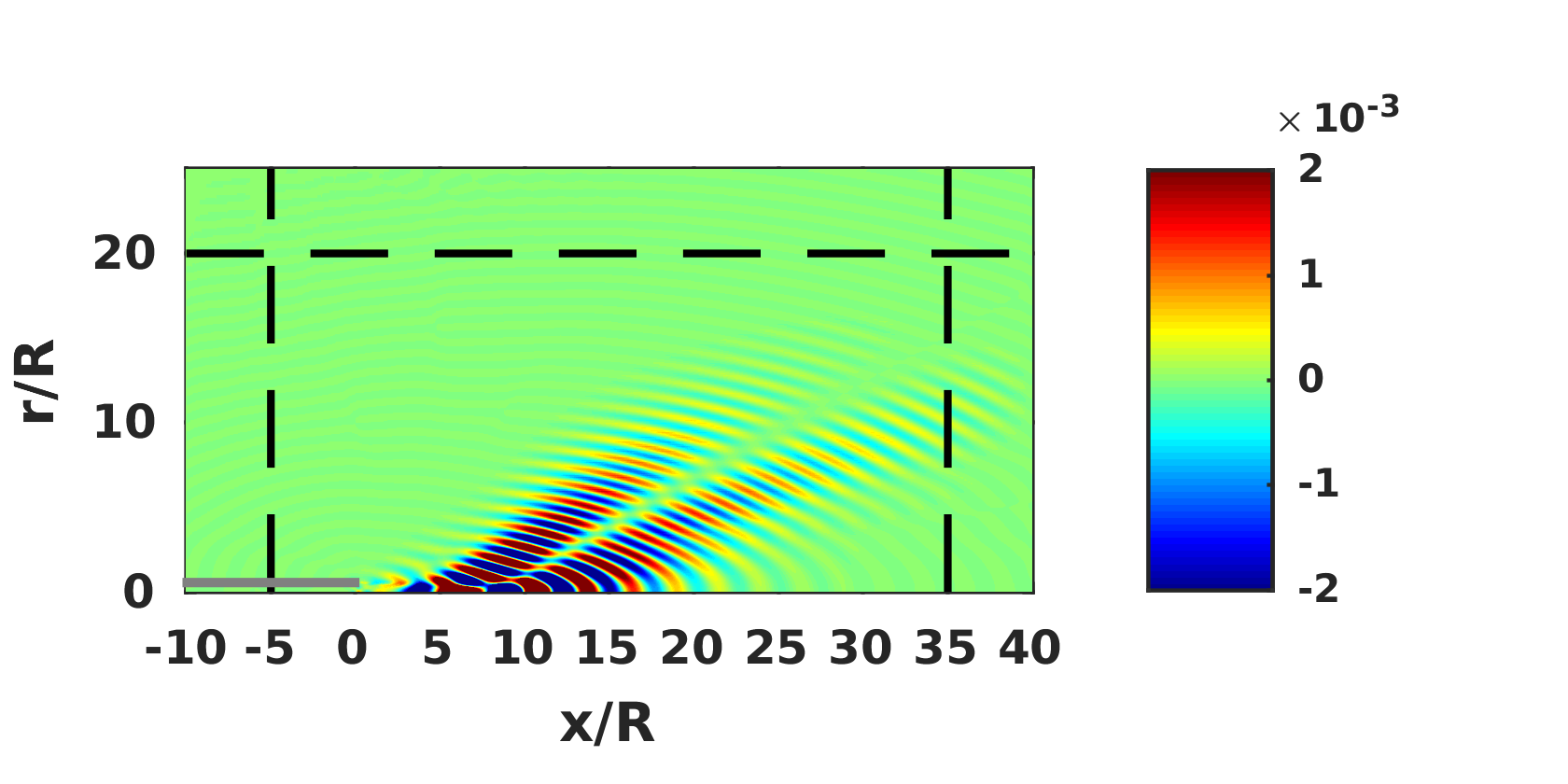} \\
    (a) $n = 1$ & (b) $n = 2$ \\ 
    \includegraphics[width=6.5cm]{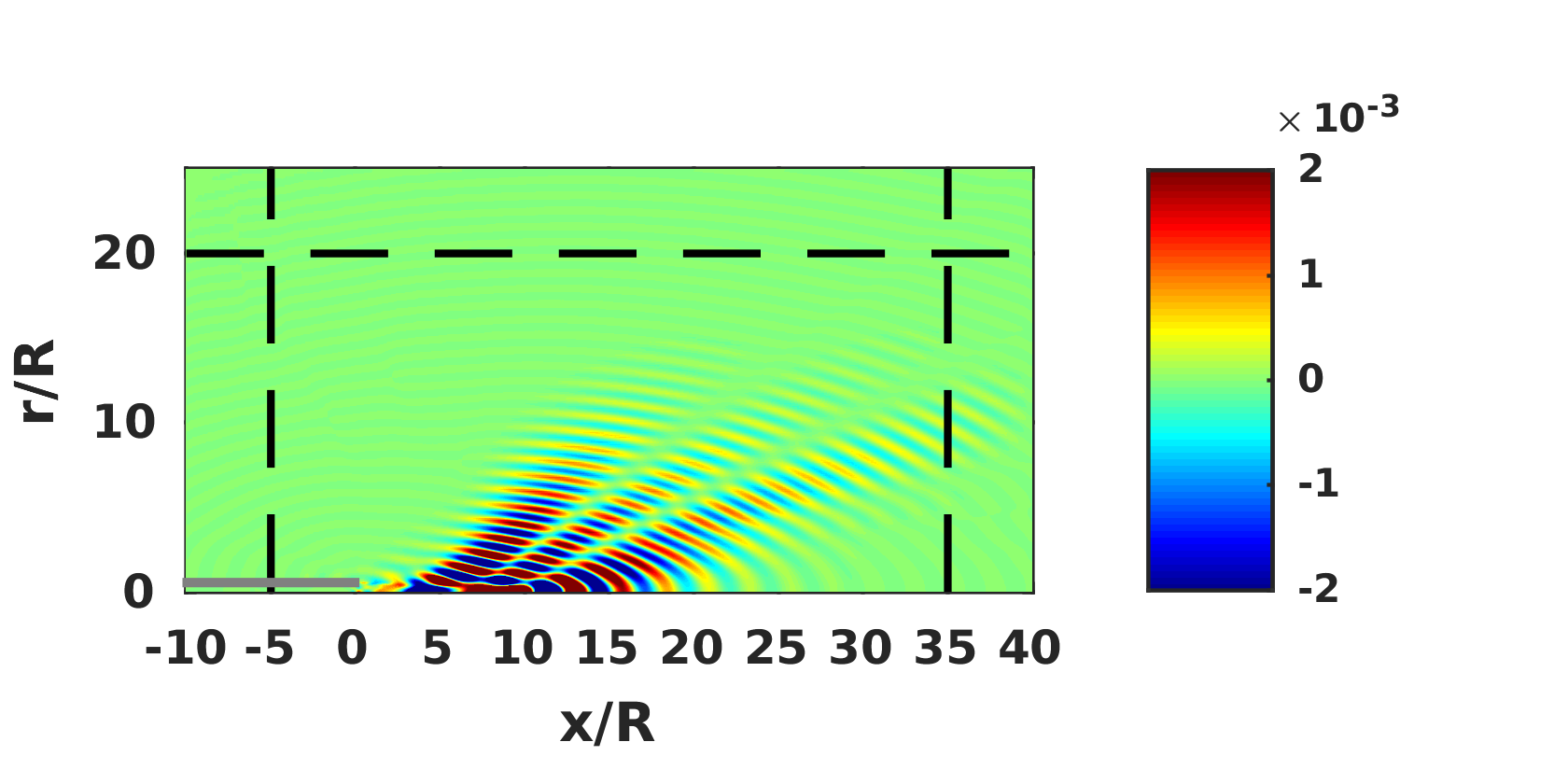} &
    \includegraphics[width=6.5cm]{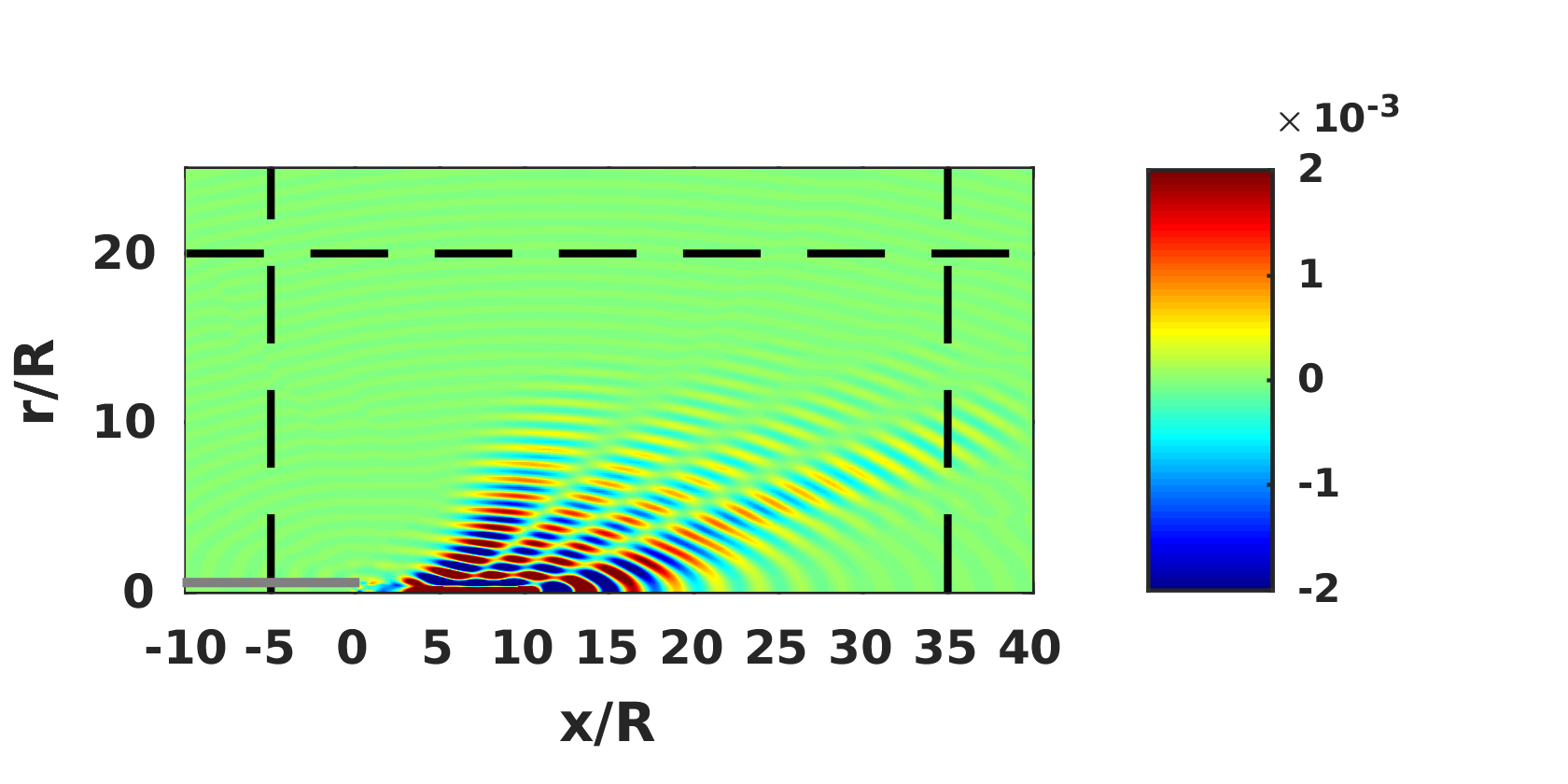} \\
		(c) $n = 3$ & (d) $n = 4$ \\ 
  \end{tabular}
  \caption{The first four unrestricted output modes of the Mach 0.9 subsonic 
	jet at forcing frequency $St = 0.59$.  Contours of the real part of 
	normalized output pressure fluctuations are shown.}
	\label{fig:output} 
\end{figure}

While the output domain is restricted to an arc at a distance of 100
jet diameters away from the nozzle exit, we can still obtain
near-field sound associated with input modes by examining the system
state after application of the resolvent operator, but before
application of the output matrix $C$.  We call these states
``unrestricted'' output modes.  These modes extend over $-5 < x/D <
35$ in the axial direction and over $0 < r/D < 20$ in the radial
direction, respectively.  The output matrix $C$ later projects them to
an arc in the far-field.  As shown in Fig.~\ref{fig:output},
unrestricted outputs resulting from these inputs follows a similar
pattern.  Black dashed lines in this figure represent the upstream,
downstream, and lateral sponge layers.  The pattern is now clear: the
optimal mode radiates a beam of acoustic radiation directed toward the
peak jet noise angle for this frequency.  Each sub-optimal output mode
is active along the edges of the preceding mode.  This creates two
beams of acoustic radiation in the first sub-optimal mode, three in
the second, and so on.

\subsection{Similarity wavepackets}
%
\subsubsection{Optimal mode}
Although the lipline wavepacket associated with optimal input mode is
well-described by a simple asymmetric function at one frequency, we
repeat the modeling procedure over a range of frequencies.  Motivated
by the self-similarity of turbulent jets, we investigate whether
similar asymmetric pseudo-Gaussian wavepacket functions can describe
our input modes over different frequencies.  Other wavepacket modeling
approaches also have yielded self-similar asymmetric bell-shaped
wavepackets over a range of frequencies~\cite{papamoschou2010,
  papamoschou2011}.  Naturally, the next task would be to construct a
universal wavepacket model, which can be widely used for a range of
frequencies, using minimum degrees of parameters; such as the first
moment (mean or centroid of a wavepacket in the axial direction)
$\mu_1$ and the standard deviation $SD$, which are respectively
defined as:
\begin{equation}
  \mu_1 = \frac{\int{x u_x(x) dx}}{\int{u_x(x) dx}}
\end{equation}
and
\begin{equation}
  SD = \sqrt{ \frac{\int{\left(x - \mu_1\right)^2 u_x(x) dx}}
	  {\int{u_x(x) dx}} }.
\end{equation}

\begin{figure}
  \centering
  \begin{tabular}{cc}
    \includegraphics[width=7cm]{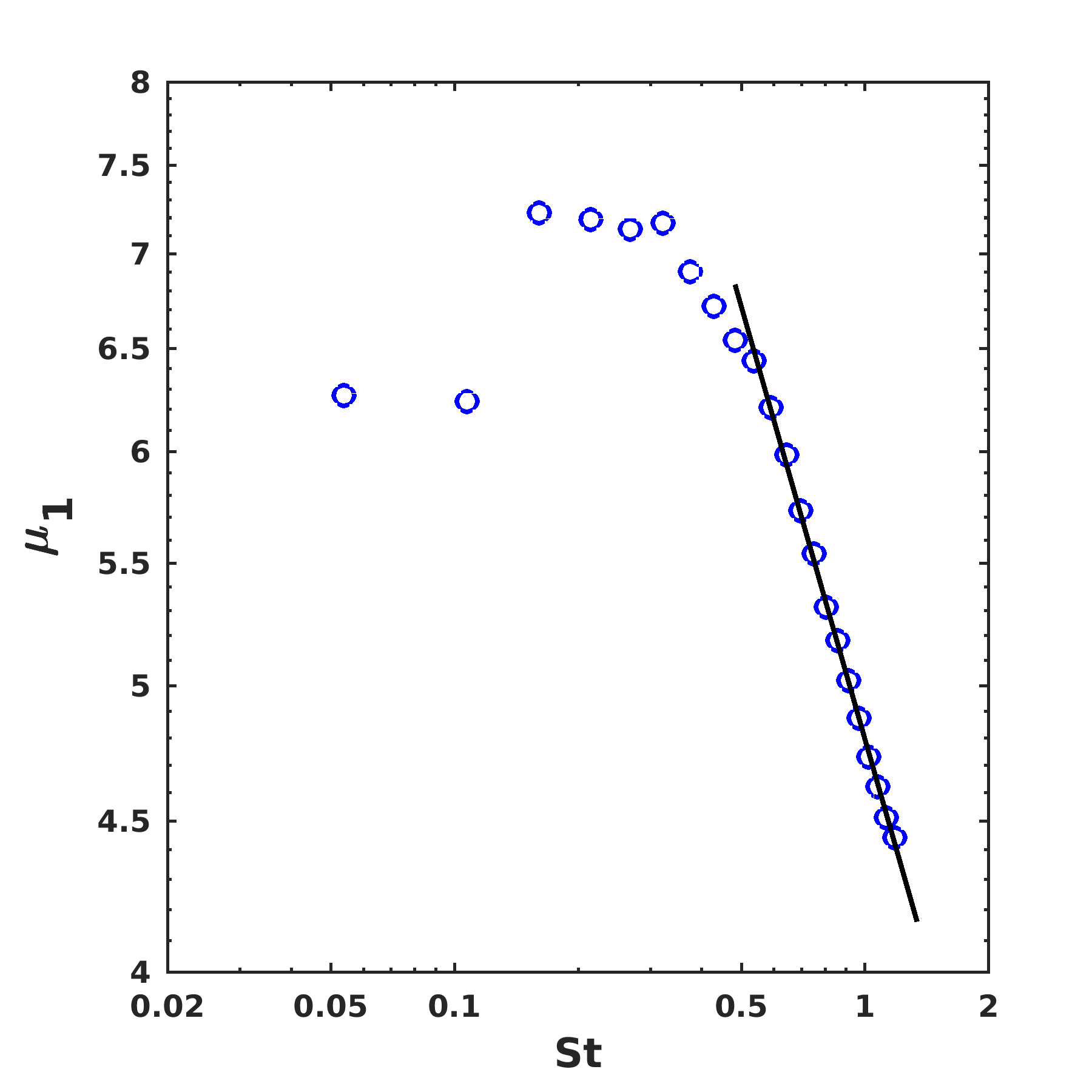} &
		\includegraphics[width=7cm]{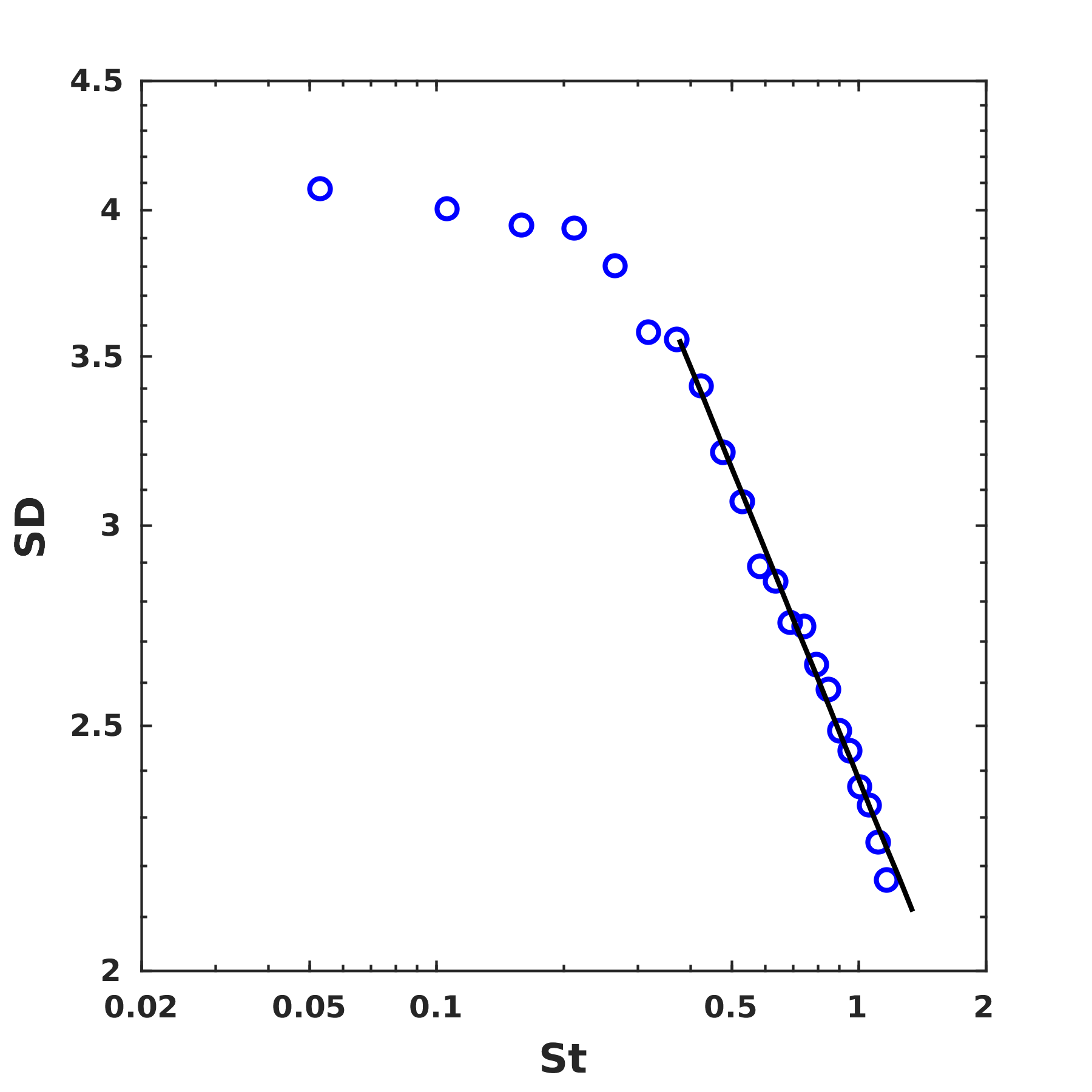} \\
		(a) & (b) \\
	\end{tabular}
	\caption{Model parameters as a function of frequencies for the lipline 
	wavepacket: (a) the mean decays and (b) the standard deviation.}
 	\label{fig:lipline_parameters}
\end{figure}

Figure~\ref{fig:lipline_parameters} shows these quantities as a function of 
frequency.  For sufficiently high frequencies, both the mean and standard 
deviation show a power-law dependence.  As presented in
Fig.~\ref{fig:lipline_parameters}(a) the mean location of the lipline 
wavepacket $\mu_1$ shifts upstream as forcing frequency increases. 
Excluding few low frequency cases, $\mu_1$ varies as $St^{-0.4858}$. 
Similarly, the standard deviation given in 
Fig.~\ref{fig:lipline_parameters}(b) follows the power-law form, though it 
decays slightly less rapidly than the mean as $St^{-0.4094}$.  We observe such 
similarities for cases over $St > 0.5$, and this agrees the result of 
theoretical approach by Papamoschou~\cite{papamoschou2011} who reported the 
self-similarity of turbulent jets for $St > 0.55$.

\begin{figure}
  \centering
  \includegraphics[width=7cm]{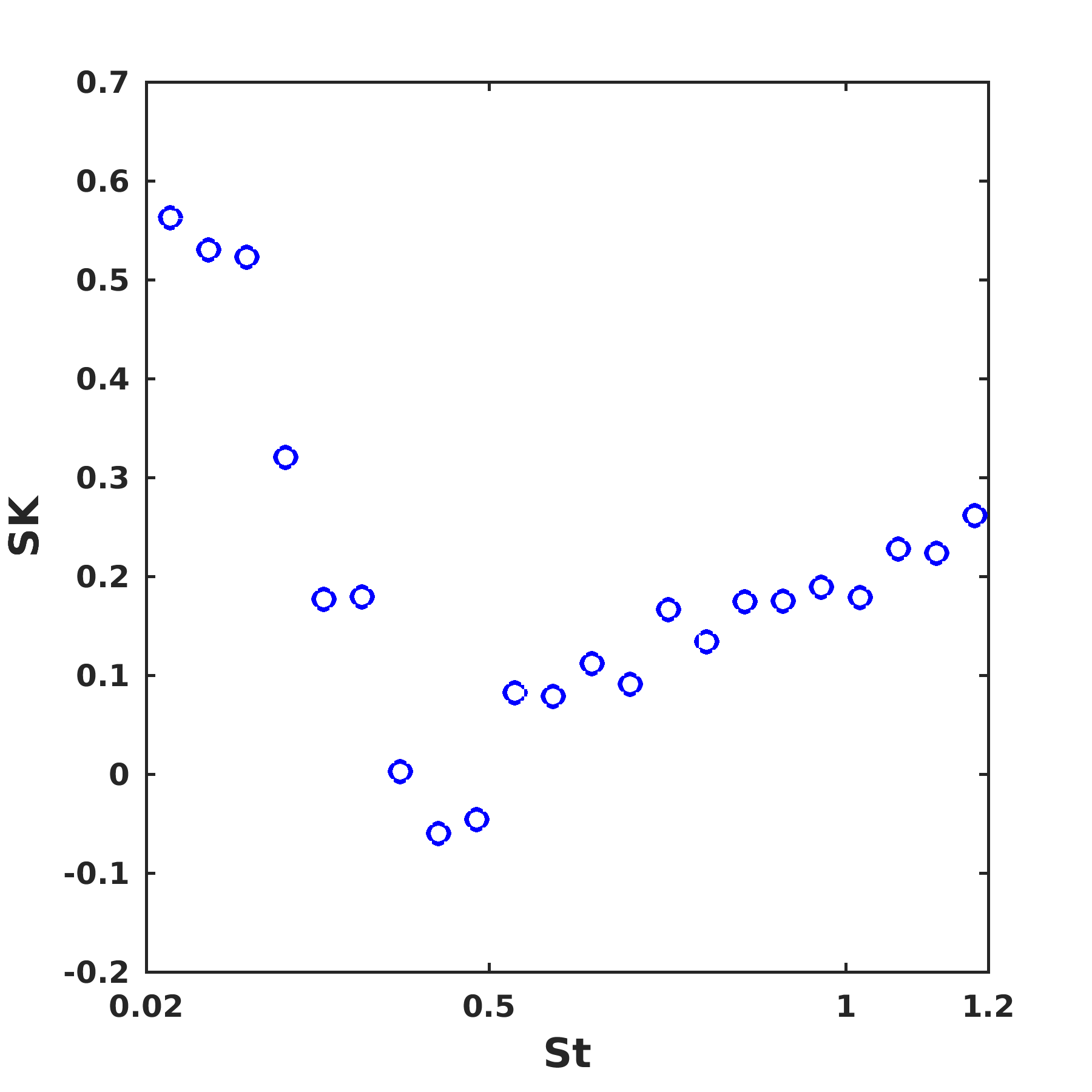}
  \caption{Skewness of the lipline wavepacket as a function of forcing 
	frequency.}
 	\label{fig:lipline_sk}
\end{figure}

Furthermore, we compute the skewness of wavepackets for each frequency defined 
by:
\begin{equation}
  SK = E\left[ \frac{\left( x - \mu_1 \right)^3}{SD^3} \right]
\end{equation}
where $E$ means the expected value, $SD$ denotes the standard deviation, and 
$\mu_1$ represents the centroid of a wavepacket in the axial direction, 
respectively.  Using this, Fig.~\ref{fig:lipline_sk} indicates 
positively skewed wavepackets for almost all frequencies as expected from long 
decaying tails.  For frequencies $St > 0.5$ wavepackets becomes more skewed 
as frequency increases, but overall, the variations of skewness remain 
small, suggesting similarity wavepackets in frequency.

\begin{figure}
  \centering
	\includegraphics[width=10cm]{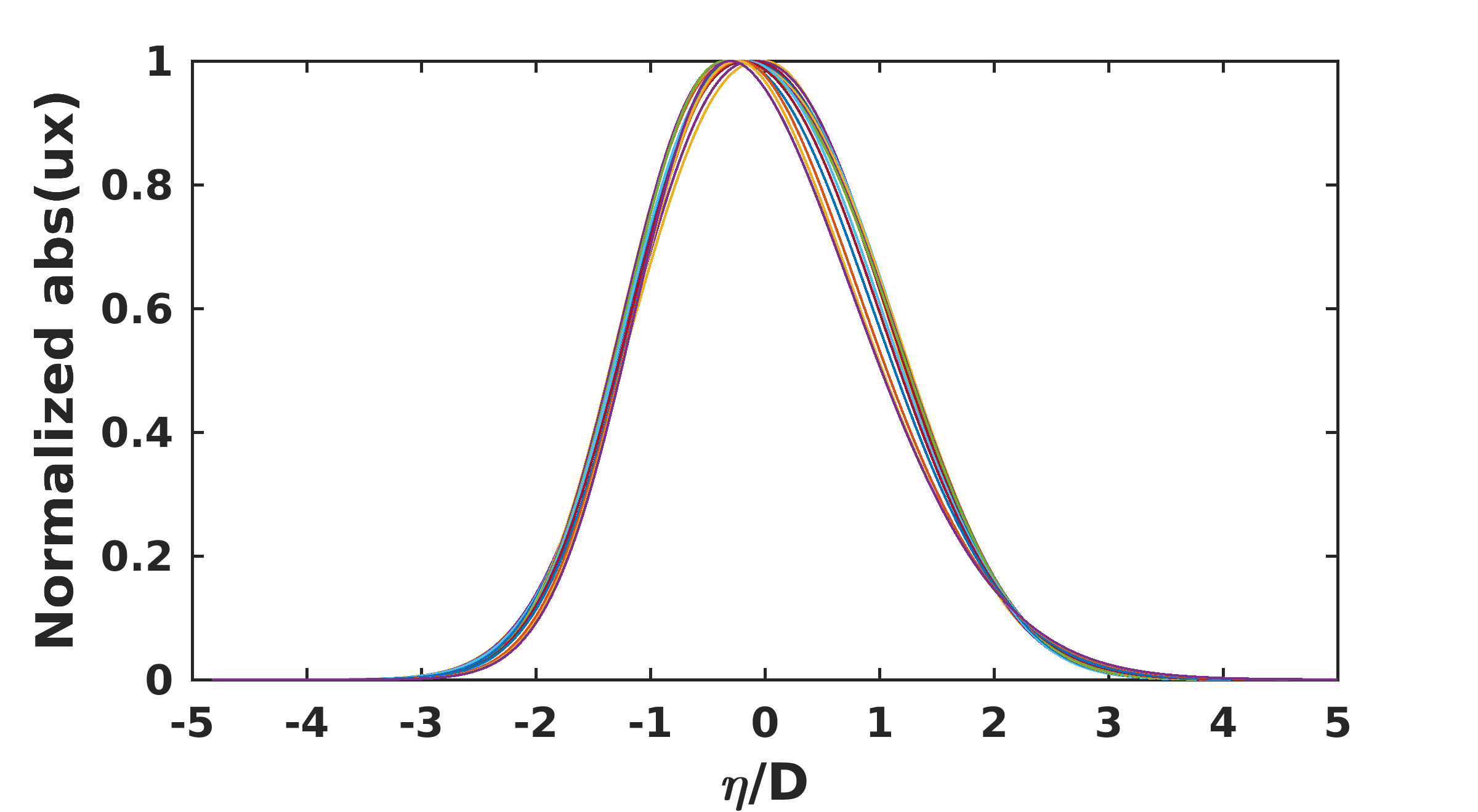}
	\caption{Collapse of scaled lipline wavepackets obtained for frequencies 
	$0.5 < St < 1.2$, suggesting simple similarity wavepacket models based on 
	the similarity variable $\eta$.}
 	\label{fig:lipline_similarity_wavepackets}
\end{figure}

Finally, we construct wavepacket models, which are functions of a new 
variable $\eta$ transformed by the mean source location $\mu_1$ and scaled 
by the standard deviation $SD$ such as:
\begin{equation}
  \eta = \frac{x - \mu_1}{SD}.
	\label{eq:similarity_variable}
\end{equation}
Here, instead of using two parameters $c_1$ and $c_2$ that control the widths 
of two parts of wavepackets separately, the standard deviation is chosen as a 
single unified parameter to describe the shape of wavepackets. 
Figure~\ref{fig:lipline_similarity_wavepackets} shows the collapse of 
wavepacket envelopes taken over a range of frequencies between $St = 0.5$ and 
$St = 1.2$. 

\subsubsection{Sub-optimal modes}
Sub-optimal modes appear to be composed of multiple wavepackets (see
Fig.~\ref{fig:input}).  The envelopes of these component wavepackets
are similar in shape to that associated with the optimal mode.
Because each sub-optimal mode conatins multiple wavepackets, modeling
them is more complicated. Moreover, as the frequency changes, a group
of wavepackets moves upstream and sometimes merges into the wavepacket
in a boundary layer developed along the nozzle wall.  It thus requires
a great care to track the same type of wavepackets and model them as
similarity wavepackets in frequency.

\begin{figure}
  \centering
	\includegraphics[width=10cm]{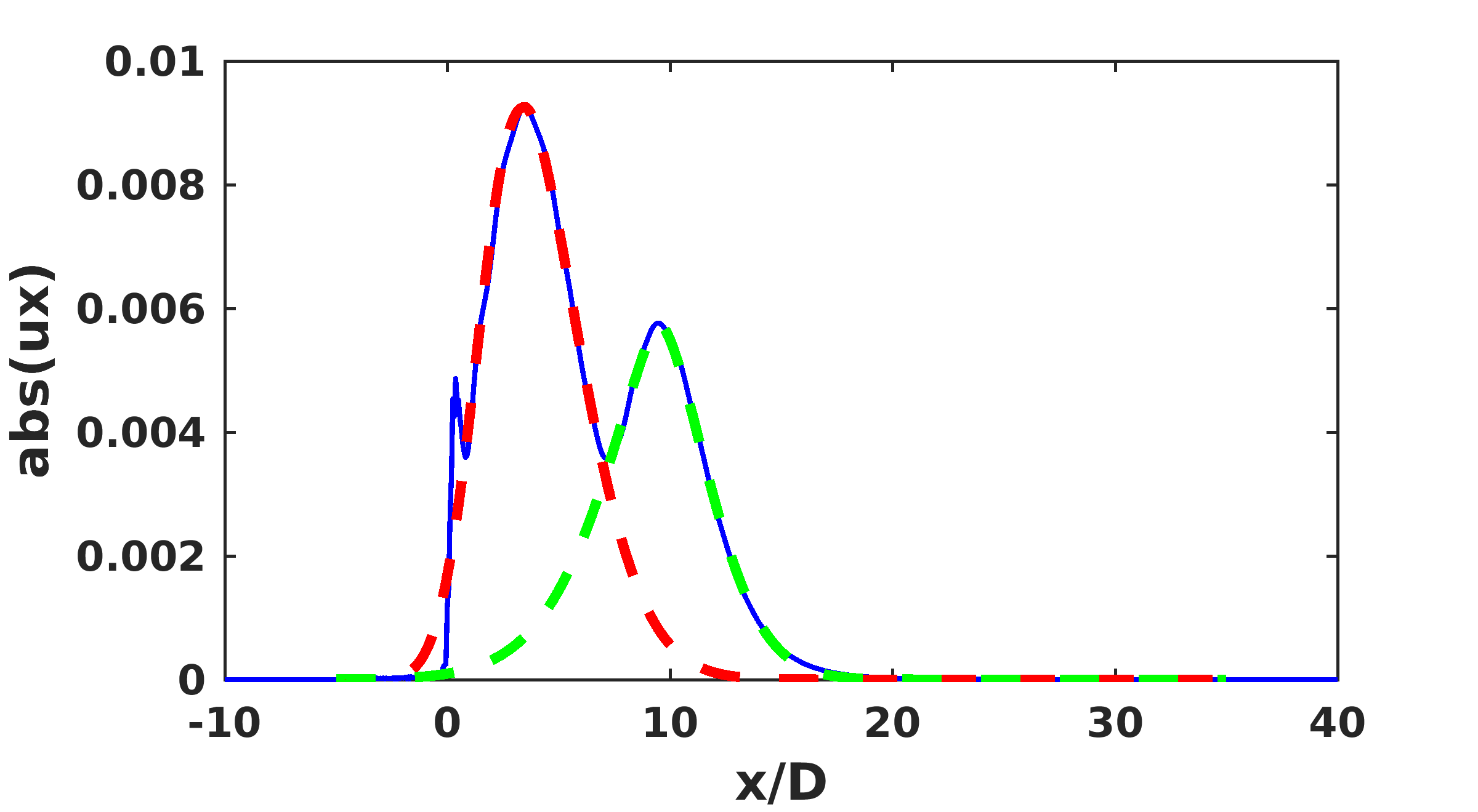} \\
	\caption{Wavepackets taken along the lipline for the first sub-optimal 
	mode ($n = 2$) at $St = 0.59$ modeled using asymmetric pseudo-Gaussian 
	functions.  W1: red dashed line and W2: green dashed line, counted from 
	the upstream.}
 	\label{fig:n2_wavepackets}
\end{figure}

In Fig.~\ref{fig:n2_wavepackets} we model wavepackets taken along the
jet lipline ($r/D = 0.5$) for the first sub-optimal mode ($n = 2$) by
asymmetric pseudo-Gaussian functions.  In contrast to the optimal
mode, the first sub-optimal mode captures two wavepackets downstream
of the nozzle exit ($x/D = 0$).  They, however, bear a similar shape
to that captured by the optimal mode, and are approximated using the
same type of asymmetric pseudo-Gaussian function given in
Eq.~\eqref{eq:pseudo_Gaussian}.  Here, the blue solid line represents
the input-mode-captured wavepacket, while the red and green dashed
lines denote modeled wavepackets.

\begin{figure}
  \centering
	\includegraphics[width=10cm]{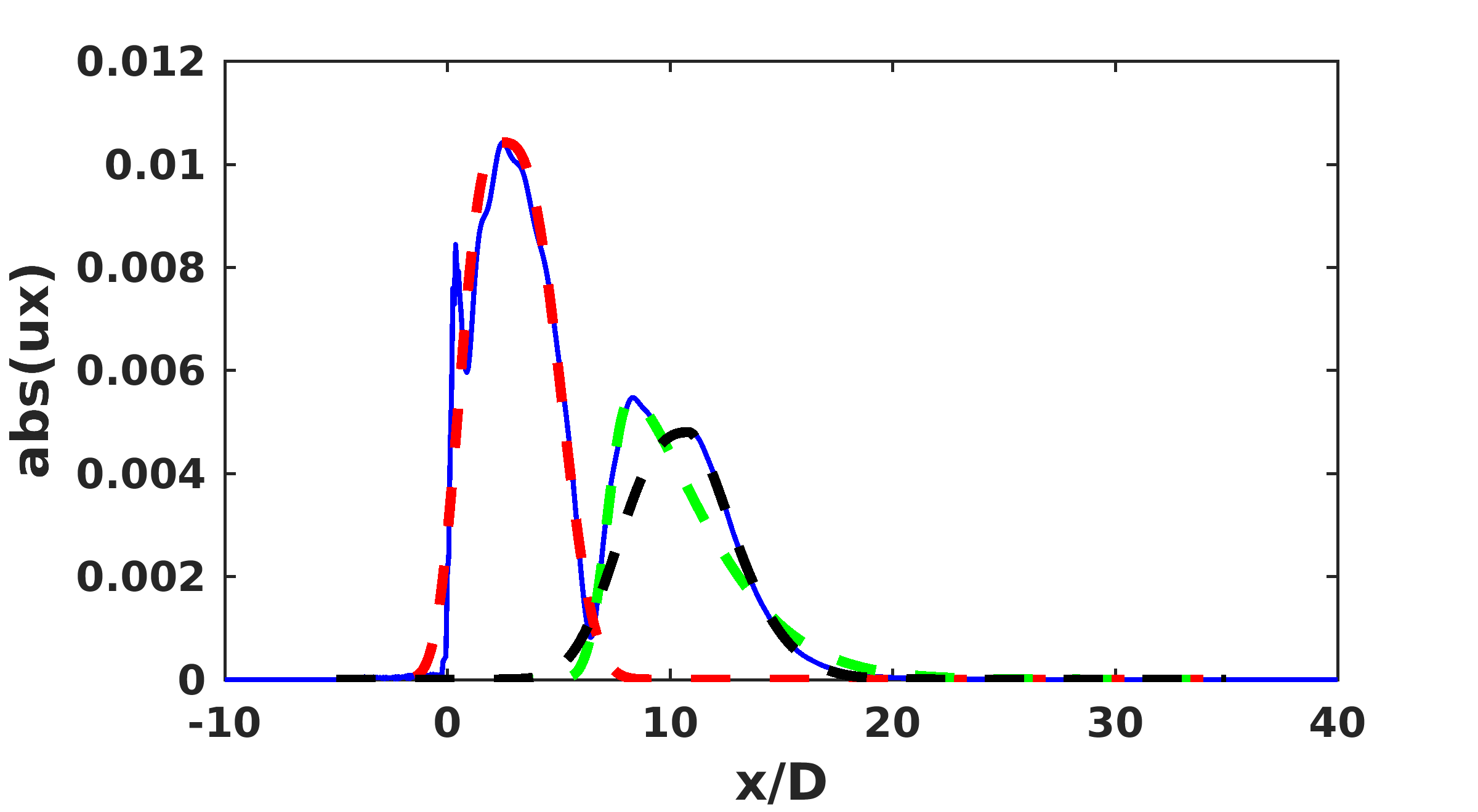} \\
	\caption{Wavepackets taken along the lipline for the second sub-optimal 
	mode (blue solid line) at $St = 0.59$ modeled using asymmetric 
	pseudo-Gaussian functions. W1: red dashed line, W2: green dashed line, and 
	W3: black dashed line, counted from the upstream.}
  \label{fig:n3_wavepackets}
\end{figure}

Fig.~\ref{fig:n3_wavepackets} shows wavepackets taken along the inner
lipline for mode number $n = 3$ (the next sub-optimal mode).  We
observe three wavepackets, denoted by $W1$, $W2$, and $W3$, measured
from the upstream.  Each wavepacket is positively skewed and
bell-shaped as in other cases.  We therefore expect that a similarity
wavepacket model could be constructed, even for this case.

The pattern shown in Figs.~\ref{fig:lipline_similarity_wavepackets}
through \ref{fig:n3_wavepackets} is approximately the same that
obtained by taking a sequence (of magnitudes) of axial derivatives of
the optimal wavepacket envelope.  Of course, the input modes are
two-dimensional, so their inter-relationship may be more complicated
than this.  Still, the sub-optimal modes appear to be associated with
dynamics in the regions where the amplitude of the preceding mode in
the sequence is undergoing the most change (i.e., where its gradient
is greatest).  In the next section, we discuss physical mechanisms
that could lead to sources that can align with such a pattern, thereby
creating noise.

\subsection{Physical origin of the sub-optimal modes}
In previous sections, we found that the optimal input mode clearly
reveals a wavepacket whose envelope corresponds to an asymmetric
pseudo-Gaussian function.  Moreover, we observed that the sub-optimal
input mode captures structures that follow upstream and downstream
edges of the optimal wavepacket.  As shown in
Figs.~\ref{fig:n2_wavepackets} and~\ref{fig:n3_wavepackets}, the
sub-optimal mode shows increasingly many humps as the mode number
increases.  The humps in each sub-optimal mode appear at locations
where the largest differences would occur, if the wavepackets in the
preceding mode were perturbed slightly in their axial position.  This
type of uncertainty in a wavepacket's axial position is known as
``jitter''~\cite{cavalieri2011,jordan2013,jordan2014}.  Jitter arises from the fact that high-speed
jets are susceptible to variations at very low frequency, much lower
than the frequencies associated with wavepackets.  On the timescales
of the wavepacket, this very low frequency variation appears to
correspond to changes in the baseflow.  The wavepacket responds to
slow variations in the base flow by changing its position (and
perhaps, shape).  While axial jitter of the optimal wavepacket leads
to a double-humped shape corresponding to the first sub-optimal mode,
axial jitter of the first sub-optimal mode produces a shape with four
humps.  As shown in Fig.~\ref{fig:n3_wavepackets}, the second
sub-optimal mode has three humps.

Alternatively, we consider axial decoherence \cite{cavalieri2014,baqui2015} as another
possible mechanism by which acoustic sources embedded in the jet
turbulence may align with the pattern of sub-optimal modes that we
find.  Entire wavepackets are almost never visible in instantaneous
snapshots of the near-field turbulent jets.  Instead, one observes
``pieces'' of wavepackets that persist over a maybe only a few
diameters before losing coherence.  Inside these windows of coherence,
fluctuations grow or decay in accordance to the overall wavepacket
envelope.  While the growth and decay of disturbances are governed by
dynamics, wavepackets have a statistical nature as turbulence drives
instability waves into and out of coherence.  An entire wavepacket,
therefore, should be thought of as the tendancy of the baseflow to
make disturbances grow or decay in accordance with instability
physics.  Because these physics do not change in time for a given
baseflow, a wavepacket is also constant and is determined by dynamics.

\begin{figure}
  \centering
  \begin{tabular}{cc}
		\includegraphics[width=6.5cm]{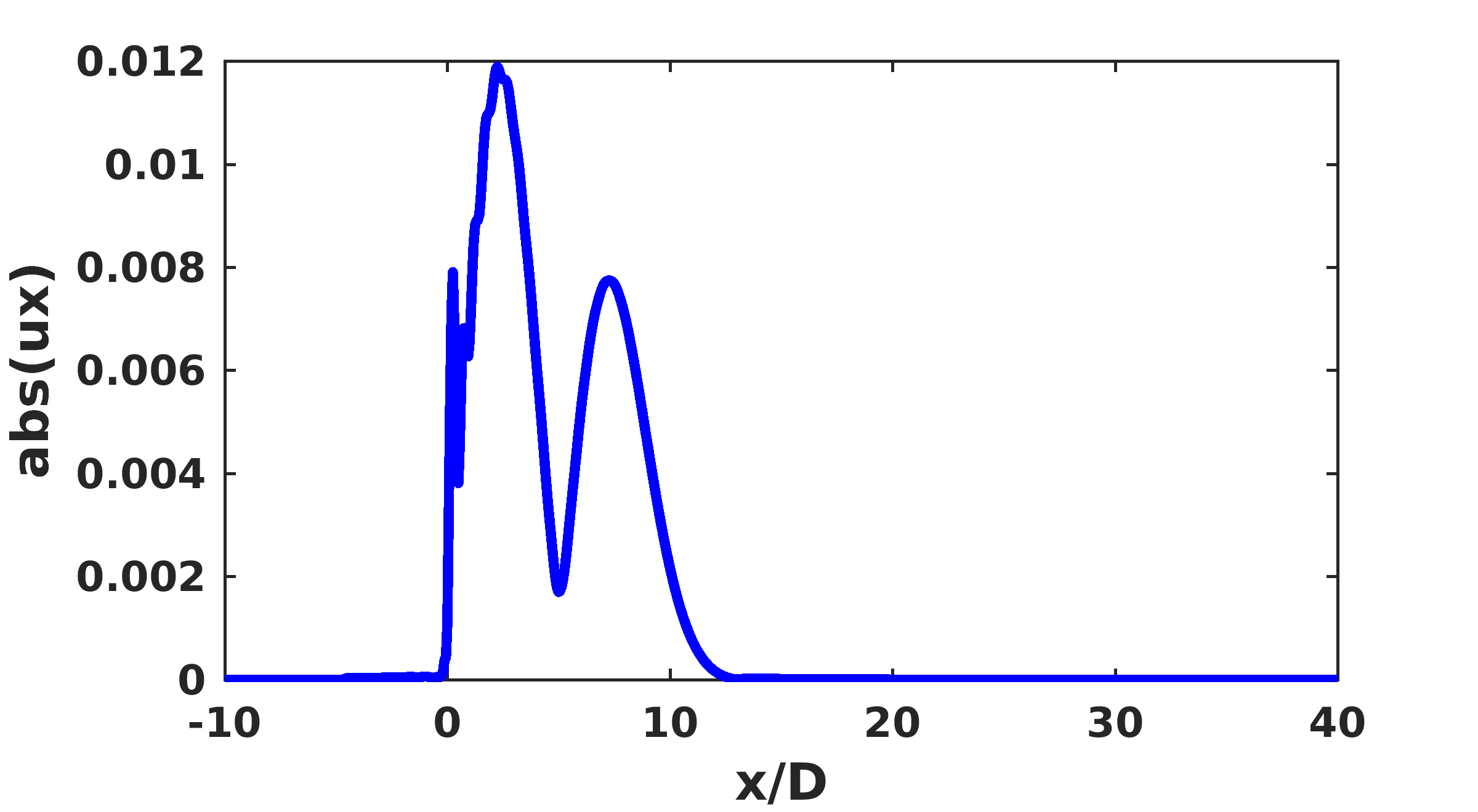} &
		\includegraphics[width=6.5cm]{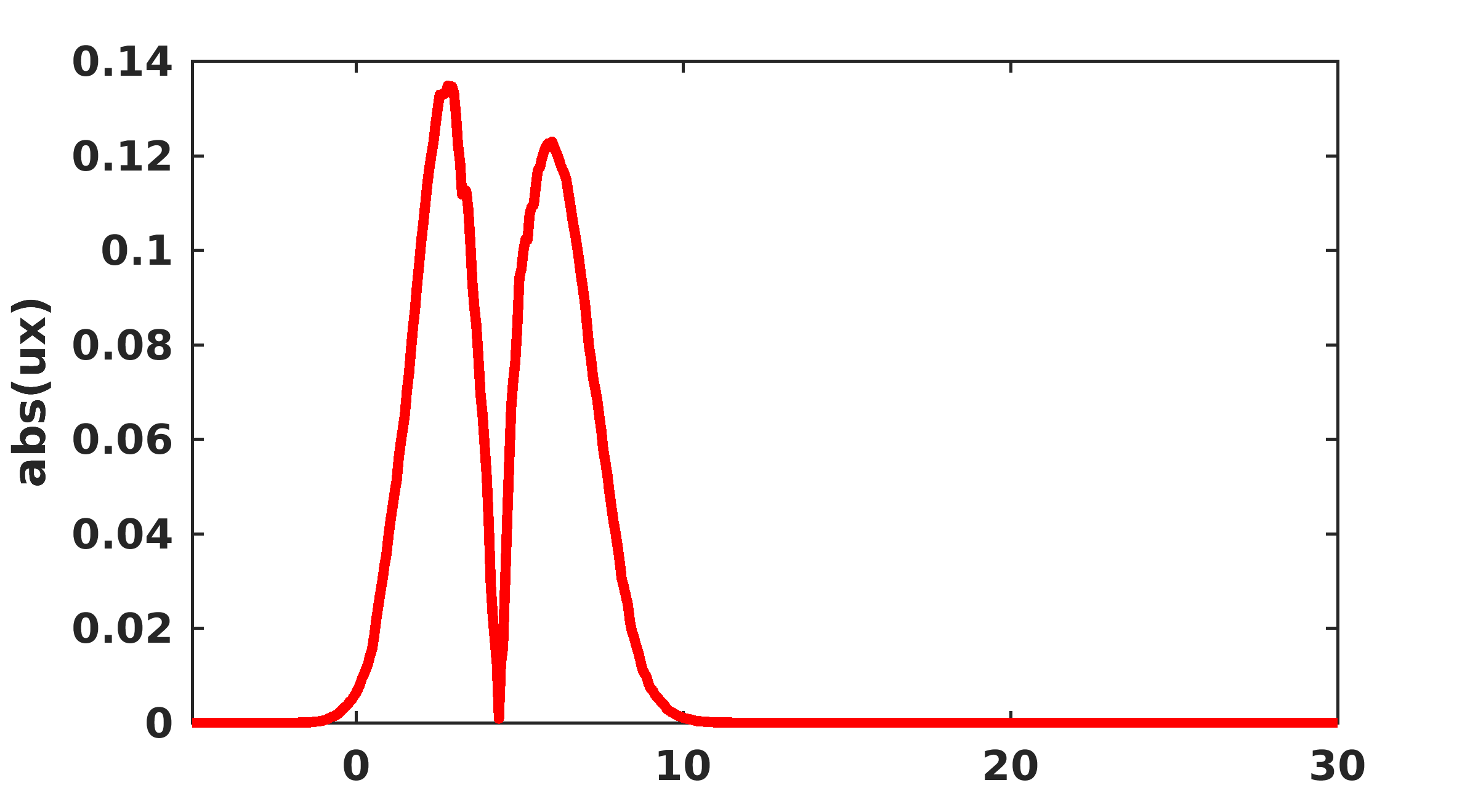} \\
		(a) $n = 2$ & (b) $n = 2$ \\
    \includegraphics[width=6.5cm]{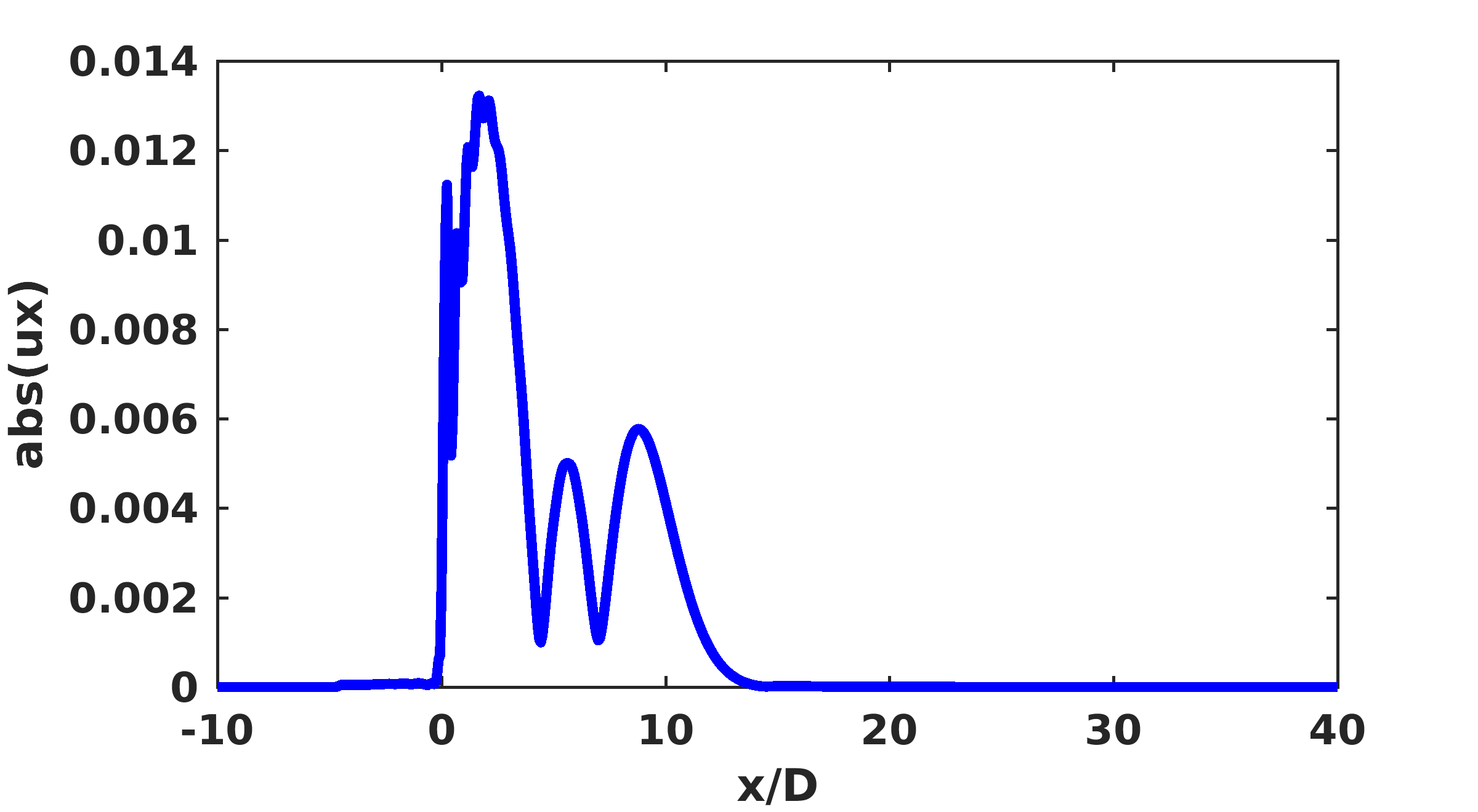} &
		\includegraphics[width=6.5cm]{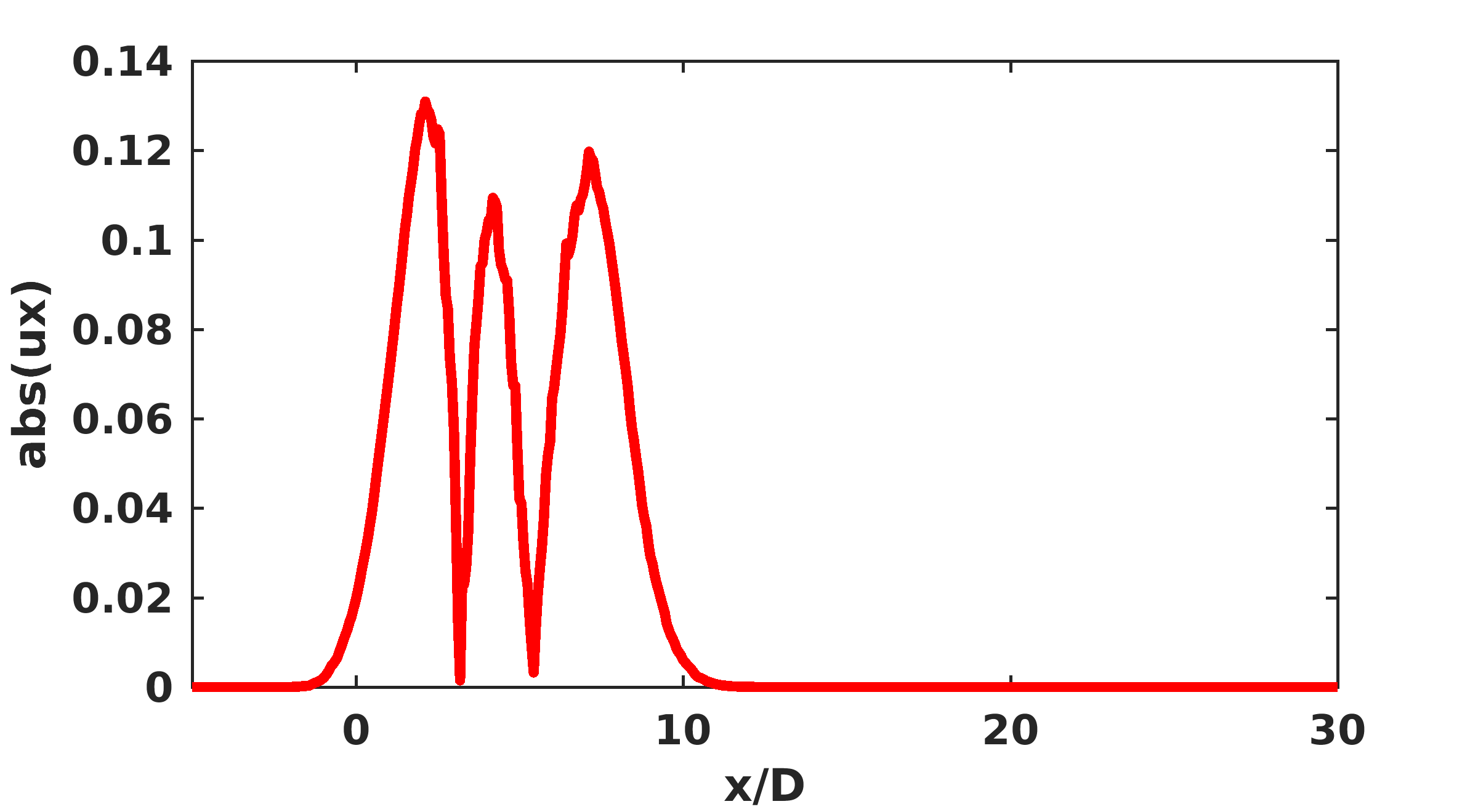} \\
		(c) $n = 3$ & (d) $n = 3$ \\
	\end{tabular}
	\caption{By taking singular value decomposition of the matrix whose 
	columns consists of a series of axial snapshots of the optimal wavepacket, 
	the left singular vectors (b,d) qualitatively reproduce the actual input 
	modes (a,c) for $St = 1.18$.}
 	\label{fig:pod}
\end{figure}

To model decoherence, we perturb the optimal wavepacket by small
random forcing, and extract short, stochastic windows of it,
positioned between $-5 \leq x/D \leq 15$.  The constant window width
$w$ is chosen to be as large as the coherence length-scales of the
axial velocities such that $w = 2D$.  Outside of a given window, we
zero all other fluctuations.  We repeat this process to build a stack
of different realizations, visiting a different part of the wavepacket
each time and reproducing the effect of its axial decoherence.
Singular value decomposition applied to this collection yields its
dominant dynamical features.  As expected, the first singular vector
recovers the original wavepacket (not shown).  The right-hand column
of Fig. \ref{fig:pod} shows the first and second singular vectors
obtained from the decomposition.  While there are differences, they
reproduce the corresponding sub-optimal modes fairly well.  In
particular, the 3rd singular vector contains three humps just like the
second sub-optimal mode.  This therefore conclude that axial
decoherence is a physical mechanism by which acoustic sources in the
jet align well with input modes predicted by I/O analysis.

We should note that unlike the connection between input and output
modes, there is not necessarily a causal relationship between axial
decoherence modes and input modes.  If it occurs, jitter may also
project a significant portion of the sources onto the input modes.
From our analysis, it seems that axial decoherence aligns even better
with the input modes, and thus is an efficient mechanism of noise
production.

\section{Discussion}
\label{sec:discussion}
In the previous section, we saw that input forcing in the form of a
wavepacket is the optimal way to generate far-field sound in a Mach 0.9
jet, according to I/O analysis.  The wavepacket envelopes identified
using our methodology agree well with theoretical models developed
from experimental observations.  Furthermore, we found that these
wavepackets exhibit self-similarity over a broad range of frequencies.
As the frequency increases, the peak of the corresponding wavepacket
moves further upstream along the jet shear layers, closer to the
nozzle.  As the wavepacket moves closer to the nozzle, its axial
extent diminishes, preserving similarity.  This explains the success
of a stochastic similarity wavepacket model~\cite{papamoschou2011} at
predicting experimental far-field acoustic spectra over a broad range
of frequencies and observer angles.  

While these results already paint a compelling picture of the physics
of jet noise generation based on wavepackets, we further test them in
this section by applying them to data obtained from a high-fidelity
large eddy simulation (LES) of a Mach 0.9 jet.  This provides
additional insight about the meaning of the I/O modes, and ultimately
about the mechanisms at play in the generation of sound in high-speed
turbulent jets.

\subsection{Large eddy simulation}

\begin{figure}
  \centering
	\includegraphics[width=14cm]{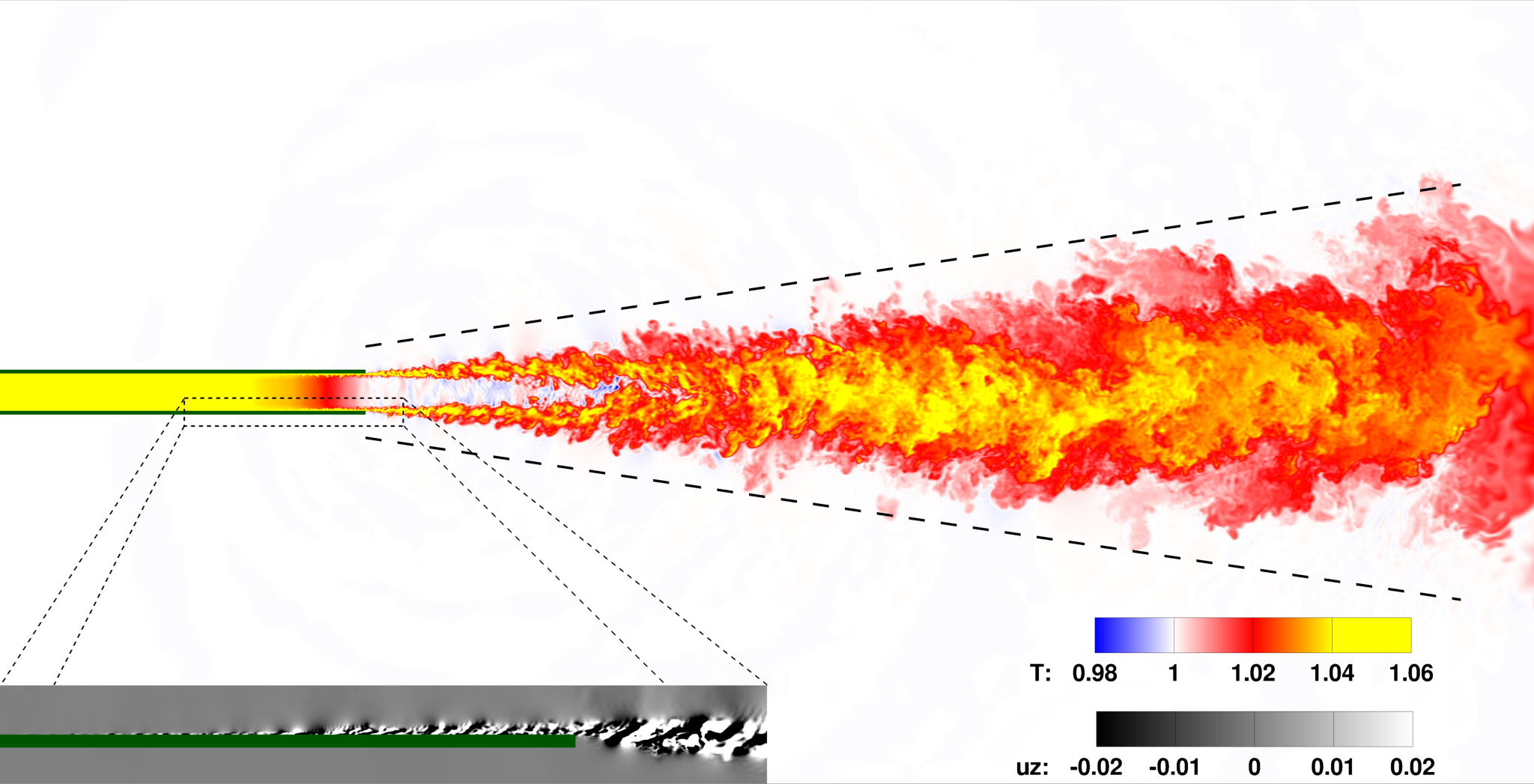}
	\caption{Contours of temperature and azimuthal velocity on a
          cross section through a snapshot taken from the LES of the
          Mach 0.9 turbulent jet.}
  \label{fig:les1}
\end{figure}

We perform large eddy simulation of a Mach 0.9 isothermal round jet
using a finite volume method on an unstructured mesh.  Our
computational domain extends 10 nozzle diameters upstream of the
nozzle exit and 50 diameters downstream.  A numerical sponge layer,
however, begins at 30 diameters downstream of the nozzle exit.  This
minimizes unphysical acoustic reflections at the outflow boundary.
The radial extent of the domain is 25 nozzle diameters.  To match the
results based on RANS in the first part of the paper, we consider a
straight cylindrical nozzle.  We simulate the entire flow inside of
the nozzle, starting at 10 diameters upstream of the nozzle exit.  In
Fig.~\ref{fig:les1}, contours of temperature on a cross section
through a snapshot taken from the LES show how the jet emerges from
the straight nozzle and develops downstream.  This figure also shows
that while the temperature of the potential core matches the ambient
temperature, the boundary layer in the nozzle and emerging shear
layers heat slightly due to viscous dissipation.  The Reynolds number
of the jet based on nozzle diameter is $Re_D = u_j D/\nu = 2 \times
10^5$, which matches that of the RANS calculations.  The unstructured
mesh used for the results reported here incorporated 62 million cells,
clustered in the turbulence-containing regions.  The nozzle lip was
resolved by approximately 800 cells in azimuthal direction.  We
repeated the LES on several different meshes to ensure that the
reported results are grid independent.  To account for turbulent
motions on scales smaller than the grid resolution, we apply the
Vreman sub-grid scale model.

The state of the boundary layer (laminar vs. turbulent) as it emerges
from the nozzle can have a significant effect on the far-field sound
produced by the jet~\cite{zaman2012,bres2012}.  To simulate realistic
nozzle interior surface roughness levels, we trip the boundary layer
by introducing low-level white noise in a zone around 7.5 diameters
upstream of the nozzle exit.  Several diameters downstream from this
position, the boundary layers along the nozzle interior become
turbulent before exiting the nozzle (see Fig.~\ref{fig:les1},
inset).

To compute far-field sound, we collect acoustic information on a
conical surface surrounding the jet, and project it to the far-field
by solving the FWH equation
\cite{lockard2000, lockard2005,bres2012}.  The sloped dashed lines in
Fig.~\ref{fig:les1} indicate the position of the conical surface
relative to the jet.  It is far enough from the jet for the flow to be
essentially irrotational, but is close enough to remain in a zone of
relatively fine resolution so that it captures acoustic waves with
frequencies up to $St \approx 5$.  Consistent with the FWH projection
scheme used in the I/O analysis, the FWH surface is open on both ends.
Comparing to results obtained using closed FWH surfaces (not shown),
we find that the absence of end caps has little effect in this case,
except at very low angles to the jet axis (less than $10^\circ$ or
greater than $170^\circ$).  This lack of effect is most likely due to
the length of our FWH surface -- it extends $30D$ downstream of the
jet axis, well past the important acoustic source containing regions
($0 < x/D < 15$).

\begin{figure}
  \centering
	\includegraphics[width=10cm]{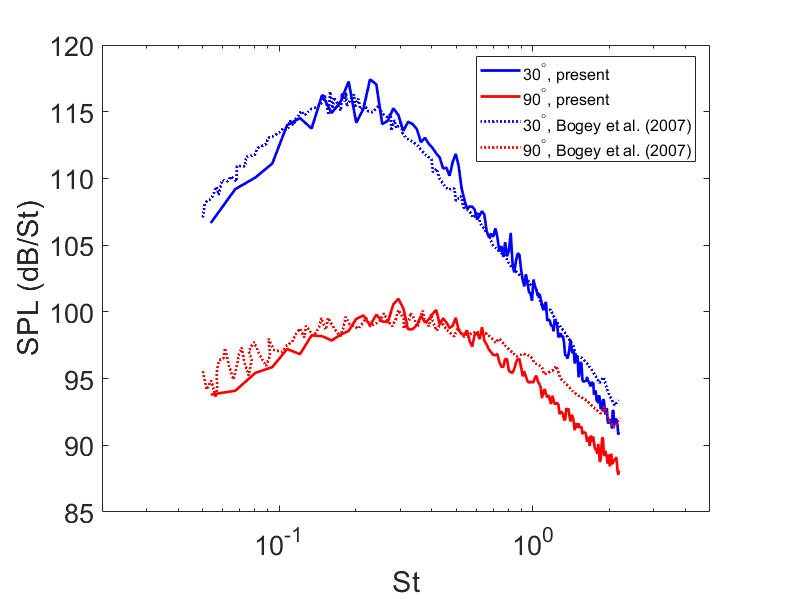}
	\caption{Far-field pressure spectra for observers located 
	$50D$ away from the nozzle exit.}
  \label{fig:farfield1}
\end{figure}

Once our simulation reaches a statistically stationary state, we run
it for an additional time of $T = 414 D/c_\infty$, recording samples
along the FWH surface at time intervals of $\Delta t = 0.05
D/c_{\infty}$ resulting in 8280 time samples total.  In terms of the
jet Strouhal number $St = f D / u_j$, this captures frequencies in the
range $0.0027 < St < 11.11$, although the LES mesh resolution begins
to affect frequecies at the high end of this range.  Because the
turbulent jet is a chaotic system, the convergence of the very lowest
frequencies should not be trusted.  Instead we apply a series of 25
Hann windows of length $82.8 D/c_{\infty}$ to the data, overlapping
with each other by 83.3\%, and average the results (Welch's
method~\cite{welch1967}).  Figure~\ref{fig:farfield1} shows far-field
pressure spectra obtained using this method for observers located
$50D$ away from the nozzle exit, at both $\phi = 30^{\circ}$ and
$90^{\circ}$ to the downstream jet axis for frequencies in the range
$0.05 < St < 2$.  The shape and levels of the spectra show excellent
agreement with both experimental measurements~\cite{bogey2007} and
previous computations~\cite{bres2015}.  This demonstrates that our
simulation produces realistic far-field noise.

\begin{figure}
  \begin{tabular}{c}
	\includegraphics[width=10cm]{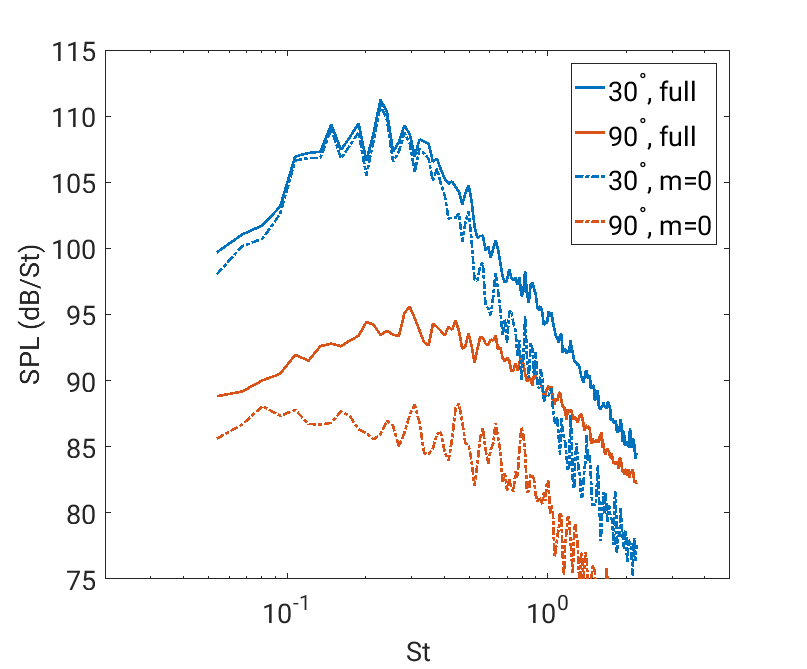} \\
	(a) \\
	\includegraphics[width=10cm]{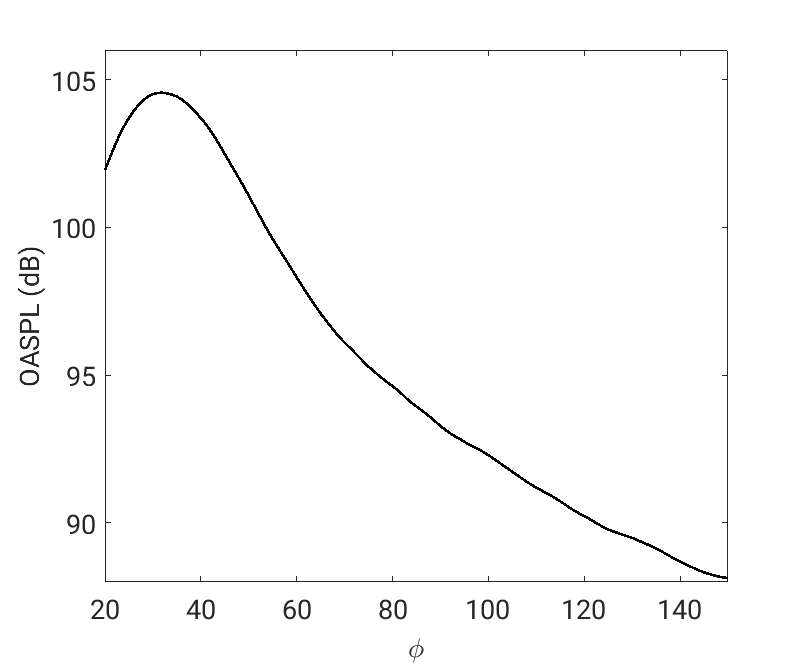} \\
	(b) \\	
	\end{tabular}
	\caption{(a) Far-field pressure spectra for observers located 
	$50D$ away from the nozzle exit, and (b) overall sound pressure 
	levels as a function of polar angle.}
  \label{fig:farfield2}
\end{figure}

Figure~\ref{fig:farfield2} shows far-field sound spectra at $100D$
from the jet exit, and overall sound pressure levels as a function of
polar angle at this same distance.  Comparing
Fig.~\ref{fig:farfield2}(a) to Fig.~\ref{fig:farfield1}, we see that
the levels decrease at distances farther from the jet as expected.
The shape of the spectra also change slightly.  This is a consequence
of the acoustic sources being extended over an axial distance in the
jet.  $50D$ is perhaps not so much larger than the $15D$ extent of the
acoustic sources.  To study the directivity of non-compact sources, it
is important to project their sound farther away.  We have chosen
$100D$ to be relatively far away from the jet, but to remain within
the realm of possible experimental measurement.

At $100D$, the peak jet noise direction is $30^{\circ}$ to the
downstream jet axis, in good agreement with experimental observations.
By taking a Fourier transform of the acoustic sources in the azimuthal
direction, we find that axisymmetric fluctuations comprise a large
part of this peak jet noise, as shown in Fig.~\ref{fig:farfield2}(a).
At higher frequencies and higher wavenumbers, however, the
contribution from axisymmetric sources diminishes in comparison to
contributions from higher azimuthal wavenumbers.

\subsection{I/O analysis of LES data}

In addition to sampling the LES on the FWH surface, we record data on
an entire regular cylindrical mesh extending over $0 < x/D < 30$ and
$0 < r/D < 9.13$.  The positions of the grid points correspond exactly
to the grid points used for the RANS calculations, and are uniformly
spaced in the axial direction, but cluster around $r/D = 0.5$ in the
radial direction.  We resolve this cylindrical domain by $481 x 180 x
64$ points in the axial, radial and azimuthal directions,
respectively.  This gives 5.5 million grid points in total, which is
much smaller than the total number of cells used in the LES.  We
interpolate from the unstructured LES mesh to the cylindrical mesh
using inverse distance weighting.  Multiplying by 40 bytes (5 double
precision numbers) per grid point, and 8280 time samples, the total
size of the database is approximately 2 Tb.

\begin{figure}
  \begin{tabular}{c}
	\includegraphics[width=12cm]{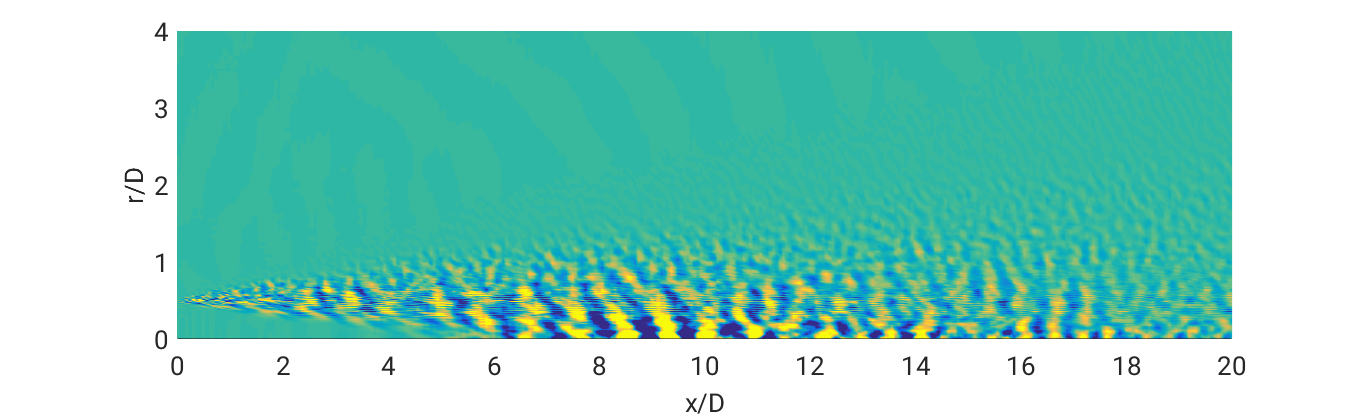} \\
	(a) \\
	\includegraphics[width=12cm]{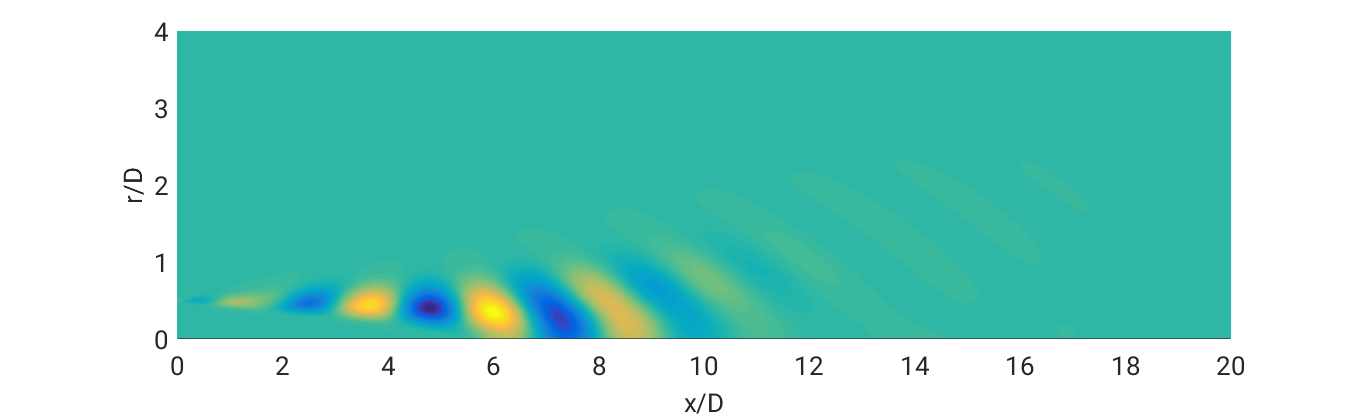} \\
	(b) \\
	\end{tabular}
	\caption{(a) The leading SPOD mode for axisymmetric fluctuations 
	from the LES and (b) the optimal input forcing from I/O analysis 
	for $m = 0$ at $St = 0.59$.}
  \label{fig:spod}
\end{figure}

From this database, we directly compute the Lighthill source
terms~\cite{lighthill1952} which are the nonlinear forcing terms
driving the velocity equations~\eqref{eq:lns_u}.  Using the same Hann
windows as in the FWH calculations, we analyze the frequency content
using Spectral Proper Orthogonal Decomposition
(SPOD)~\cite{towne2018,schmidt2017}.  Figure~\ref{fig:spod}(a) shows
the leading SPOD mode for axisymmetric fluctuations at $St = 0.59$.
The Lighthill sources involve spatial derivatives since they include
the divergence of the fluctuating Reynolds stress tensor.  Because of
this, stochastic elements at high wavenumbers tend to be amplified.
Still, Fig.~\ref{fig:spod}(a) reveals significant order at this
frequency, indicating the presence of non-compact sources.  We would
expect this order to be enhanced if SPOD analysis were applied to a
quantity like pressure which tends to be more smooth in space.  We
should note that, in exactly the same way that acoustic analogies may
be written in terms of different
variables~\cite{curle1955,goldstein2008}, different formulations of
I/O analysis are possible.  Some reformulations may yield source terms
with desirable properties such as smoothness in space.  A particularly
interesting formulation may be the one based on the decomposition of
Doak~\cite{doak1972} recently adapted for jet
noise~\cite{unnikrishnan2018}, since it optimally separates the
aerodynamic and acoustic components of pressure within the jet
itself. Nevertheless, the Lighthill source corresponds exactly to the
forcing in our I/O analysis so we focus upon it to evaluate our modes.

Figure \ref{fig:spod}(b) shows the optimal input forcing from I/O
analysis at the same frequency and azimuthal wavenumber.  At first
glance, it seems dissimilar to the SPOD mode.  To quantify this, we
project the leading SPOD mode onto the orthonormal basis formed by the
first 20 input modes at this frequency such that
\begin{equation}
  \hat{q}_{LES} = \sum_n a_n \hat{q}_{in,n},
\end{equation}
where $\hat{q}_{LES}$ is the Fourier transformed Lighthill source and
$\hat{q}_{in,n}$ are the orthonormal input modes.  The coefficients
$a_n$ may be found simply using the following inner product with
quadrature weights accounting for the non-uniform mesh.
\begin{equation}
a_n = \left< \hat{q}_{LES}, \hat{q}_{in,n} \right>.
\end{equation}

\begin{figure}
  \begin{tabular}{c}
	  \includegraphics[width=8cm]{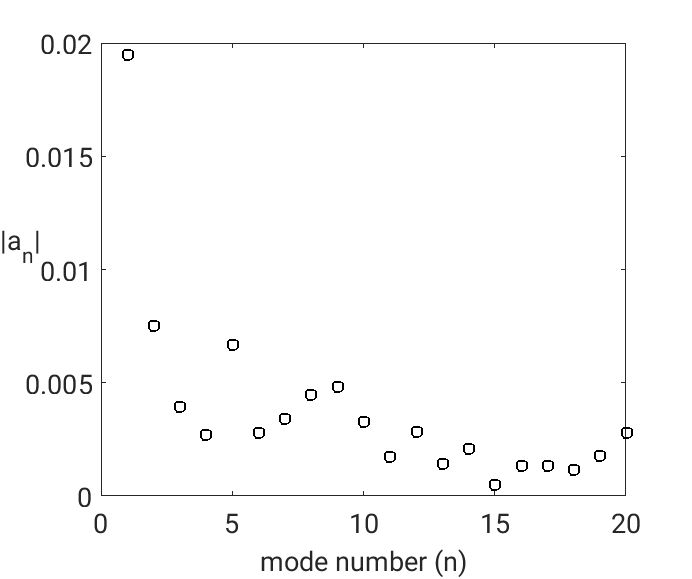} \\
		(a) \\
    \includegraphics[width=8cm]{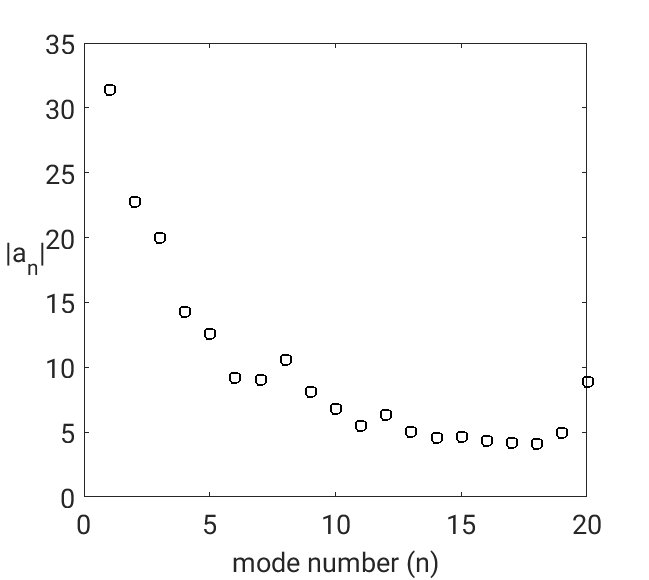} \\
    (b) \\
  \end{tabular}
	\caption{(The coefficients $a_n$ vs. mode number (a) from projecting
	the SPOD mode alone onto the basis of 20 input modes and (b) from
	projecting each Fourier-transformed Hann window and averaging
	the coefficients.}
  \label{fig:projection}
\end{figure}

Figure~\ref{fig:projection}(a) show the
coefficients $a_n$ vs. mode number.  Apparently, very little energy of
the leading SPOD mode is captured by the input modes.  This, however,
is a natural consequence of the fact that very little of the
aerodynamic energy contained in the turbulent fluctuations in a jet is
ever radiated as sound~\cite{dowling1983}.  While SPOD captures the
aerodynamic energy contained in the Lighthill source terms, it filters
out the aeroacoustically relevant dynamics, in this case.  The spatial
separation of near-field inputs from far-field outputs in our I/O
framework, however, means that our input modes are connected with
dynamics in the jet that produce noise.  If we instead first project
each Fourier-transformed Hann window onto the same basis of 20 input
modes, and then average the coefficients $a_n$, we find that they
increase by three orders of magnitude (see Fig.
\ref{fig:projection}(b)) compared to those obtained from projecting
the SPOD mode alone.  Indeed, the input modes are quite active in this
jet.  Moreover, the input coefficients follow a regular pattern,
decreasing with input mode number, suggesting that a model for this
decay may be possible.  While we leave this modeling for a future
study, we may interpret our input modes physically, however, as
filters for the acoustically important dynamics embedded in the jet
turbulence.

\begin{figure}
  \centering
  \includegraphics[width=10cm]{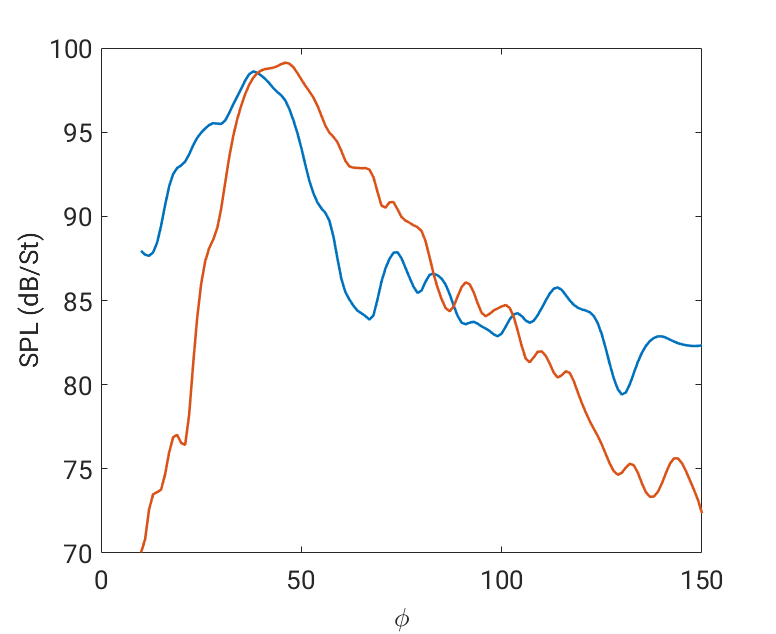} \\
	\caption{The reconstructed far-field sound levels by superposing
	corresponding output modes.}
  \label{fig:projection_farfield}
\end{figure}

Using the coefficients $a_n$, Fig.~\ref{fig:projection_farfield} shows the
reconstructed far-field sound levels obtained by superposing
corresponding output modes as a function of polar angle (red curve).
We compare this to the far-field sound levels from LES (blue curve).  Only 20 input
modes do a fairly good job at predicting both the levels and
directivity of the resulting far-field sound.  Evidently, the input
modes do in fact capture the dynamics in the jet that produce sound.

\begin{figure}
  \begin{tabular}{c}
	\includegraphics[width=12cm]{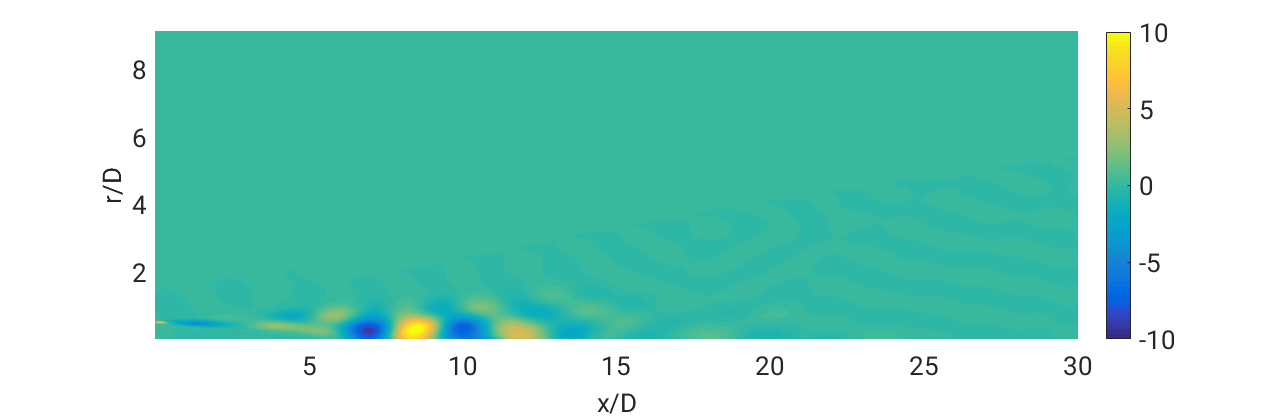} \\
	(a) \\
	\includegraphics[width=12cm]{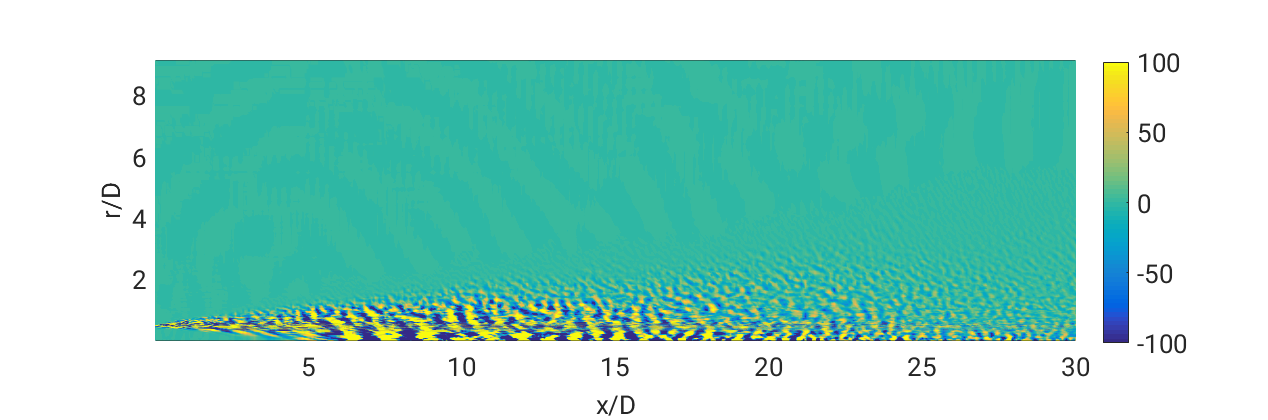} \\
	(b) \\
	\end{tabular}
	\caption{(a) The reconstructed input forcing for I/O analysis
          using the coefficients $a_n$; (b) The real part of the
          Fourier coefficients obtained at frequency $St = 0.59$ for
          one Hann window of the LES data.}
  \label{fig:recon}
\end{figure}

\begin{figure}
  \centering
	\includegraphics[width=10cm]{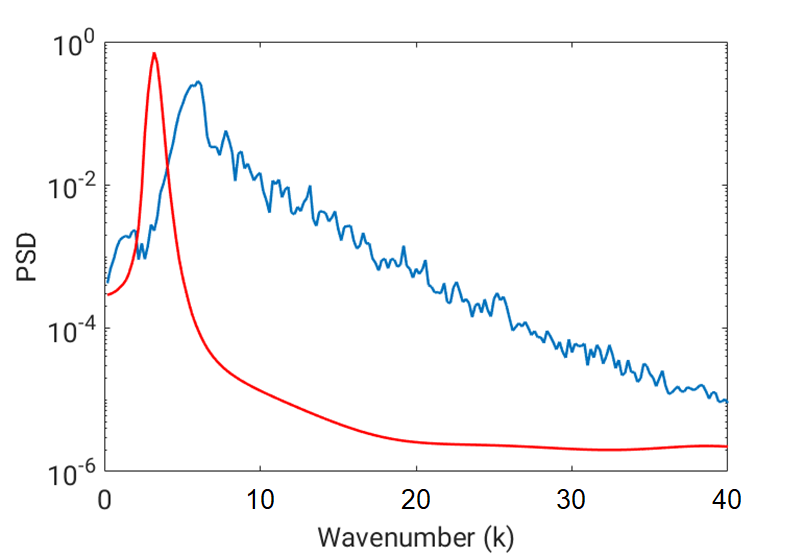} \\
	\caption{The spectra obtained from both the LES data (blue 
	line) and the reconstructed input (red line).}
  \label{fig:wavenumbers}
\end{figure}

In a similar fashion, Fig.~\ref{fig:recon}(a) shows the x-component
of the reconstructed input forcing using the coefficients $a_n$.  We
compare this to a single Hann window of the LES data at the same
frequency, shown in Fig.~\ref{fig:recon}(b) (this is different from
the leading SPOD mode).  First, we note that the amplitude of the
reconstructed input forcing is less than the amplitude of the Fourier
coefficients computed from the LES data, even though the coefficients
$a_n$ are much larger than those obtained previously.  Secondly, while
both the LES data and the reconstructed input show coherence, there
appears to be a mismatch in the wavenumbers between the two.  We
quantify this difference by applying Fourier transforms in the axial
direction to the two fields and average the resulting wavenumber
spectra in the radial direction.  Figure~\ref{fig:wavenumbers} shows
the spectra obtained from both the LES data (blue line) and the
reconstructed input (red line).  Although the LES spectra is more
broadbanded than the reconstructed input, it contains a sharp peak at
$k = 5.3$, from which we can estimate a convection velocity of $u_c =
\omega/k \approx 0.7 u_j$.  The reconstructed input is focused,
however, on a wavenumber that is precisely half that obtained from the
LES.  Its convection velocity, therefore, is $u_c = 1.4 u_j$, and
because of this can emit acoustic radiation directly.  Physically, our
input modes are connected to the radiating tail of a wavepacket
represented in wavenumber space~\cite{freund2001,jordan2013}.

We observe a similar spatial subharmonic relationship between LES
wavepackets and the I/O modes over all frequencies.  The dominant
wavenumber contained in the I/O modes forcing to be precisely half
that extracted from the LES data.  While I/O analysis is linear, it
incorporates all axial wavenumbers simultaneously as a global method,
and so resolves physical waveforms that may contain subharmonic
components.  In the LES, these wavenumbers could couple together in
time as well, through the nonlinearity of the forcing terms.
Alternatively, the stochastic nature of the turbulent fluctuations in
the jet could intermittently align to perturb the coherent structures
contained in the jet turbulence in just the right way so that they
emit sound.  The optimal way to perturb wavepackets observed in the
LES is given by the reconstructed input forcing shown in Fig.
\ref{fig:recon}(a).

\begin{figure}
  \centering
	\includegraphics[width=10cm]{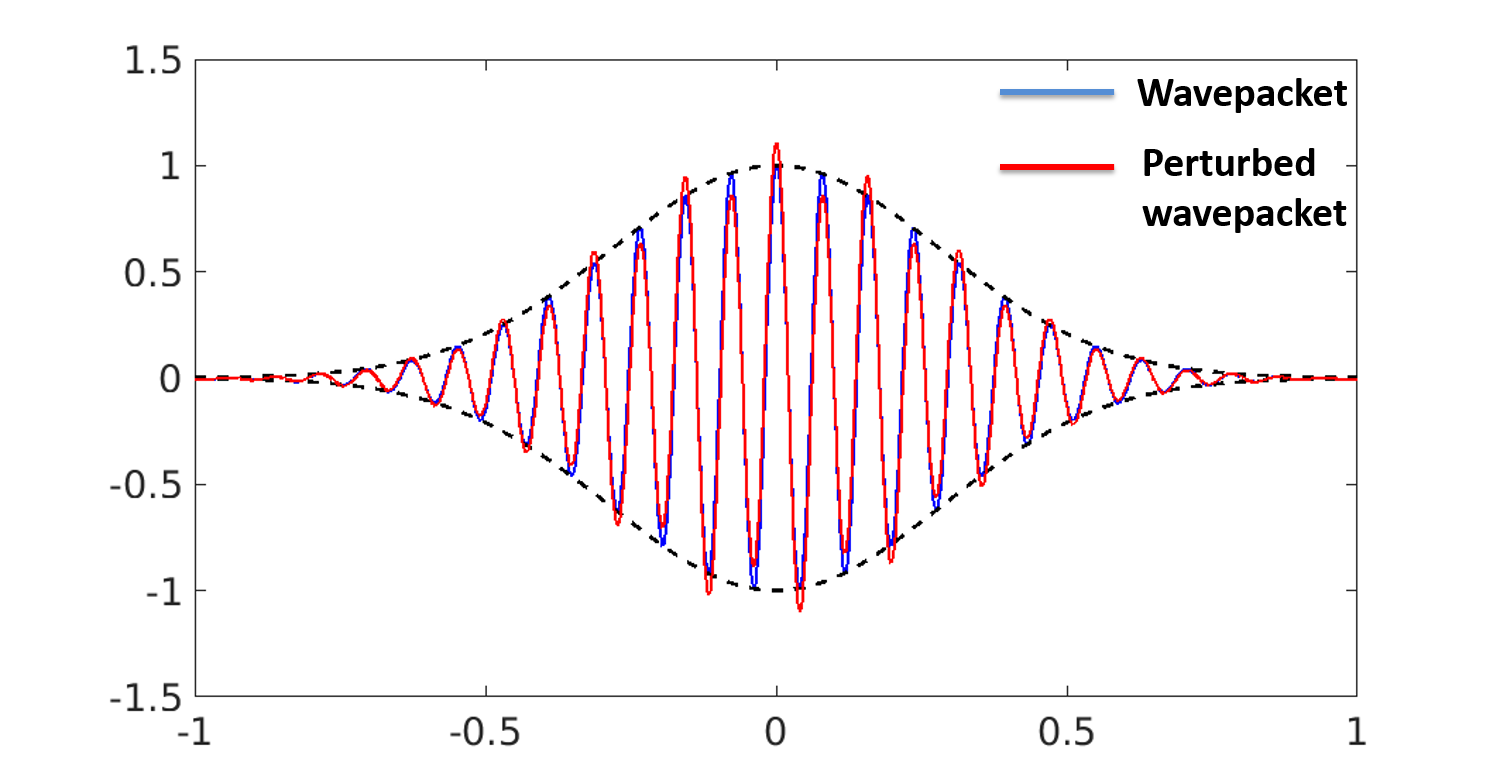} \\
	\caption{A Gaussian wavepacket perturbed by spatially
          subharmonic forcing.}
  \label{fig:badhairday}
\end{figure}

Whether or not the forcing comes from nonlinearity or stochasticity,
the fact that our input modes correspond to small-amplitude, spatially
subharmonic perturbations to dominant instability wavepackets suggests
the physical mechanism for noise generation shown in Fig.
\ref{fig:badhairday}.  In this figure, subharmonic forcing is applied to
a Gaussian wavepacket so that the peak amplitude of every other wave
is slightly amplified or diminished.  In terms of coherent structures
in the jet, this would correspond to slight offsets in their exact
positions.  In other words, in subsonic jets, it is the jitter of the
components of a wavepacket that is responsible for noise.  This type
of jitter is exactly the mechanism observed in compressible mixing by
Wei and Freund, 2006~\cite{wei2006}.  In their study, they applied adjoint-based
optimal control to reduce the sound generated from the mixing layer by
11 dB.  Remarkably, the coherent flow features (large scale vortices)
of the controlled and uncontrolled simulations remained nearly
identical.  A slight jitter in the exact positions of the large scale
flow features accounted for the only difference between the two
simulations.

\begin{figure}
  \centering
	\includegraphics[width=10cm]{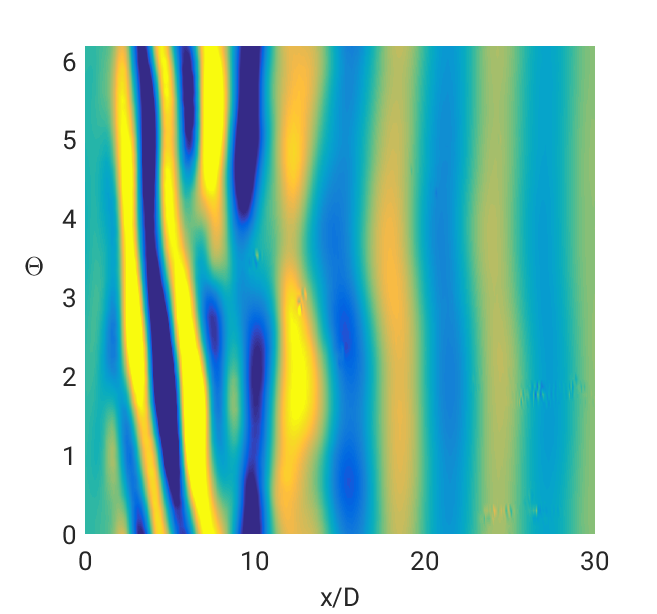} \\
	\caption{The real part of the Fourier coefficients at a
          frequency of $St = 0.2$ on the conical FWH surface.}
  \label{fig:unwrapped}
\end{figure}

To show that the subharmonic relationship between the instability
wavepackets and their acoustic radiation exists in the LES data, we
consider the pressure on the FWH surface used to compute far-field
sound.  Figure~\ref{fig:unwrapped} shows the real part of the Fourier
coefficients at a frequency of $St = 0.2$ on the conical surface,
which has been unwrapped in the azimuthal direction.  Compared to the
the Lighthill source shown in Fig.~\ref{fig:spod}(a), the pressure
field is much more smooth.  Because the FWH surface is so close to the
jet, it captures near-field pressure fluctuations created
aerodynamically by an instability wavepacket.  This wavepacket is
evident on the upstream part of the FWH surface, close to the nozzle
exit.  Since they are associated with aerodynamic effects, these
upstream waves do not radiate to the far-field, however, and die out
exponentially in the radial direction.  The downstream portion of the
FWH surface captures waves with twice the wavelength.  At the same
frequency, these waves do radiate to the far-field and are responsible
for the majority of the sound emitted by the jet at this frequency.
In this particular realization, the upstream aerodynamic wavepacket
upstream appears to be of helical type, whereas the noise radiated is
primarily axisymmetric, further suggesting a separation between
aerodynamic and aeroacoustic components of the jet dynamics.

\section{Conclusions}
\label{sec:conclusions}
In the present study we explore in detail the physics of sound
generation mechanisms in a Mach 0.9 turbulent jet using coherent modes
generated by I/O analysis.  In the spirit of Lighthill's acoustic
analogy, a crucial aspect of our I/O formulation is the spatial
separation of acoustic sources from the far-field noise they produce.
Different from acoustic analogies, however, I/O analysis allows us to
decompose the sources in terms of acoustically important dynamics,
rather than describing them only in terms of statistical
cross-correlations.  Using our method, we find that the optimal
sound-producing dynamics correspond to non-compact coherent sources in
the form of wavepackets.  In terms of their envelope, our wavepackets
agree remarkably well with wavepackets measured in the near-field of
turbulent jets in both laboratory experiments and simulations, and to
those used in semi-empirical models.  Rather than stopping at the
near-field, however, I/O analysis provides a method whereby sound is
traced back to its origins embedded in the jet turbulence.  The
wavepackets shown in this paper were obtained along the lipline of the
jet.

Furthermore, we find that our wavepackets remain similar in shape over
a range of frequencies.  This behavior agrees well with the stochastic
similarity wavepacket model \cite{papamoschou2011}, and may help explain
some of its success.  Rather than being inversely proportional to the
frequency, however, we find that our wavepackets scale approximately
as $St^{-0.5}$.  Clearly, though, the optimal input modes predicted by
I/O analysis are associated with self-similar instability wavepackets,
and can be represented with a model of very low complexity.

For a Mach 0.9 jet, we additionally obtain a spectrum of sub-optimal
modes that have nearly the same gain as the optimal mode.  These
sub-optimal modes follow a pattern that aligns well with modes
obtained from a simple model axial decoherence of a wavepacket.
Furthermore, the acoustic output of the sub-optimal involve
increasingly wide angles of far-field directivity with mode number.
The sub-optimal modes therefore play a key role in noise radiation at
high angles to the downstream jet axis.  Physically, the input modes
from I/O analysis seem to correspond to decoherence modes and at the
same time provide insight into the effect of decoherence on far-field
noise radiation.  It may be interesting to introduce a model of
decoherence in place of axial uncertainty (jitter) in similarity
wavepacket models of jet noise.

Projecting high-fidelity LES data onto the basis of input modes, we
find that the input modes are indeed quite active.  Because they are
filters for aeroacoustically important dynamics, they should, however,
be applied directly to the LES data before any type of averaging or
decomposition is applied.  When this is done, the reconstructed output
acoustics match the levels and peak direction obtained from LES.  This
indicates that the input modes do indeed capture the acoustically
relevant dynamics in the jet.  Moreover, only a limited number of
input modes are necessary to do this, which suggests that these
noise-producing dynamics are low-dimensional in nature.  

It is important to note that, relative to the energetically dominant
motions in the jet, the acoustically important dynamics associated
with our input modes have small energy.  While the envelopes of our
input modes correspond exactly to wavepacket envelopes associated with
instability waves, the wavenumber of our input modes has a subharmonic
relationship with the wavenumbers associated with Kelvin-Helmholtz
instability.  Our input modes are associated with the radiating
supersonic tail of a wavepacket in wavenumber space.  While this tail
is created by the shape of the envelope in physical space, the fact
that the wavenumber of our input modes is exactly half that of the
Kelvin-Helmholtz instability educed by SPOD, suggests a physical
mechanism whereby waves inside the packet are perturbed in an
alternating pattern.  This also explains why we recover the same
envelope as instability wavepackets such as those computed by PSE: our
input modes correspond to small, spatially subharmonic perturbations
to instability waves.  This type of noise-producing pattern has been
observed previously in low Reynolds number simulations of compressible
mixing layers~\cite{wei2006}.

\section*{Acknowledgments}
The simulation presented in this paper was made possible by a grant of
computational resources at the Argonne Leadership Computing Facility
through the INCITE program.  The authors also wish to thank
Prof. M. R. Jovanovi\'c for useful discussions on an early version of this work.

\appendix*
\section{Verification of the linearized FWH solver}
\label{appA}
We start from the permeable surface FWH equation for a stationary source in 
a quiescent medium~\cite{lockard2000,lockard2005,bres2012}, which is given by:
\begin{equation}
  \label{eq:fwh}
  \left( \frac{\partial^2}{\partial t^2} 
  - c_{\infty}^2 \frac{\partial^2}{\partial x_i \partial x_j} \right)  
  \left[ \left( \rho - \rho_{\infty} \right) H \left( S \right) \right] = 
  \frac{\partial}{\partial t} \left[ Q_n \delta \left( S \right) \right] -
  \frac{\partial}{\partial x_i} \left[ F_i \delta \left( S \right) \right] -
  \frac{\partial^2}{\partial x_i \partial x_j} 
	\left[ T_{ij} H \left( S \right) \right], 
\end{equation}
where the monopole source terms $Q_n$, dipole source terms $F_i$, and 
the Lighthill stress tensor $T_{ij}$ are respectively defined as:
\begin{subequations}
\begin{align}
  Q_n &=& \rho u_i \hat{n}_i, \\
  F_i &=& \left( P_{ij} + \rho u_i u_j \right) \hat{n}_j, \\
  T_{ij} &=& \rho u_i u_j + P_{ij} - c_{\infty}^2 
	\left( \rho - \rho_{\infty} \right) \delta_{ij}.	
\end{align}
\label{eq:sources}
\end{subequations}
In the above equations the function $S$ defines the surface so that the 
solution to Eq.~\eqref{eq:fwh} is sought outside of the surface $S = 0$, 
and in this regard the Heaviside function $H$ becomes unity for $S > 0$ and 
zero for $S < 0$.  Furthermore, $\hat{n}_i$ represents a unit outward normal 
vector to the surface $S = 0$, and $u_i$ is the local fluid velocities on 
the surface $S = 0$.  The total density is given by $\rho$ while the ambient 
properties are denoted by the subscript $\infty$.  Then, perturbation 
properties may be distinguished by the superscript $'$ such as the density 
perturbation is written as $\rho' = \rho - \rho_{\infty}$.  Note in the 
Lighthill stress tensor the compressive stress tensor $P_{ij}$ is defined as 
$P_{ij} = \left( p - p_{\infty} \right) \delta_{ij}$ after neglecting the 
viscous term.

To perform the Fourier analysis, Eq.~\eqref{eq:fwh} may be re-written in 
more convenient form as:
\begin{equation}
  \left( \frac{\partial^2}{\partial x_j \partial x_j} + k^2 \right) 
  \left[ \left( c_{\infty}^2 \tilde{\rho}' \right) H \left( S \right) \right] = 
  - \mathrm{i}\omega \tilde{Q}_n \delta \left( S \right) +
  \frac{\partial}{\partial x_i} 
	\left[ \tilde{F}_i \delta \left( S \right) \right] -
  \frac{\partial^2}{\partial x_i \partial x_j} 
	\left[ \tilde{T}_{ij} H \left( S \right) \right], \\[4pt]
\label{eq:fwh2}
\end{equation}
where $k$ is the wavenumber given by $k = \omega / c_{\infty}$.  In fact, we 
may replace the term $c_{\infty}^2 \tilde{\rho}'$ by the pressure perturbation 
$p'$ and rewrite Eq.~\eqref{eq:fwh2} in a pressure-based form since the 
density perturbations are small outside of the source region~\cite{spalart2009}. 
The integral solution is then given by: 
\begin{equation}
  \tilde{p}' \left( \bm{x},\omega \right) = -
	\int_S \tilde{F}_i\left( \bm{y},\omega \right) 
	\frac{\partial G \left( \bm{x},\bm{y} \right)}
	{\partial y_i} dS -
	\int_S \mathrm{i}\omega \tilde{Q}_n\left( \bm{y},\omega \right)			
	G\left( \bm{x},\bm{y} \right) dS,
  \label{eq:fwh_soln}
\end{equation}
assuming that the volume-distributed source terms are negligibly small.  
Here, the three-dimensional free-space Green's function $G$ is given by:
\begin{equation}
  G\left( \bm{x},\bm{y} \right) = 
	\frac{-e^{-\mathrm{i}kr}}{4 \pi r}
	\label{eq:green}
\end{equation}
where $r$ represents the distance between the source $\bm{y}$ and the observer 
$\bm{x}$ such that $r = | \bm{x} - \bm{y} |$.  Using the chain rule, the 
derivative of the Green's function with respect to the source $\bm{y}$ may 
be then evaluated as
\begin{equation}
  \frac{\partial G}{\partial y_i} =
	\frac{\partial G}{\partial r}
	\frac{\partial r}{\partial y_i},
\end{equation} 
and in cylindrical coordinates we use here, 
$\frac{\partial r}{\partial y_i}$ is simply:
\begin{equation}
  \frac{\partial r}{\partial y_i} =
	-\hat{r}_i =
	\frac{x_i - y_i}{|x_i - y_i|}.
\end{equation} 

Now we linearize the integral solution in Eq.~\eqref{eq:fwh_soln} to 
implement a FWH solver within I/O analysis framework.  Since 
outside of the hydrodynamic region $r/D > 3$, local properties are mostly 
perturbation properties, we may neglect the second-order terms in the dipole 
source term $F_i$ and rewrite it as:
\begin{equation}
  F_i = P_{ij}\hat{n}_j.
\end{equation} 
The linear monopole source term $Q_n$ remains unchanged as it is in 
Eq.~\eqref{eq:sources}.

For a FWH projection surface, we select a straight cylinder with radius 
$r/D = 6$ whose axis lies along the jet centerline.  The projection surface 
is thus located much closer to the jet than the Kirchhoff surface in earlier 
study~\cite{jeun2016} was, and yet it is far enough to enclose all noise 
sources.  The cylindrical projection surface extends from $x/D = -5$ to 
$x/D = 35$ axially, and an outflow disk is neither closed nor equipped with 
end-caps.  We still expect that spurious modes will not contaminate the 
acoustic response since the projection surface is sufficiently long and 
lean, and this assumption is tested for the following two cases.    

\subsection{Case 1: Monopole}
In this appendix we validate the FWH formulation that are directly 
implemented within I/O analysis framework as a linear operator 
inside the matrix $C$, by testing a simple case for which an analytic 
solution exists.  We consider a point monopole source located at the 
origin.  The far-field pressure at $r$ radiated from the source is exactly 
written as:
\begin{equation}
  \frac{p_{exact}(r,\omega)}{p_{ref}} = 
	\frac{e^{-\mathrm{i} k r}}{4 \pi r},
	\label{eq:monopole}
\end{equation} 
using the free-space Green's function given in Eq.~\eqref{eq:green}.  
Being consistent with notations used in the main sections, the 
distance between the source $\bm{y}$ and the observer $\bm{x}$ is given by 
$r = \left|\bm{x} - \bm{y}\right|$.  The reference pressure is set to be 
$p_{ref} = 10^{-6}$.  Furthermore, $\omega$ represents the frequency and $k$ 
denotes the wavenumber such that $k = 2 \pi / \lambda$.  The wavelength 
$\lambda$ is given as $1D$ and the speed of sound $c = 1/1.5 = 0.6667$. 
Note that sound radiated from a monopole is omnidirectional so the exact 
solution above is independent of an azimuthal angle $\theta$.  The velocity 
field of a monopole source is obtained by substituting this into the 
conservation of momentum, and then used in the linearized FWH formulation 
given in Eq.~\eqref{eq:fwh_soln}.
 
\begin{figure}
  \centering
	\begin{tabular}{cc}
	  \includegraphics[width=8cm]{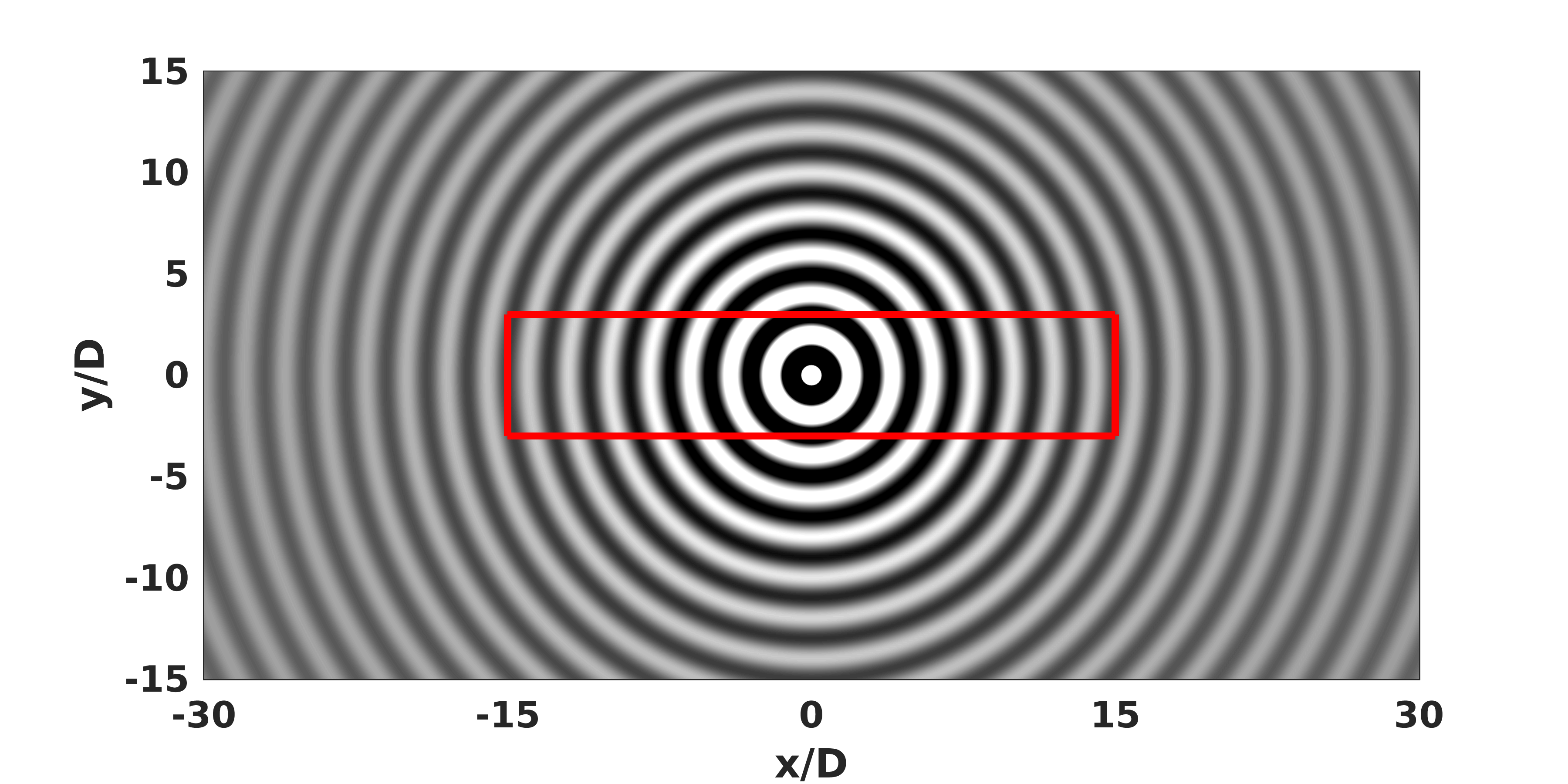} &
	  \includegraphics[width=4cm]{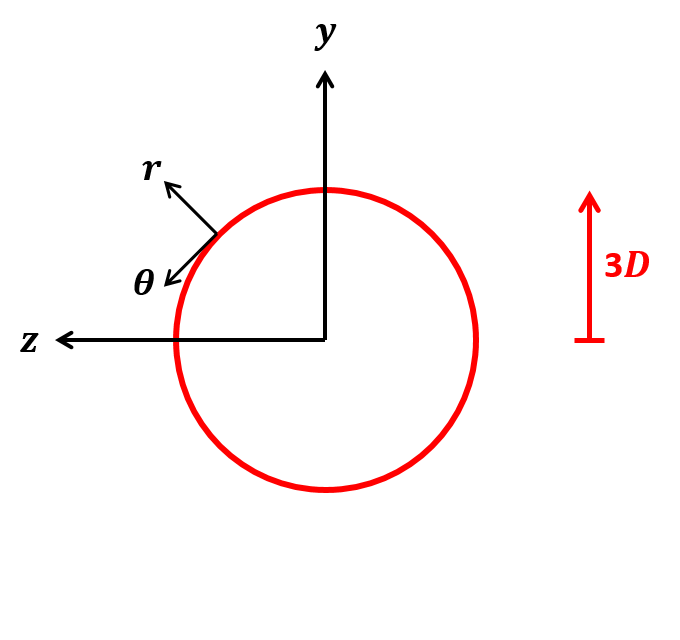} \\
	  (a) & (b) \\
	\end{tabular}
	\caption{(a) A straight cylindrical FWH surface (a red-outlined 
	rectangle) placed in a monopole field on $xy$-plane cross-section 
	at $z = 0$.  The length of the projection surface is $30D$, and 
	the diameter of its cross-section is $3D$.  (b) A cross-sectional 
	view of the projection surface on $yz$-plane is zoomed-in. 
	The azimuthal angle $\theta$ is measured in a counter-clockwise 
	direction.}
  \label{fig:fwh_surface_monopole} 
\end{figure}

\begin{figure}
  \begin{center}
  \begin{tabular}{cc}
	  \includegraphics[width=6.5cm]{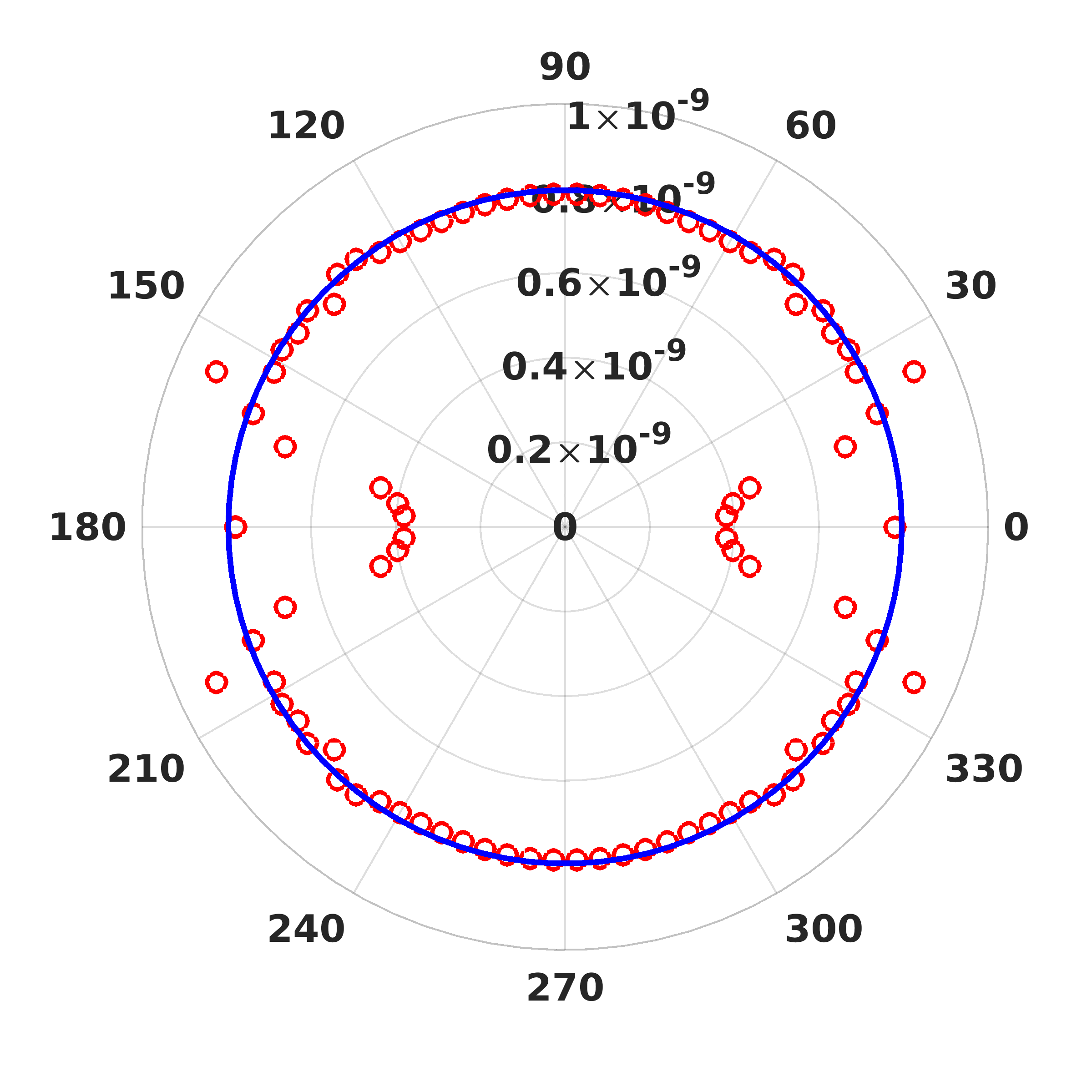} &
    \includegraphics[width=6.5cm]{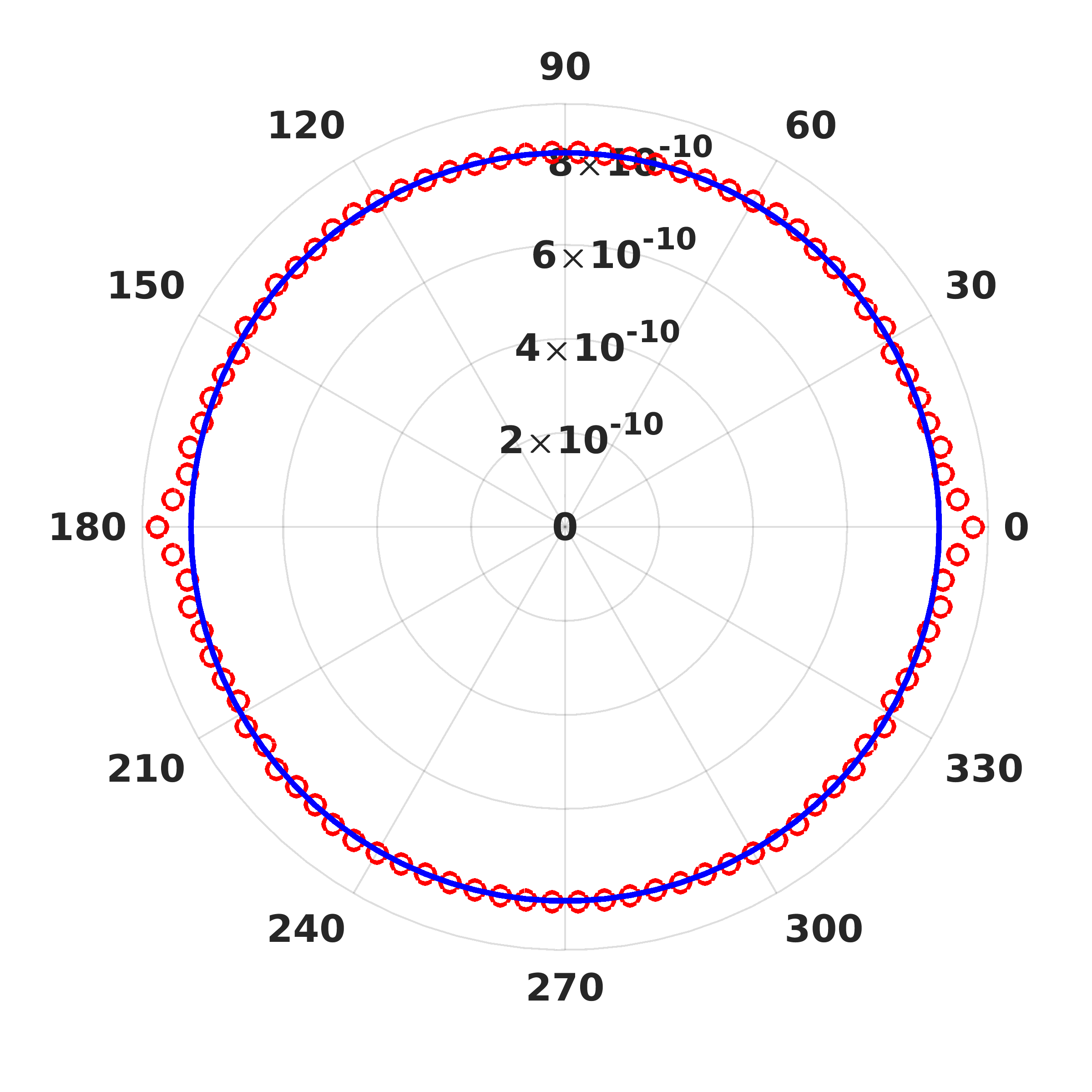} \\
    (a) Open outflow disks & (b) Closed outflow disks \\
  \end{tabular}
  \end{center}
  \caption{Acoustic far-field measured at 100 diameters away 
	from a monopole centered at the origin using the projection 
	surfaces whose outflow disks are (a) left open and (b) closed 
	with end-caps, respectively.  In each figure blue solid line 
	represents the exact solution, and red markers are computed 
	using a FWH formulation implemented inside I/O 
	analysis framework.}
  \label{fig:monopole_field}
\end{figure}

We place a straight cylindrical FWH projection surface so that its axis 
lies along the $x$-axis as shown in Fig.~\ref{fig:fwh_surface_monopole}.  
We choose the projection surface with radius $R_s = 3D$, which extends from 
$x = -15D$ to $15D$ so that its length $L_s = 30D$.  The number of grid 
points in the axial, radial, and azimuthal directions are given by 
$N_x = 481$, $N_r = 125$, and $N_\theta = 64$, respectively.  The grid are 
uniform in the streamwise and azimuthal directions but stretched in the 
radial direction.  Meanwhile, to assess the effects of open outflow 
disk, we test FWH surfaces equipped with and without 
end-caps~\cite{shur2005}, when computing the far-field pressure fields.  The 
results are summarized in Fig.~\ref{fig:monopole_field}. 

In Fig.~\ref{fig:monopole_field} the blue solid lines correspond to the 
exact far-field sound pressure, whereas red symbols represent solutions 
computed using the FWH solver implemented as a linear operator.  If they 
are left open, the numerical solutions at small radiation angles 
($-20^\circ < \phi < 20^\circ$ and $160^\circ < \phi < 220^\circ$) deviate 
from the exact solution but agree fairly well elsewhere as shown in 
Fig.~\ref{fig:monopole_field}(a).  In contrast, by closing the outflow 
disks the projection surface, Fig.~\ref{fig:monopole_field}(b) reproduces 
the exact solution at almost all observer angles.    

\begin{figure}
  \centering
	\includegraphics[width=10cm]{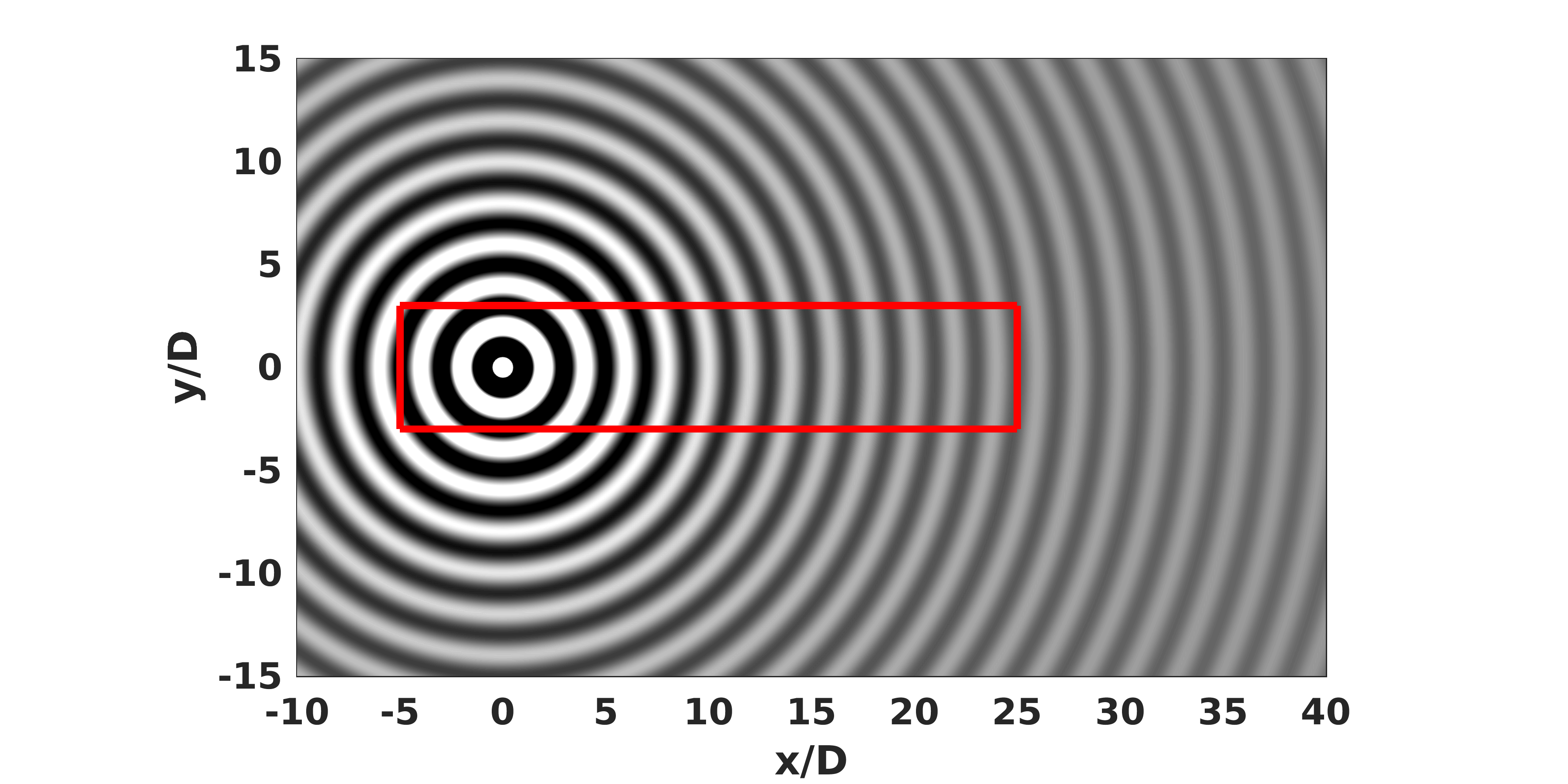} \\
	\caption{A straight cylindrical FWH surface (a red-outlined 
	rectangle) is now asymmetric about a point monopole centered 
	at the origin.  The length and diameter of the projection 
	surface remain unchanged as in the previous case.}
  \label{fig:fwh_monopole_5D} 
\end{figure}

\begin{figure}
  \begin{center}
  \begin{tabular}{cc}
	  \includegraphics[width=6.5cm]{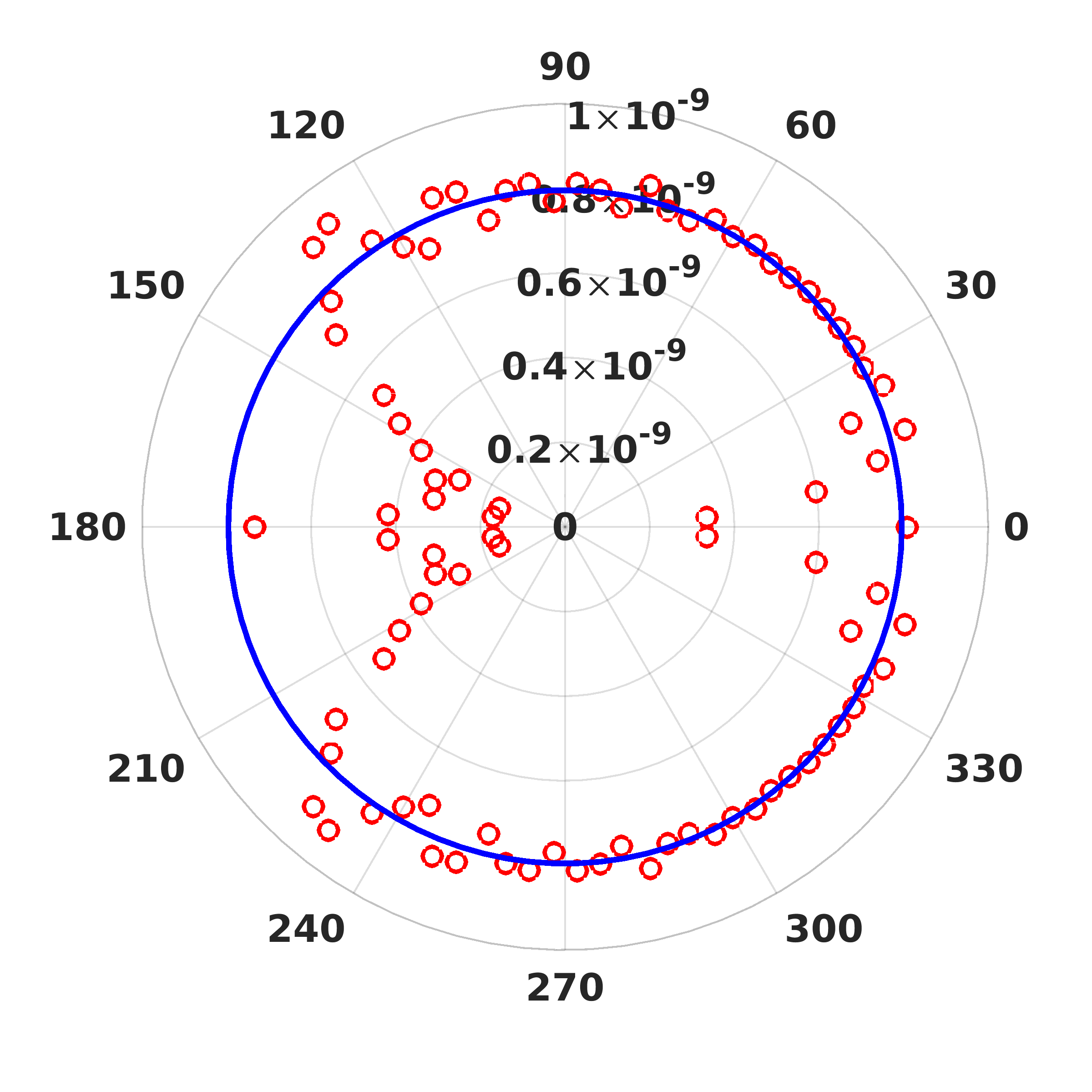} &
    \includegraphics[width=6.5cm]{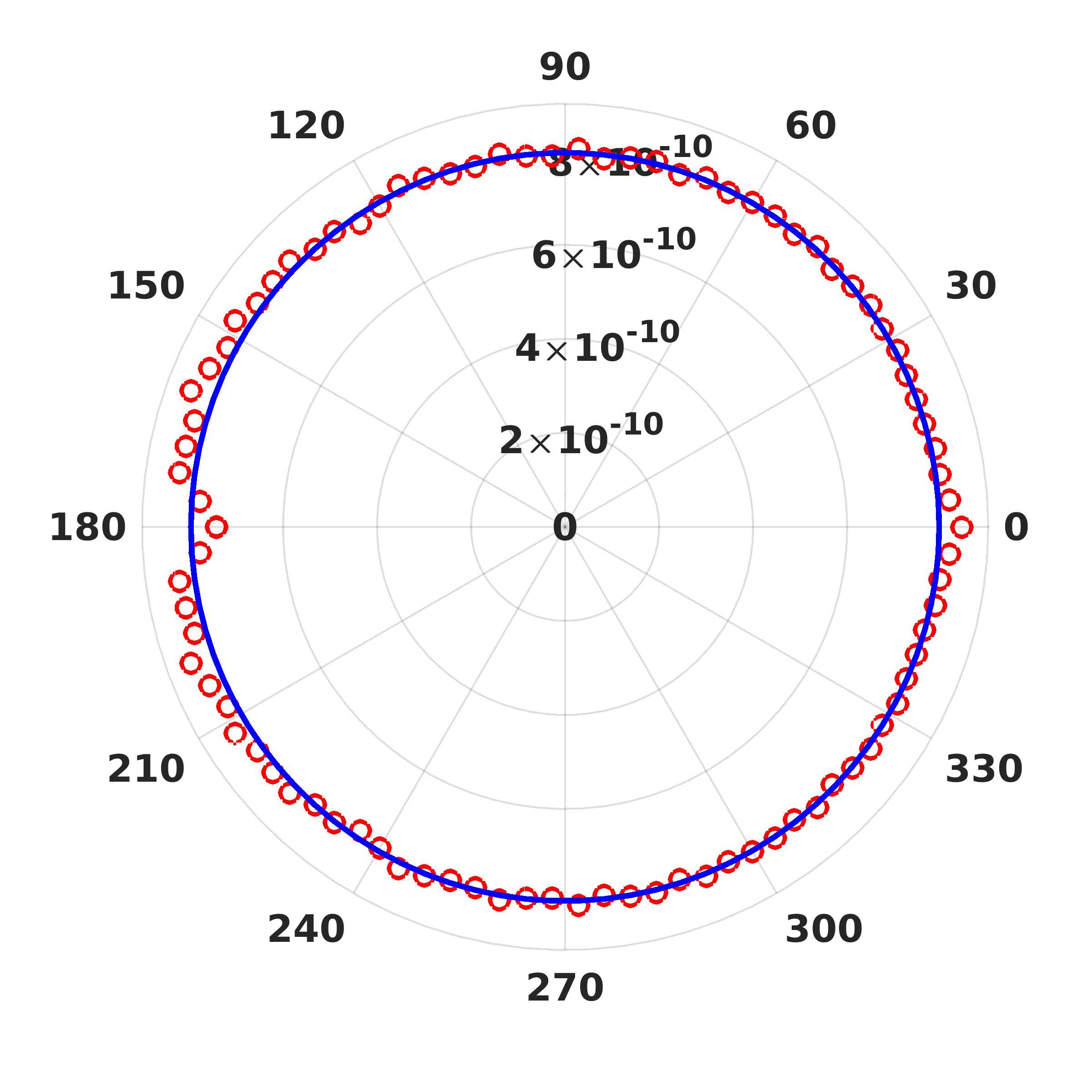} \\
    (a) Open outflow disks & (b) Closed outflow disks \\
  \end{tabular}
  \end{center}
  \caption{Acoustic far-field measured at 100 diameters away 
	from a monopole placed at 5 diameters away from the left-end 
	of the FWH projection surface.  Pressures (red markers) 
	are predicted using (a) open and (b) closed outflow disks, 
	respectively, and compared to the analytic solution 
	represented by blue solid lines.}
  \label{fig:monopole_field2}
\end{figure}

Now, a monopole source is still placed at the origin, but the FWH 
projection is centered at $x/D = 10D$ to mimic the flow configuration of 
turbulent jets we consider in this dissertation.  Considering that in 
I/O analysis while turbulent jets enter to a quiescent fluid at 
$x/D = 0$, the numerical domain extends from $x/D = -10$ to 40, new 
projection surface is not symmetric any longer about an acoustic source in 
streamwise direction.  The schematic view of a point monopole source located 
at the origin and an asymmetric projection surface is given in 
Fig.~\ref{fig:fwh_monopole_5D}.  Under this condition, the linearized 
FWH solver recovers the far-field pressure very closely to the analytic 
solution at radiation angles of our interests as shown in 
Fig.~\ref{fig:monopole_field2} regardless of the types of outflow disks.  
We lose the symmetry of the pressure field and compromise some accuracy at 
small observer angles, but the results are still within acceptable accuracy.  

\subsection{Case 2: Dipole}
\begin{figure}
  \centering
	\includegraphics[width=10cm]{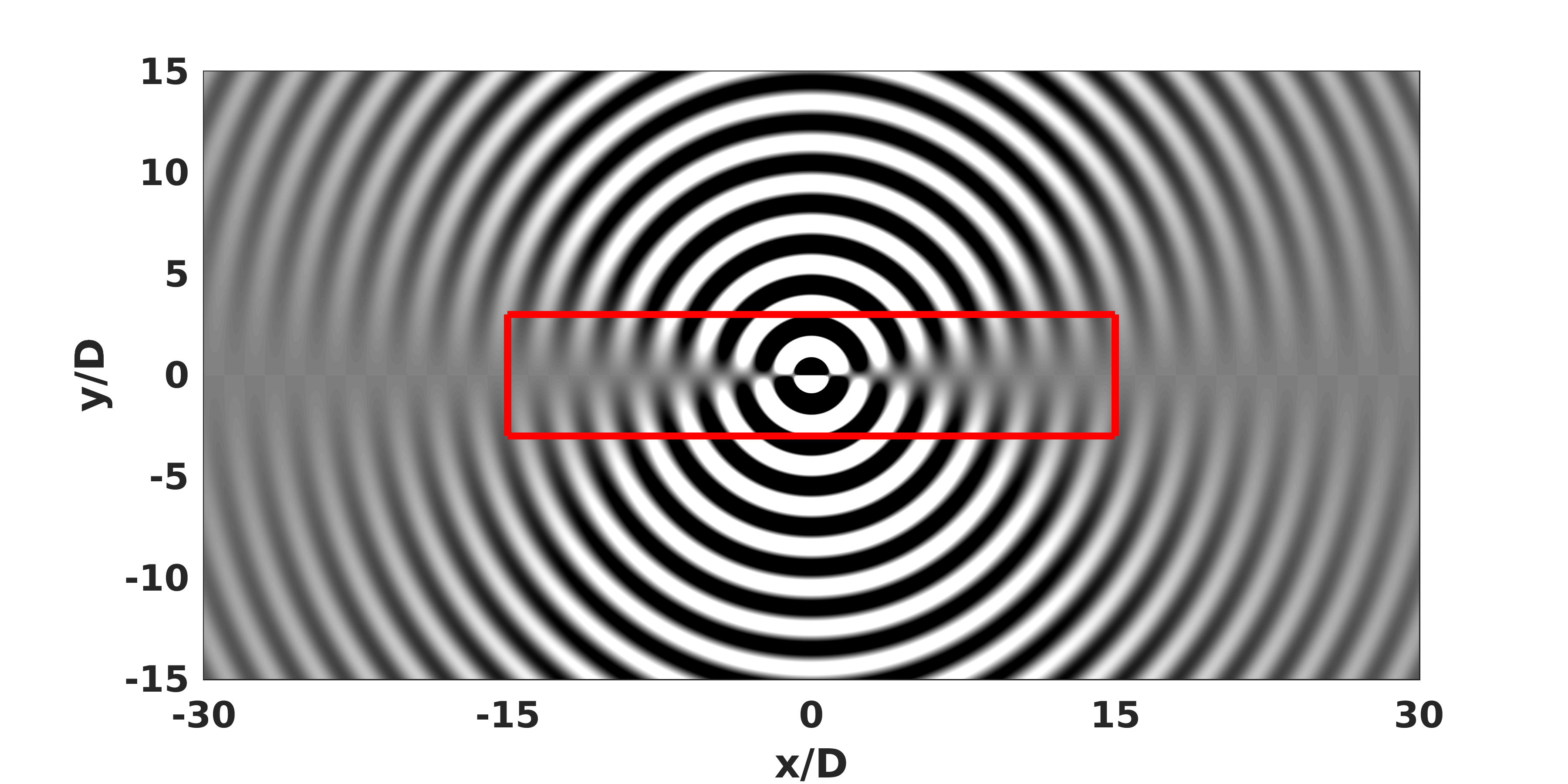}
  \caption{A straight cylindrical FWH surface for sound 
	radiation of a dipole centered at the origin.}
  \label{fig:fwh_surface_dipole} 
\end{figure}

As seen in Eq.~\ref{eq:fwh_soln}, the far-field sound is represented 
by surface integrals of monopole and dipole sources.  In this sense we also 
test the linearized FWH solver with sound radiation from a dipole source.  
We consider a dipole source located along the $y$-axis as shown in 
Fig.~\ref{fig:fwh_surface_dipole}.  The analytic solution for a pressure 
field is then derived by differentiating the free-space Green's 
function~\eqref{eq:green} in $y$ such as:
\begin{equation}
  \frac{p_{exact}(r,\omega)}{p_{ref}} 
	= \frac{\partial G}{\partial y}
	= \frac{\partial G}{\partial r} \frac{\partial r}{\partial y}
	= \frac{e^{-\mathrm{i} k r}}{4 \pi r}
	\left(-\mathrm{i} k - \frac{1}{r} \right)
	\frac{\partial r}{\partial y}.
	\label{eq:dipole}
\end{equation} 
In cylindrical coordinates we use here, $\frac{\partial r}{\partial y}$ is 
conveniently computed as $\frac{y}{r}$.  The reference pressure remain 
unchanged as $p_{ref} = 10^{-6}$.  The wavelength $\lambda$ is still given 
as $1D$, but at this time we change the speed of sound to $c = 1/0.9 = 1.111$. 

\begin{figure}
  \begin{center}
  \begin{tabular}{cc}
	  \includegraphics[width=6.5cm]{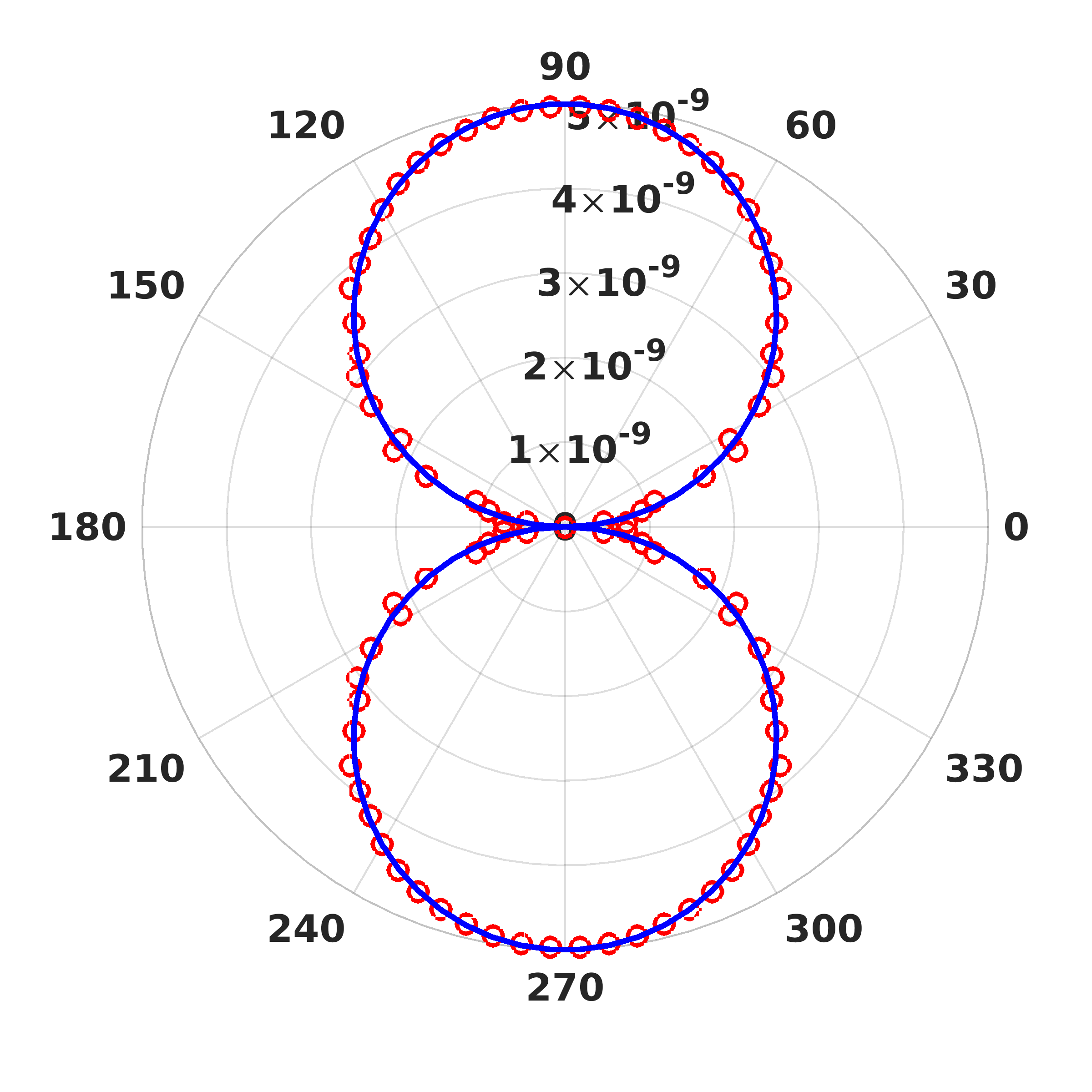} &
    \includegraphics[width=6.5cm]{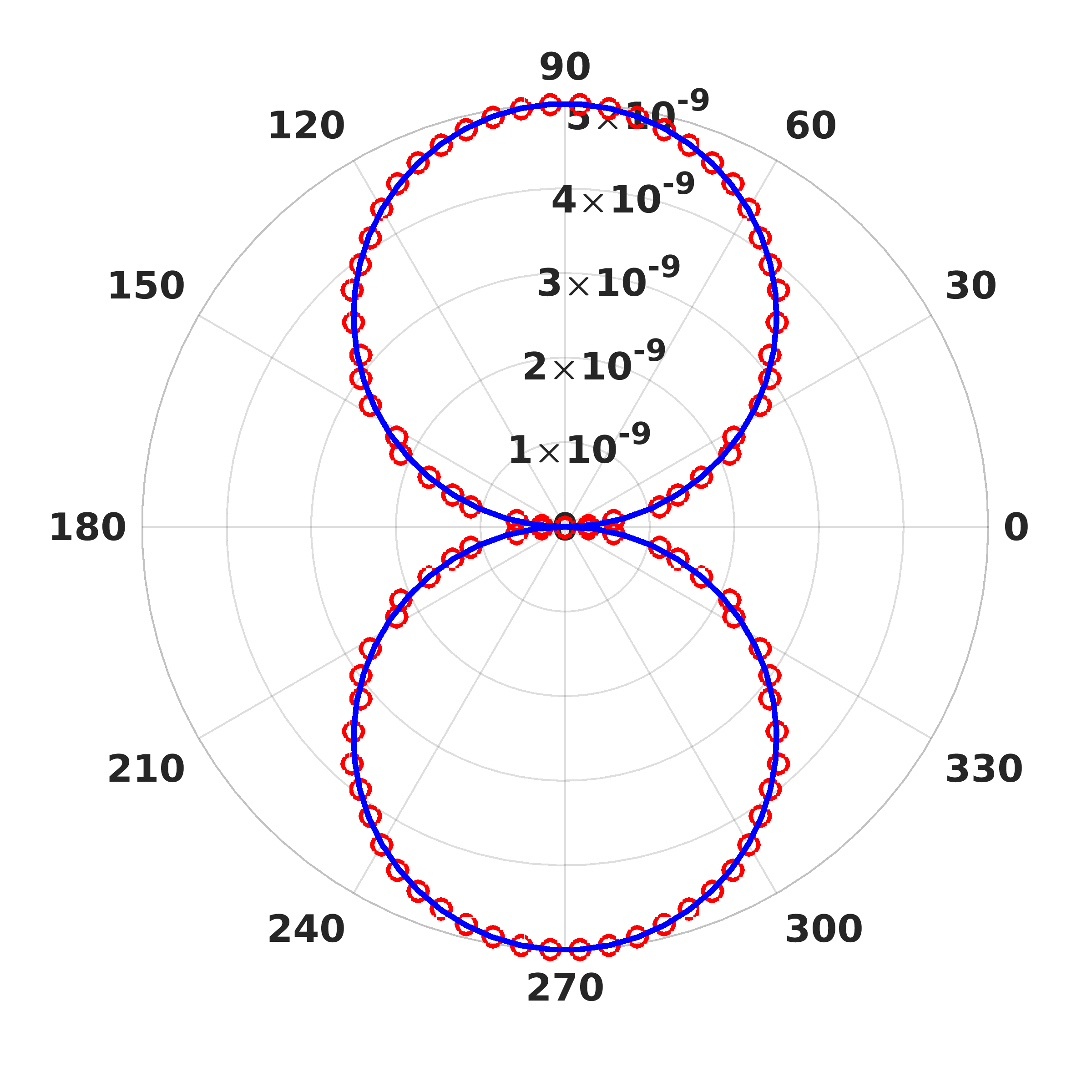} \\
    (a) Open outflow disks & (b) Closed outflow disks \\
  \end{tabular}
  \end{center}
  \caption{Dipole field is measured in terms of pressure at 
	100 diameters away from the origin.  The outflow disks of 
	the projection surfaces are (a) left open and (b) closed with 
	an end-cap.  In each figure blue solid line represents the 
	exact solution, and red markers are computed using a FWH 
	formulation implemented inside I/O analysis 
	framework.}
  \label{fig:dipole_field}
\end{figure}

We use the same FWH projection surface, which was used in the previous 
section to test sound radiation from a monopole acoustic source.  Again, a 
cylindrical projection surface may be either open or closed at 
$x/D = 30$.  The dipole fields obtained using different outflow disk options on the 
projection surfaces are compared to the exact solution (denoted by blue 
solid lines) in Fig.~\ref{fig:dipole_field}.  The far-field acoustic 
predictions in both cases show reasonably good agreements with the exact 
solution except for the observers very close to the $x$-axis.  

\begin{figure}
  \begin{center}
  \begin{tabular}{cc}
	  \includegraphics[width=6.5cm]{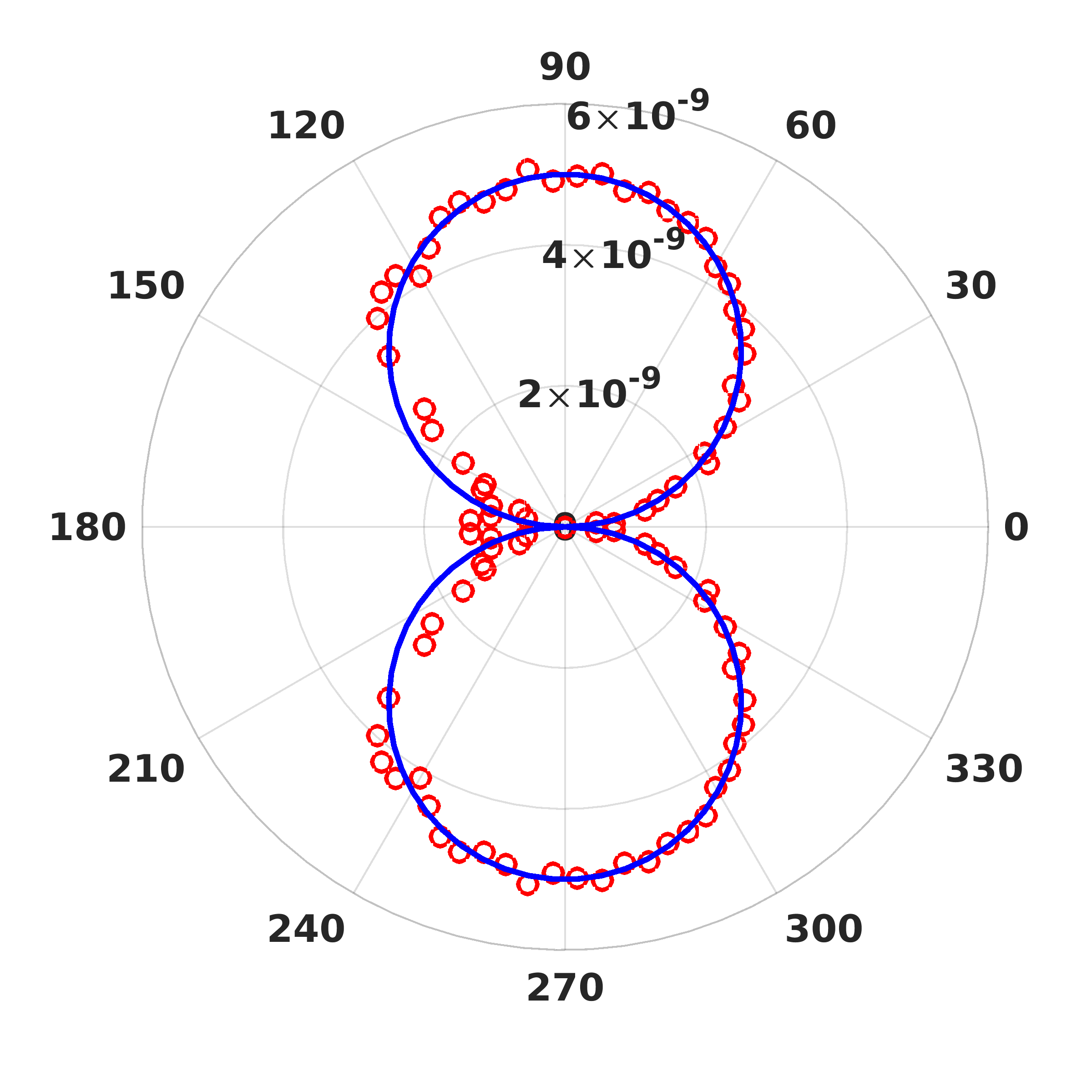} &
    \includegraphics[width=6.5cm]{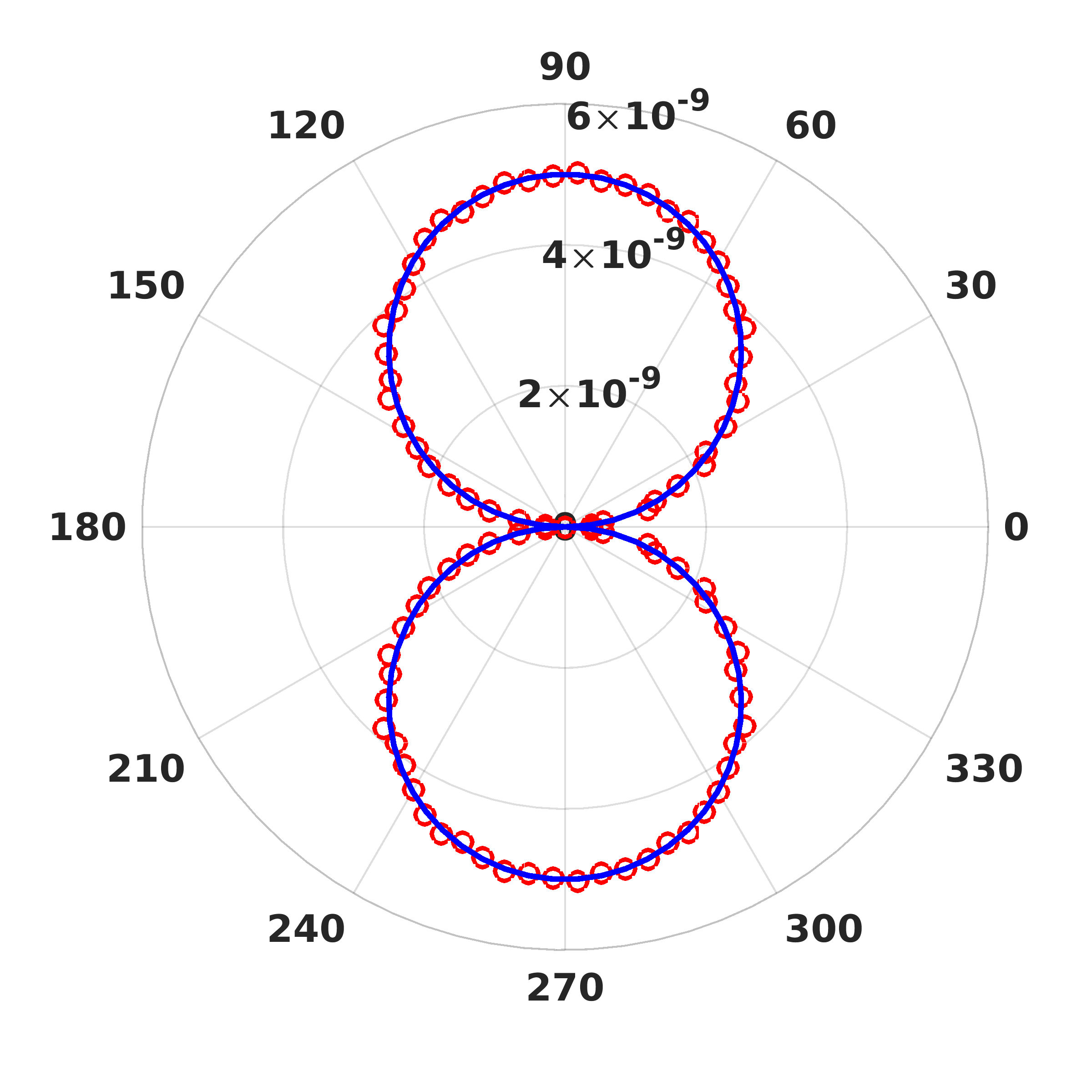} \\
    (a) Open outflow disks & (b) Closed outflow disks \\
  \end{tabular}
  \end{center}
  \caption{Acoustic far-field measured at 100 diameters away 
	from a dipole placed at 5 diameters away from the left-end 
	of the FWH projection surface along $y$-axis.  Pressures 
	(red markers) are predicted using (a) open and (b) closed 
	outflow disks, respectively, and compared to the analytic 
	solution represented by blue solid lines.  Equipped with 
	end-caps at both ends of the projection surface, the 
	predictions are significantly enhanced.}
  \label{fig:dipole_field2}
\end{figure}

Similarly in the case of a point monopole, we investigate the effect of 
asymmetric projection surface about a point dipole centered at the origin.  
In Fig.~\ref{fig:dipole_field2}, the predictions by the linearized FWH 
solver are in good agreement with the analytic solution.  Errors at small 
angles are decreased significantly with end-caps at $x/D = -5$ and 
$x/D = 25$ even with the asymmetric projection surface. 

In sum, for turbulent jets, regions of low radiation angles correspond to 
regions that are close to the jet centerline, and we do not expect much 
output sound there.  Based on the tests given in the previous and the 
present sections, we thus conclude that for a long and lean enough 
cylindrical projection surface, the FWH formulation implemented inside the 
input matrix $C$ would work even without end-cap treatments.  From this, a 
linearized FWH solver employs an open FWH projection surface, for 
convenience.

\end{document}